\documentclass{aa}  

\usepackage{graphicx}
\usepackage{txfonts}
\usepackage[labelfont=bf]{caption}
\usepackage{subcaption}
\usepackage{gensymb}
\usepackage[usenames, dvipsnames]{xcolor}
\usepackage{longtable}
\usepackage{colortbl}
\usepackage{lscape}
\usepackage{xfrac} 
\usepackage{float}
\usepackage{hyperref}
\hypersetup{
    colorlinks,
    citecolor=blue,
    linkcolor=blue
    }

\newcommand{\ratio}{\eta}
\newcommand{\numberOfGRGsOei}{2060 }
\newcommand{\numberOfGRGsTotal}{3341 }

\begin{document}

\title{Measuring the giant radio galaxy length distribution with the LoTSS}

   \author{Martijn S.S.L. Oei
          \inst{1}
          \and
          Reinout J. van Weeren
          \inst{1}
          \and
          Aivin R.D.J.G.I.B. Gast
          \inst{2}
          \and
          Andrea Botteon
          \inst{1}
          \and
          Martin J. Hardcastle
          \inst{3}
          \and
          Pratik Dabhade
          \inst{4}
          \and
          Tim W. Shimwell
          \inst{1}
          \and
          Huub J.A. R\"ottgering
          \inst{1}
          \and
          Alexander Drabent
          \inst{5}
          }

   \institute{Leiden Observatory, Leiden University, Niels Bohrweg 2, NL-2300 RA Leiden, The Netherlands\\
              \email{\url{oei@strw.leidenuniv.nl}}
              \and
              Somerville College, University of Oxford, Woodstock Road, Oxford OX2 6HD, United Kingdom
              \and
              Centre for Astrophysics Research, University of Hertfordshire, College Lane, Hatfield AL10 9AB, United Kingdom
              \and
              Observatoire de Paris, LERMA, Coll\`ege de France, CNRS, PSL University, Sorbonne University, 75014 Paris, France
              \and
              Th\"uringer Landessternwarte, Sternwarte 5, 07778 Tautenburg, Germany
             }

   \date{\today}

 
\abstract
{
Many massive galaxies launch jets from the accretion disk of their central black hole, but only ${\sim}10^3$ instances are known in which the associated outflows form giant radio galaxies (GRGs, or giants): luminous structures of megaparsec extent that consist of atomic nuclei, relativistic electrons, and magnetic fields.
Large samples are imperative to understanding the enigmatic growth of giants, and recent systematic searches in homogeneous surveys constitute a promising development.
For the first time, it is possible to perform meaningful precision statistics with GRG lengths, but a framework to do so is missing.
}
{
We measured the intrinsic GRG length distribution by combining a novel statistical framework with a LOFAR Two-metre Sky Survey (LoTSS) sample of freshly discovered giants.
In turn, this allowed us to answer an array of questions on giants.
For example, we can now assess how rare a 5 Mpc giant is compared with one of 1 Mpc, and how much larger --- given a projected length --- the corresponding intrinsic length is expected to be.
Notably, we can now also infer the GRG number density in the Local Universe.
}
{
We assumed the intrinsic GRG length distribution to be Paretian (i.e. of power-law form) with tail index $\xi$, and predicted the observed distribution by modelling projection and selection effects.
To infer $\xi$, we also systematically searched the LoTSS for hitherto unknown giants and compiled the largest catalogue of giants to date.
}
{
We show that if intrinsic GRG lengths are Pareto distributed with index $\xi$, then projected GRG lengths are also Pareto distributed with index $\xi$.
Selection effects induce curvature in the observed projected GRG length distribution: angular length selection flattens it towards the lower end, while surface brightness selection steepens it towards the higher end.
We explicitly derived a GRG's posterior over intrinsic lengths given its projected length, laying bare the $\xi$ dependence.
We also discovered \numberOfGRGsOei giants within LoTSS DR2 pipeline products; our sample more than doubles the known population.
Spectacular discoveries include the largest, second-largest, and fourth-largest GRG known ($l_\mathrm{p} = 5.1\ \mathrm{Mpc}$, $l_\mathrm{p} = 5.0\ \mathrm{Mpc}$, and $l_\mathrm{p} = 4.8\ \mathrm{Mpc}$), the largest GRG known hosted by a spiral galaxy ($l_\mathrm{p} = 2.5\ \mathrm{Mpc}$), and the largest secure GRG known beyond redshift 1 ($l_\mathrm{p} = 3.9\ \mathrm{Mpc}$).
We increase the number of known giants whose angular length exceeds that of the Moon from 10 to 23; among the discoveries is the angularly largest known radio galaxy in the Northern Sky, which is also the angularly largest known GRG ($\phi = 2\degree$).
Combining theory and data, we determined that intrinsic GRG lengths are well described by a Pareto distribution, and measured the index $\xi = -3.5 \pm 0.5$.
This implies that, given its projected length, a GRG's intrinsic length is expected to be just $15\%$ larger.
Finally, we determined the comoving number density of giants in the Local Universe to be $n_\mathrm{GRG} = 5 \pm 2\ \left(100\ \mathrm{Mpc}\right)^{-3}$.
}
{
We developed a practical mathematical framework that elucidates the statistics of giant radio galaxy lengths.
Through a LoTSS search, we also discovered \numberOfGRGsOei new giants.
By combining both advances, we determined that intrinsic GRG lengths are well described by a Pareto distribution with index $\xi = -3.5 \pm 0.5$, and that giants are truly rare in a cosmological sense: most clusters and filaments of the Cosmic Web are not currently home to a giant.
Thus, our work yields new observational constraints for analytical models and simulations featuring radio galaxy growth.
}

\maketitle

\section{Introduction}
When gas, dust, and stars accrete onto a supermassive black hole (SMBH) in the centre of a galaxy, collimated jets arise along the Kerr rotation axis that blast some of the infalling material into the intergalactic medium (IGM) \citep[e.g.][]{Blandford11974}.
In this process, the ejecta dissolve into a relativistic plasma that drags along a magnetic field and glows in synchrotron light.
The resulting luminous structure is called a radio galaxy (RG); the central black hole that has generated it an active galactic nucleus (AGN).

It is increasingly clear that RGs and their AGN play an important role in galaxy evolution and cosmology.
By heating the interstellar medium (ISM) or even expelling it from their host galaxies through galactic superwinds, AGN quench star formation \citep[e.g.][]{DiMatteo12005}.
\citet{Beckmann12017} have shown that AGN-induced star formation quenching is most pronounced in massive galaxies.
There is also compelling evidence that the accompanying RGs provide the energy necessary to stop \citep[e.g.][]{McNamara12012, Yang12019} bremsstrahlung-mediated cooling flows \citep{Fabian11984} in clusters of galaxies.
In the absence of cooling flows, the intra-cluster medium (ICM) remains dilute and hot, and galaxies in the centres of clusters are denied infalling gas that could otherwise reignite star formation.
Cosmological simulations that include this RG feedback indeed resolve \citep[e.g.][]{Croton12006} the overprediction of baryonic masses and luminosities of central cluster galaxies that early simulations found.
Finally, RGs may be responsible for magnetising the IGM that pervades the filaments of the Cosmic Web \citep[e.g.][]{Vazza12017}.

Despite the emerging picture that RGs trace quenched star formation, inhibit cooling flows, and magnetise filaments, our knowledge of them is far from complete.
Concerning geometry, a major unknown is the exact connection between the morphology of RGs and the pressure field of the ambient IGM, especially in filaments and cluster outskirts.
Another question is whether small and large RGs come from the same initial population, or whether their growth is driven by distinct physical processes.
Finally, we do not know how large can RGs become, and, more generally, how many RGs there are of each length.

To test RG growth models that answer these and other questions, it is imperative to study the subpopulation of most spatially extreme RGs: the giant radio galaxies (GRGs).
The defining feature of giants is that their proper lengths --- when projected onto the celestial sphere --- exceed some threshold $l_\mathrm{p,GRG}$, which is canonically chosen as 0.7 Mpc or 1 Mpc.
If $l_\mathrm{p,GRG} = 0.7\ \mathrm{Mpc}$, then the preceding literature describes a total of 1281 giants.

In recent years, several studies have successfully searched for giants in systematic, wide-area surveys such as the NRAO VLA Sky Survey \citep[NVSS;][]{Condon11998} and the LOFAR Two-metre Sky Survey \citep[LoTSS;][]{Shimwell12017}.
A combination of manual (i.e. visual) and automated searches \citep{Solovyov12011, Solovyov12014, Amirkhanyan12016, Proctor12016, Dabhade12017, Dabhade12020October} in the NVSS yielded 313 new giants (24\% of the aforementioned literature population).
Meanwhile, \citet{Dabhade12020March} discovered 225 new giants (17\% of this same population) in the LoTSS DR1 \citep{Shimwell12019}, whose survey footprint is 80 times smaller than NVSS's.
Such searches have the advantage of introducing almost homogeneous selection effects throughout the survey footprint, which can potentially be modelled and thus corrected for during any subsequent statistical inference.

In this work this idea comes to fruition, by conducting a precision analysis of the intrinsic giant radio galaxy length distribution.
To do so, we require two ingredients.
First, in Sect.~\ref{sec:theory}, we develop a statistical framework that allows one to answer probabilistic questions regarding both large samples of giants and individual specimina.
Then, in Sect.~\ref{sec:LoTSSDR2GRGSearch}, we describe our LoTSS DR2 \citep{Shimwell12022} GRG search campaign and the trove of previously unknown giants it has yielded; moreover, we describe the assemblage of the most complete GRG catalogue to date.
In Sect.~\ref{sec:results}, combining theory and data, we infer the tail index parameter that describes the intrinsic GRG length distribution, which constrains future models and simulations aimed at understanding RG growth.
In Sect.~\ref{sec:discussion}, we discuss caveats of the current work and give recommendations for future extensions, before we present conclusions in Sect.~\ref{sec:conclusions}.

We assume a concordance inflationary $\Lambda$ cold dark matter cosmology with parameters $\mathfrak{M}$ from \citet{Planck12020}; that is to say $\mathfrak{M} = \left(h = 0.6766, \Omega_\mathrm{BM,0} = 0.0490, \Omega_\mathrm{M,0} = 0.3111, \Omega_{\Lambda,0} = 0.6889\right)$, where $H_0 \coloneqq h \cdot 100\ \mathrm{km\ s^{-1}\ Mpc^{-1}}$.
We define the spectral index $\alpha$ such that it relates to flux density $F_\nu$ at frequency $\nu$ as $F_\nu \propto \nu^\alpha$, and define giants using threshold $l_\mathrm{p,GRG} \coloneqq 0.7\ \mathrm{Mpc}$.
Regarding the terminology, we use `angular length' where others use `largest angular size' (LAS), and `projected proper length' where others use `largest linear size' (LLS).\footnote{An object's size can refer to either its one-, two-, or three-dimensional extent.
We therefore consider `angular size' to be ambiguous terminology; we propose that `angular length' better captures one-dimensionality.
Naturally, an object's `length' is understood to be its total one-dimensional extent, so that the qualifier `largest' seems superfluous.
We further remark that `length' is synonymous with, but more succinct than, `linear size'.
In cosmology, one must distinguish between proper and comoving lengths, especially when objects are not gravitationally bound --- like GRG lobes.
In this work, we consider two types of proper lengths: intrinsic proper lengths and projected proper lengths.
}

\section{Theory}
\label{sec:theory}
To measure the intrinsic GRG length distribution, we must first establish a suitable statistical framework.
In this section, we provide a summary of the theory developed in Appendix~\ref{sec:supplementaryMaterialPowerLawModel}.
Following Occam's razor, we construct a model with minimal assumptions that provides new insight into the GRG phenomenon and the detection biases inherent to systematic search campaigns.

\subsection{Intrinsic proper length}
\label{sec:intrinsicProperLength}
Firstly, we assume that giants and non-giant RGs share a common length distribution.\footnote{Luckily, this assumption turns out to be irrelevant in the forthcoming GRG-only expressions, which are this work's focus.}
In particular, because power laws are abundant in Nature, we assume that the intrinsic proper length random variable (RV) $L$ has a Pareto distribution with tail index $\xi < -1$ and support from $l_\mathrm{min} > 0$ onwards.
If an RV is Pareto distributed, then the relative occurrence of two possible outcomes equals their ratio raised to a power: the tail index $\xi$.
In astrophysics, Pareto distributions describe the kinetic energies of freshly accelerated electrons in large-scale structure and supernova shocks \citep[e.g.][]{Kirk11987}, the initial masses of main-sequence stars \citep[e.g.][]{Kroupa12001}, and the luminosities of gamma-ray bursts \citep[e.g.][]{Bloom12001}, to name a few examples.
Previous works \citep[e.g.][]{Andernach12021} have already hinted at the approximate validity of a Pareto distribution description for GRG lengths.
By comparing our final model to observations, as discussed in Section~\ref{sec:results} and visualised in Fig.~\ref{fig:ECDFCompleteness}, we demonstrate that this assumption is indeed a powerful approximation in the current case.

The probability density function (PDF) $f_L: \mathbb{R} \to \mathbb{R}_{\geq 0}$ thus becomes
\begin{align}
    f_L\left(l\right) = \begin{cases}
    0 & l \leq l_\mathrm{min},\\
    \frac{-(\xi+1)}{l_\mathrm{min}} \left(\frac{l}{l_\mathrm{min}}\right)^\xi & l > l_\mathrm{min}.
    \end{cases}
\end{align}
We refer the reader to Appendix~\ref{ap:intrinsicLengthDistribution} for a derivation of this expression and a demonstration of its connection to the literature's most common parametrisation.
\begin{figure*}
    \centering
    \begin{subfigure}{.49\textwidth}
    \centering
    \includegraphics[width=\textwidth]{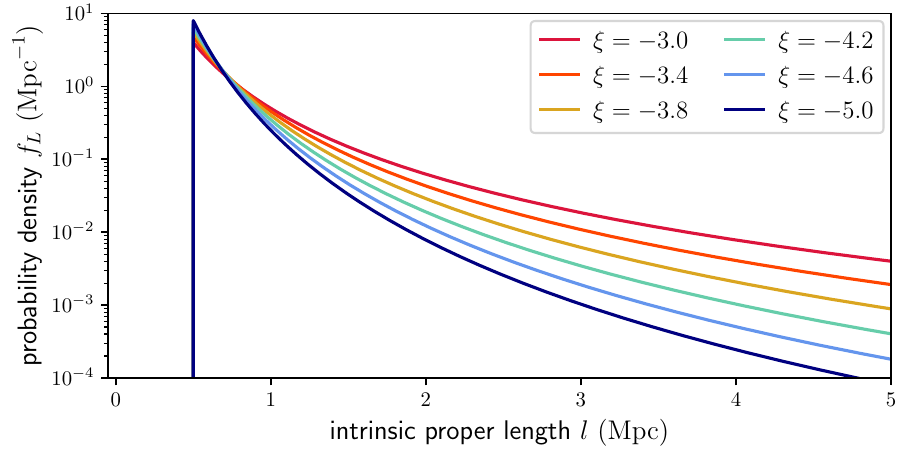}
    \end{subfigure}
    \begin{subfigure}{.49\textwidth}
    \centering
    \includegraphics[width=\textwidth]{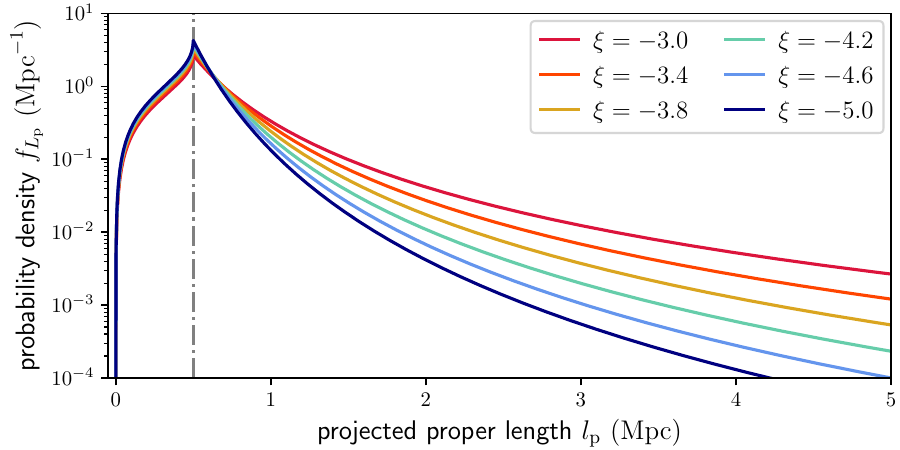}
    \end{subfigure}
    \caption{
    PDFs of radio galaxy intrinsic proper lengths $L$ and projected proper lengths $L_\mathrm{p}$.
    If the intrinsic lengths $L$ are Pareto distributed above some cut-off $l_\mathrm{min}$, then their projections on the sky $L_\mathrm{p}$ are also Pareto distributed above this cut-off. The tail indices are the same.
    We show the PDFs $f_L$ \textit{(left)} and $f_{L_\mathrm{p}}$ \textit{(right)} for $l_\mathrm{min} = 0.5\ \mathrm{Mpc}$, $l_\mathrm{max} = \infty$ (see Appendix~\ref{sec:supplementaryMaterialPowerLawModel}) and various tail indices $\xi$.
    The support of $f_L$ starts at $l_\mathrm{min}$, which is marked by the vertical grey line in the right panel.
    }
    \label{fig:marginals}
\end{figure*}\noindent

\subsection{Projected proper length}
From the distribution of intrinsic lengths and the assumption of random radio galaxy orientations, we now derive the distribution of projected lengths.
This distribution is more easily compared to observations, which usually lack inclination angle information.
\subsubsection{Distribution for RGs}
To model length and orientation, we consider a vector $\vec{L} \in \mathbb{R}^3$ (of length $L \coloneqq ||\vec{L}||_2$) for each RG.
In accordance with the IAU Solar System convention for positive poles, the unit vector $\hat{L} \in \mathbb{S}^2$ marks the direction from which the central Kerr black hole is seen rotating in anticlockwise direction.\footnote{$\mathbb{S}^2$ is the unit two-sphere.}
We define the inclination angle $\Theta$ as the angle between $\vec{L}$ and a vector parallel to the line of sight pointing towards the observer.\footnote{When $\Theta = 0$, the positive pole's jet points towards us, and the black hole is seen rotating in anticlockwise direction; when $\Theta = 90\degree$, the jets lie in the plane of the sky; when $\Theta = 180\degree$, the positive pole's jet points away from us, and the black hole rotates in clockwise direction.}
Observations that allow one to measure the orientation of the RG axis in 3D are time-intensive, and so usually only the RG length projected onto the plane of the sky is known.

Geometrically, we model RGs as line segments --- as if they were `thin sticks', with vanishing volumes --- so that the projected proper length RV $L_\mathrm{p}$ relates to $L$ and $\Theta$ through
\begin{align}
    L_\mathrm{p} = L \sin{\Theta}.
\end{align}
We assume that $\hat{L}$ is drawn from a uniform distribution on $\mathbb{S}^2$, so that the PDF $f_\Theta: [0, \pi] \to \mathbb{R}_{\geq 0}$ becomes $f_\Theta\left(\theta\right) = \frac{1}{2}\sin{\theta}$.
Since $L$ and $\Theta$ are independent, we find in Appendix~\ref{ap:projectedLengthDistribution} that the PDF of $L_\mathrm{p}$, $f_{L_\mathrm{p}}: \mathbb{R} \to \mathbb{R}_{\geq 0}$, is
\begin{align}
    f_{L_\mathrm{p}}\left(l_\mathrm{p}\right) = \begin{cases}
    0 & \text{if } l_\mathrm{p} \leq 0,\\
    \frac{-(\xi+1)}{l_\mathrm{min}}\frac{l_\mathrm{p}}{l_\mathrm{min}}I\left(\xi-1,\frac{l_\mathrm{p}}{l_\mathrm{min}}\right) & \text{if } 0 < l_\mathrm{p} \leq l_\mathrm{min},\\
    \frac{\left(\xi+1\right)^2}{l_\mathrm{min}}\frac{\sqrt{\pi}}{4}\left(\frac{l_\mathrm{p}}{l_\mathrm{min}}\right)^\xi \frac{\Gamma\left(-\frac{\xi}{2}-\frac{1}{2}\right)}{\Gamma\left(-\frac{\xi}{2}+1\right)} & \text{if } l_\mathrm{p} > l_\mathrm{min},
    \end{cases}
\label{eq:PDFProjectedProperLength}
\end{align}
where
\begin{align}
    I\left(a,b\right)\coloneqq \int_1^\infty \frac{\eta^a\ \mathrm{d}\eta}{\sqrt{\eta^2-b^2}} \text{ for } a < 0, \vert b \vert < 1.
\end{align}
We note that for $l_\mathrm{p} > l_\mathrm{min}$, the projected proper length has a Pareto distribution with the same tail index as the intrinsic proper length distribution.
We compare $f_L$ and $f_{L_\mathrm{p}}$ in Fig.~\ref{fig:marginals}.

\subsubsection{Distribution for giants}
For giants specifically (i.e. RGs such that $L_\mathrm{p} > l_\mathrm{p,GRG}$, where $l_\mathrm{p,GRG}$ is some constant threshold; in this work, $l_\mathrm{p,GRG} \coloneqq 0.7\ \mathrm{Mpc}$), the projected proper length distribution becomes a Pareto distribution with tail index $\xi$ again:
\begin{align}
    f_{L_\mathrm{p}\ \vert\ L_\mathrm{p} > l_\mathrm{p, GRG}}\left(l_\mathrm{p}\right) =& \begin{cases}
    0 & \text{if } l_\mathrm{p} \leq l_\mathrm{p, GRG}\\
    \frac{-\left(\xi+1\right)}{l_\mathrm{p, GRG}}\left(\frac{l_\mathrm{p}}{l_\mathrm{p, GRG}}\right)^\xi & \text{if } l_\mathrm{p} > l_\mathrm{p, GRG}.
    \end{cases}
\label{eq:projectedProperLengthPDF}
\end{align}
In other words, for giants, projection retains the Paretianity of lengths.
A measurement of the tail index of the projected length distribution is immediately also a measurement of the tail index of the intrinsic length distribution.

The survival function, which gives the probability that a GRG has a projected proper length exceeding $l_\mathrm{p}$, takes on a particularly simple form:
\begin{align}
    \mathbb{P}\left(L_\mathrm{p} > l_\mathrm{p}\ \vert\ L_\mathrm{p} > l_\mathrm{p, GRG}\right) = \left(\frac{l_\mathrm{p}}{l_\mathrm{p, GRG}}\right)^{\xi+1}.
\label{eq:projectedProperLengthGRGSF}
\end{align}
The mean projected proper length of giants is the expectation value of $L_\mathrm{p}\ \vert\ L_\mathrm{p} > l_\mathrm{p,GRG}$:
\begin{align}
    \mathbb{E}[L_\mathrm{p}\ \vert\ L_\mathrm{p}>l_\mathrm{p,GRG}] = l_\mathrm{p,GRG} \frac{\xi+1}{\xi+2},
\end{align}
which is only defined when $\xi < -2$.
For example, when $\xi = -4$, $\mathbb{E}[L_\mathrm{p}\ \vert\ L_\mathrm{p}>l_\mathrm{p,GRG}] = \frac{3}{2} l_\mathrm{p,GRG}$, which becomes $1.05\ \mathrm{Mpc}$ for $l_\mathrm{p,GRG} \coloneqq 0.7\ \mathrm{Mpc}$ and $1.5\ \mathrm{Mpc}$ for $l_\mathrm{p,GRG} \coloneqq 1\ \mathrm{Mpc}$.

Appendix~\ref{ap:projectedLengthDistributionGRGs} provides a derivation for all three expressions.

\subsection{Deprojection factor}
The deprojection factor, $D \coloneqq \frac{L}{L_\mathrm{p}} = \sin^{-1}{\Theta}$, quantifies how much larger intrinsic lengths are compared with projected lengths.
The PDF of $D$, $f_D: \mathbb{R} \to \mathbb{R}_{\geq 0}$, is
\begin{align}
f_D\left(d\right) = \begin{cases}
0 &\text{if $d \leq 1$;}\\
\frac{1}{d^2\sqrt{d^2-1}} &\text{if $d > 1$.}
\end{cases}
\end{align}
The mean deprojection factor $\mathbb{E}[D] = \frac{\pi}{2}$.
Deprojection factors can become arbitrarily large under the current model, because projected lengths can become arbitrarily small.
As discussed in Sect.~\ref{sec:movingBeyondLineSegmentProjection}, this is not a very realistic set-up.
In reality, an RG's projected length is bounded from below by its lobes, which have a non-vanishing volume and thus extend along all three spatial dimensions.
Upon projection, the projected length therefore cannot shrink beyond some lower limit that depends on the lobe geometry.
In Appendix~\ref{ap:deprojectionFactor}, we show that by enriching the conventional stick-like geometry with spherical lobes, deprojection factors indeed become bounded.

\subsection{Intrinsic proper length posterior and its moments}
Because an RG's intrinsic length is more physically informative than its projected length, we ideally obtain the former.
In this subsection, we quantify what a measurement $L_\mathrm{p} = l_\mathrm{p}$ already reveals about $L$.

\begin{figure}
    \centering
    \begin{subfigure}{\columnwidth}
    \includegraphics[width=\columnwidth]{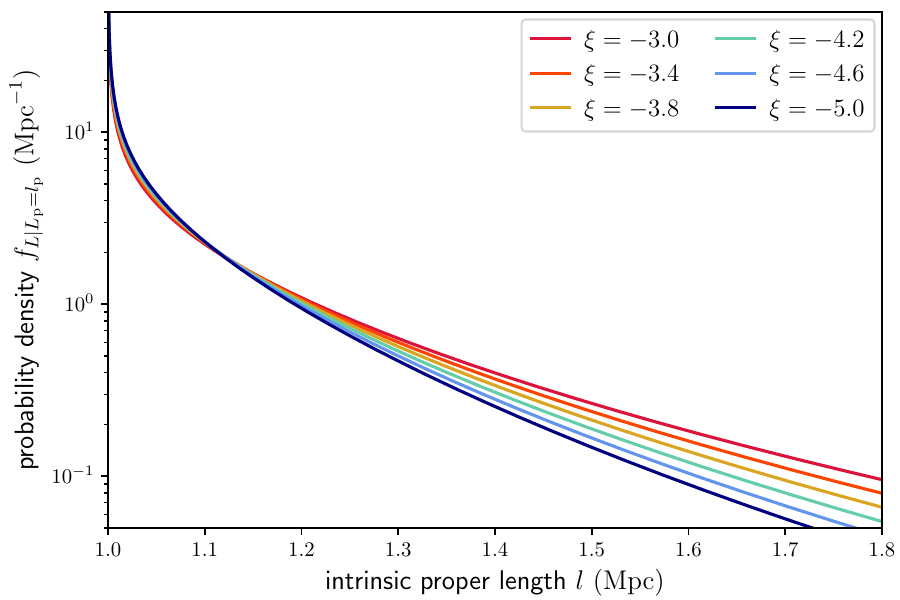}
    \end{subfigure}
    \begin{subfigure}{\columnwidth}
    \includegraphics[width=\columnwidth]{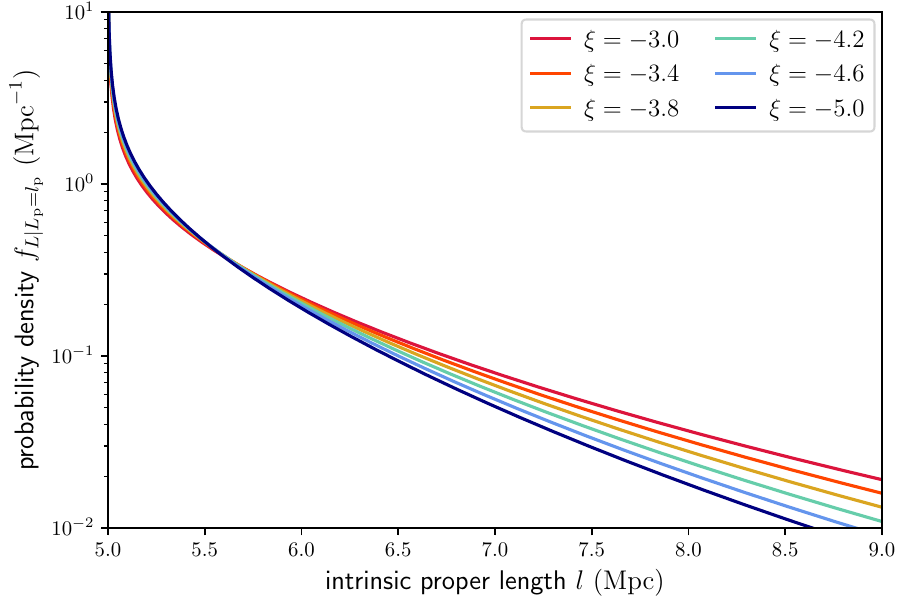}
    \end{subfigure}
    \caption{
    Posterior PDFs of intrinsic lengths for a given projected length $L\ \vert\ L_\mathrm{p} = l_\mathrm{p}$.
    If tail index $\xi$ is known, then an RG's $l_\mathrm{p}$ fixes the probability distribution over its possible $l$.
    This distribution is strongly skewed, and the same for all $l_\mathrm{p}$ --- save for horizontal translation and vertical scaling.
    We illustrate this point by showing posterior PDFs for giants with two different $l_\mathrm{p}$.
    \textit{Top:} $l_\mathrm{p} = 1\ \mathrm{Mpc}$.
    \textit{Bottom:} $l_\mathrm{p} = 5\ \mathrm{Mpc}$.
    For $\xi = -4$, the posterior mean $\mathbb{E}[L\ \vert\ L_\mathrm{p} = l_\mathrm{p}] = 1.13\ l_\mathrm{p}$ and the posterior standard deviation $\sqrt{\mathbb{V}[L\ \vert\ L_\mathrm{p} = l_\mathrm{p}]} = 0.23\ l_\mathrm{p}$ (see Table~\ref{tab:posteriorMoments}).
    }
    \label{fig:posteriorPDF}
\end{figure}

We first note that the projected length bounds the intrinsic length from below.
The intrinsic length can be much larger, however, but this is improbable for two reasons: large lengths are rarer than small lengths, although how drastic this effect is depends on $\xi$; in addition, viewing directions with large inclination angles are uncommon.
The best we can do is to construct a posterior distribution for $L$ given $L_\mathrm{p} = l_\mathrm{p}$.
This posterior has a concise analytic form.
If $l_\mathrm{p} > l_\mathrm{min}$, which is the relevant case for giants, the distribution of $L\ \vert\ L_\mathrm{p} = l_\mathrm{p}$ is
\begin{align}
    f_{L\vert L_\mathrm{p} = l_\mathrm{p}}\left(l\right) =
    \begin{cases}
    0 & \text{if } l \leq l_\mathrm{p}\\
    \frac{-\xi}{2^{1+\xi}\pi}\frac{\Gamma^2\left(-\frac{\xi}{2}\right)}{\Gamma\left(-\xi\right)}\frac{1}{l_\mathrm{p}}\frac{1}{\sqrt{\left(\frac{l}{l_\mathrm{p}}\right)^2 - 1}}\left(\frac{l}{l_\mathrm{p}}\right)^{\xi-1} & \text{if } l > l_\mathrm{p}.
    \end{cases}
\label{eq:PDFProperLengthGivenProjectedProperLength}
\end{align}
For $l \gg l_\mathrm{p}$, $f_{L\vert L_\mathrm{p} = l_\mathrm{p}}\left(l\right) \propto \left(\frac{l}{l_\mathrm{p}}\right)^{\xi-2}$: the posterior PDF tends to a power law in $l$ with index $\xi - 2$.
In Fig.~\ref{fig:posteriorPDF}, we visualise the posterior PDF for several values of $\xi$.
\noindent
Clearly, to evaluate Eq.~\ref{eq:PDFProperLengthGivenProjectedProperLength}, one must choose $l_\mathrm{p}$ --- however, the shape of the distribution is the same for all choices.
We illustrate this by comparing the PDF for a comparatively small GRG ($l_\mathrm{p} = 1.0\ \mathrm{Mpc}$; top panel) to the PDF for Alcyoneus\footnote{Alcyoneus is the projectively longest giant known to date \citep{Oei12022Alcyoneus}.} ($l_\mathrm{p} = 5.0\ \mathrm{Mpc}$; bottom panel).

The posterior mean is
\begin{align}
\mathbb{E}\left[L\ \vert\ L_\mathrm{p} = l_\mathrm{p}\right] = l_\mathrm{p} \cdot \frac{-\xi}{2^{2\xi+3}\pi} \frac{\Gamma^4\left(-\frac{\xi}{2}\right)}{\Gamma^2\left(-\xi\right)};
\end{align}
the posterior variance is
\begin{align}
    \mathbb{V}\left[L\ \vert\ L_\mathrm{p} = l_\mathrm{p}\right] = l_\mathrm{p}^2\left(\frac{\xi}{\xi + 1} - \frac{\xi^2}{2^{4\xi + 6}\pi^2}\frac{\Gamma^8\left(-\frac{\xi}{2}\right)}{\Gamma^4\left(-\xi\right)}\right).
\end{align}
Both mean and standard deviation scale linearly in $l_\mathrm{p}$: the projection effect is a multiplicative noise source.
In Table~\ref{tab:posteriorMoments}, we provide explicit values for various $\xi$.
\begin{center}
\captionof{table}{
Intrinsic proper length posterior mean and standard deviation in multiples of projected proper length $l_\mathrm{p}$, given for various tail indices $\xi$.
}
\begin{tabular}{c | c c}
& $\mathbb{E}\left[L\ \vert\ L_\mathrm{p} = l_\mathrm{p}\right]\left(\xi\right)\ \ \left(l_\mathrm{p}\right)$ & $\sqrt{\mathbb{V}\left[L\ \vert\ L_\mathrm{p} = l_\mathrm{p}\right]\left(\xi\right)}\ \ \left(l_\mathrm{p}\right)$ \\
\hline\\[-5pt]
$\xi = -2$ & $\frac{4}{\pi} \approx 1.27$ & $\sqrt{2 - \frac{16}{\pi^2}} \approx 0.62$\\[5pt]
$\xi = -3$ & $\frac{3\pi}{8} \approx 1.18$ & $\sqrt{\frac{3}{2} - \frac{9\pi^2}{64}} \approx 0.33$\\[5pt]
$\xi = -4$ & $\frac{32}{9\pi} \approx 1.13$ & $\sqrt{\frac{4}{3} - \frac{1024}{81\pi^2}} \approx 0.23$\\[5pt]
$\xi = -5$ & $\frac{45\pi}{128} \approx 1.10$ & $\sqrt{\frac{5}{4} - \frac{2025\pi^2}{16384}} \approx 0.17$\\[5pt]
\label{tab:posteriorMoments}
\end{tabular}
\end{center}
Higher moments exist up to order $\lceil -\xi \rceil$; because the PDF $f_{L\ \vert\ L_\mathrm{p} = l_\mathrm{p}}\left(l\right)$ is strongly skewed, such moments do further specify the distribution.

It is important to note that it is formally incorrect to statistically deproject RGs by drawing samples from deprojection factor $D$ and multiplying them with some measurement $L_\mathrm{p} = l_\mathrm{p}$, or even more crudely, by multiplying the latter with $\mathbb{E}[D]$.
The reason that renders such approaches invalid is that $L_\mathrm{p} = L \sin{\Theta}$ and $D = \sin^{-1}{\Theta}$ are not independent RVs.
We refer to Appendix~\ref{ap:posterior} for an explicit proof of this fact, and for derivations of this subsection's expressions.

\subsection{GRG inclination angle}
\begin{figure}
    \centering
    \includegraphics[width=\columnwidth]{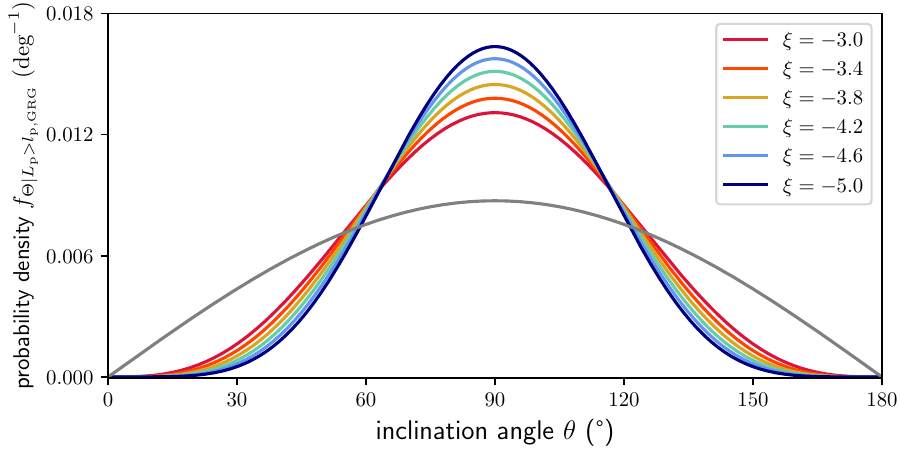}
    \caption{
    PDFs of GRG inclination angles $\Theta\ \vert\ L_\mathrm{p} > l_\mathrm{p,GRG}$ (\textit{colours}) and RG inclination angles $\Theta$ (\textit{grey}).
    Giants more often have orientations close to the sky plane.
    The strength of this tendency is governed by $\xi$, with larger $\xi$ meaning more dispersion.
    }
    \label{fig:PDFGRGInclinationAngles}
\end{figure}

Radio galaxies with jets that make a small angle with the plane of the sky are more likely to have a projected length exceeding $l_\mathrm{p,GRG}$ than those with jets that make a large angle with the plane of the sky.
For this reason, the inclination angle distribution of giants is different from that of RGs: it is more peaked around $\theta = 90\degree$.
More precisely, the PDF $f_{\Theta\ \vert\ L_\mathrm{p} > l_\mathrm{p,GRG}}: [0,\pi] \to \mathbb{R}_{\geq 0}$ of the GRG inclination angle $\Theta\ \vert\ L_\mathrm{p} > l_\mathrm{p,GRG}$ has the general form
\begin{align}
    f_{\Theta\ \vert\ L_\mathrm{p} > l_\mathrm{p,GRG}}(\theta) = \frac{\left(1 - F_L\left(\frac{l_\mathrm{p,GRG}}{\sin{\theta}}\right)\right) f_\Theta\left(\theta\right)}{1 - F_{L_\mathrm{p}}(l_\mathrm{p,GRG})}.
\end{align}
Under our Pareto distribution assumption for $L$, this concretises to
\begin{align}
    f_{\Theta\ \vert\ L_\mathrm{p} > l_\mathrm{p,GRG}}(\theta) = \frac{2}{-(\xi+1)\sqrt{\pi}}\frac{\Gamma\left(-\frac{\xi}{2}+1\right)}{\Gamma\left(-\frac{\xi}{2}-\frac{1}{2}\right)}\sin^{-\xi}{\theta}.
\end{align}
We note that $f_{\Theta\ \vert\ L_\mathrm{p} > l_\mathrm{p,GRG}}(\theta) \propto \sin^{-\xi}{\theta}$; the factor in front serves only as a normalisation constant.
The distribution is independent of the choice of $l_\mathrm{p,GRG}$ and depends on a single parameter: $\xi$.
We visualise $f_{\Theta\ \vert\ L_\mathrm{p} > l_\mathrm{p,GRG}}(\theta)$ in Fig.~\ref{fig:PDFGRGInclinationAngles}.
\noindent
Appendix~\ref{ap:GRGInclinationAngle} contains a brief derivation.

\subsection{GRG angular length}
The model predicts the distribution of GRG angular lengths in the Local Universe up to comoving distance $r_\mathrm{max}$.
The GRG angular length RV $\Phi\ \vert\ L_\mathrm{p} > l_\mathrm{p,GRG}$ relates to the GRG projected proper length RV $L_\mathrm{p}\ \vert\ L_\mathrm{p} > l_\mathrm{p,GRG}$ and the comoving distance RV $R$ as
\begin{align}
    \Phi\ \vert\ L_\mathrm{p} > l_\mathrm{p,GRG} = L_\mathrm{p}\ \vert\ L_\mathrm{p} > l_\mathrm{p,GRG} \cdot  \frac{1+z\left(R\right)}{R}.
\end{align}
(We note that this relation is valid only in a \emph{flat} Friedmann--Lema\^itre--Robertson--Walker (FLRW) universe.)
We also assume that the GRG number density is constant in the Local Universe.
The PDF of $\Phi\ \vert\ L_\mathrm{p} > l_\mathrm{p,GRG}$ has useful analytic forms under two different idealisations.

In a Euclidean universe, $z(R) = 0$, and the minimal GRG angular length $\phi_\mathrm{GRG} = \frac{l_\mathrm{p,GRG}}{r_\mathrm{max}}$.
Then
\begin{align}
f_{\Phi\ \vert\ L_\mathrm{p} > l_\mathrm{p,GRG}}(\phi) = \begin{cases}
0 & \text{if } \phi \leq \phi_\mathrm{GRG}\\
-3 \frac{\xi + 1}{\xi + 4}\cdot \frac{1}{\phi_\mathrm{GRG}}\left(\left(\frac{\phi}{\phi_\mathrm{GRG}}\right)^\xi - \left(\frac{\phi}{\phi_\mathrm{GRG}}\right)^{-4}\right) & \text{if } \phi > \phi_\mathrm{GRG},
\end{cases}
\end{align}
which is valid as long as $\xi \neq -4$.

\begin{figure}
    \centering
    \begin{subfigure}{\columnwidth}
    \includegraphics[width=\columnwidth]{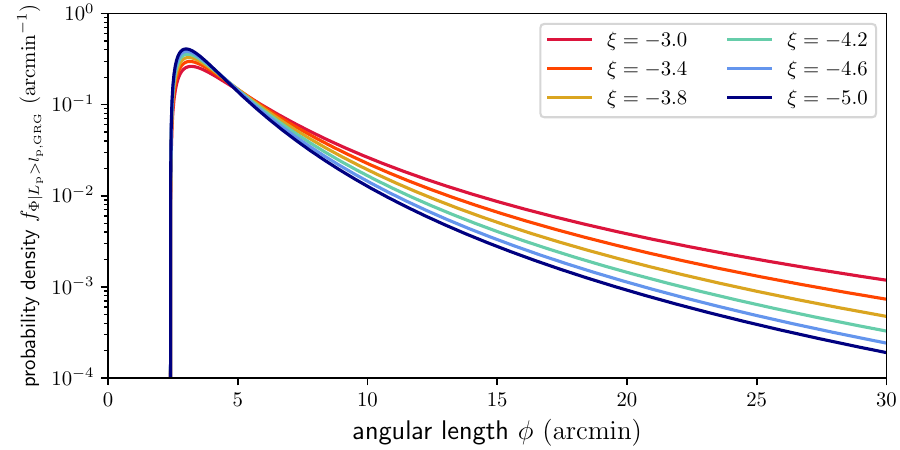}
    \end{subfigure}
    \begin{subfigure}{\columnwidth}
    \includegraphics[width=\columnwidth]{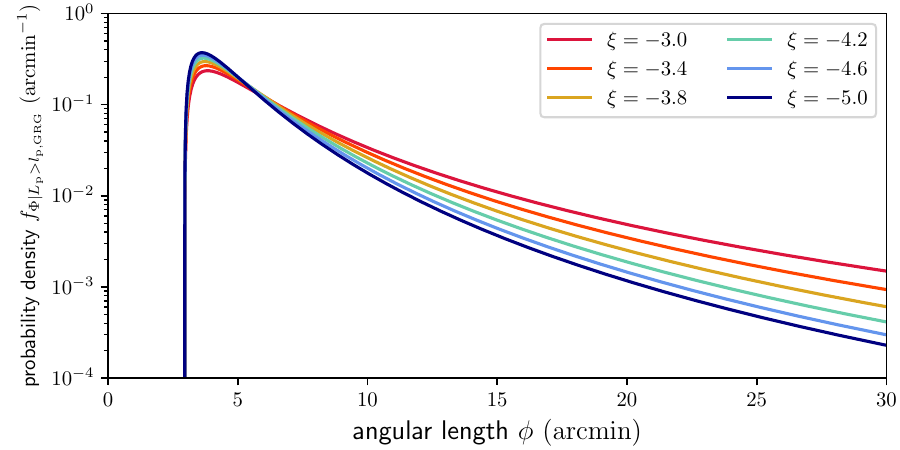}
    \end{subfigure}
    \caption{
    PDFs of GRG angular lengths $\Phi\ \vert\ L_\mathrm{p} > l_\mathrm{p,GRG}$.
    We fix $r_\mathrm{max} = 1\ \mathrm{Gpc}$ (and $l_\mathrm{p,GRG} = 0.7\ \mathrm{Mpc}$), and vary $\xi$.
    \textit{Top:} Euclidean universe. \textit{Bottom:} expanding universe at low redshifts.
    }
    \label{fig:PDFGRGAngularLength}
\end{figure}

In an expanding universe at low redshifts, the Hubble--Lema\^itre law $z(R) \approx \frac{R}{d_H}$ holds; the Hubble distance $d_H \coloneqq \frac{c}{H_0}$.
In this case,
\begin{align}
&f_{\Phi\ \vert\ L_\mathrm{p} > l_\mathrm{p,GRG}}(\phi) =\nonumber\\
&\begin{cases}
0 & \text{if } \phi \leq \phi_\mathrm{GRG} + \frac{l_\mathrm{p,GRG}}{d_H}\\
\frac{-3\left(\xi+1\right)\phi^\xi}{r_\mathrm{max}^3 l_\mathrm{p,GRG}^{\xi+1}} \int_{\frac{1}{r_\mathrm{max}} + \frac{1}{d_H}}^{\frac{\phi}{l_\mathrm{p,GRG}}} \frac{\mathrm{d}k}{k^{\xi+1}\left(k - \frac{1}{d_H}\right)^4} & \text{if } \phi > \phi_\mathrm{GRG} + \frac{l_\mathrm{p,GRG}}{d_H}.
\end{cases}
\end{align}
Figure~\ref{fig:PDFGRGAngularLength} shows GRG angular length PDFs under both idealisations.
\noindent
The PDFs undergo a minor shift upon changing universe type but are otherwise similar.
For most current-day applications, it will therefore be unnecessary to calculate an even more refined version of $f_{\Phi\ \vert\ L_\mathrm{p} > l_\mathrm{p,GRG}}(\phi)$.
Appendix~\ref{ap:angularLength} contains derivations and details.

\subsection{Maximum likelihood estimation of the tail index}
The GRG projected proper length distribution features just one parameter of physical interest: the tail index $\xi$.
If observational selection effects are negligible, one can directly use maximum likelihood estimation (MLE) on GRG data to infer $\xi$.
In particular, we consider a set of projected lengths $\{L_{\mathrm{p},1}, ..., L_{\mathrm{p},N}\} \sim L_\mathrm{p}\ \vert\ L_\mathrm{p} > l_\mathrm{p, GRG}$ from $N$ giants.
Appendix~\ref{ap:MLE} shows that the maximum likelihood estimate of $\xi$ is the RV $\xi_\mathrm{MLE}$, given by
\begin{align}
    \xi_\mathrm{MLE} &= -\frac{N}{\sum_{i=1}^N \ln{\frac{L_{\mathrm{p},i}}{l_\mathrm{p,GRG}}}} - 1.
\label{eq:MLE}
\end{align}

\subsection{Observed projected proper length}
\subsubsection{General considerations}
\label{sec:generalConsiderations}
In the preceding theory, we have ignored observational selection effects that favour some projected proper lengths over others.
In practice, several such effects occur; the importance of each varies per survey and (G)RG search campaign within it.
One of them is the bias against physically long RGs that the interferometer's largest detectable angular scale can induce.\footnote{
For example, the Faint Images of the Radio Sky at Twenty Centimeters (FIRST) survey used the Very Large Array (VLA) in B-configuration, leading it to detect angular scales of at most two arcminutes.
By contrast, the largest angular scale of the LoTSS --- the survey relevant to this work --- is about a degree.
(For the $6''$ and $20''$ resolutions, the shortest baseline is 100 metres; for the $60''$ and $90''$ resolutions, the shortest baseline is 68 metres.)
As virtually all giants are of subdegree angular length, we need not consider this bias in our case.}
As a result, the projected proper length of an observed RG might not be adequately modelled through RV $L_\mathrm{p}$.
Instead, we must introduce a new RV $L_\mathrm{p,obs}$.

We define the completeness $C: \mathbb{R}_{>0} \times \mathbb{R}_{>0} \to \left[0,1\right]$ at $(l_\mathrm{p}, z_\mathrm{max})$ to be the fraction of all RGs with projected proper length $l_\mathrm{p}$ in the cosmological volume up to $z_\mathrm{max}$ that is detected in a particular RG search campaign.
Then, assuming that the distribution of $L_\mathrm{p}$ does not evolve with redshift between $z = z_\mathrm{max}$ and $z = 0$ (i.e. $\xi$ remains constant),
\begin{align}
C\left(l_\mathrm{p}, z_\mathrm{max}\right) = \frac{\int_0^{z_\mathrm{max}}p_\mathrm{obs}\left(l_\mathrm{p},z\right) r^2\left(z\right) E^{-1}\left(z\right)\ \mathrm{d}z}{\int_0^{z_\mathrm{max}}r^2\left(z\right)E^{-1}\left(z\right)\ \mathrm{d}z},
\label{eq:completeness}
\end{align}
where $p_\mathrm{obs}(l_\mathrm{p},z)$ is the probability that an RG of projected proper length $l_\mathrm{p}$ at cosmological redshift $z$ is detected through the campaign, and $r\left(z\right)$ is the comoving radial distance at cosmological redshift $z$.
In a \emph{flat} FLRW universe, the dimensionless Hubble parameter $E$ is
\begin{align}
    E\left(z\right) \coloneqq& \frac{H\left(z\right)}{H_0} = \sqrt{\Omega_\mathrm{R,0} \left(1+z\right)^4 + \Omega_\mathrm{M,0} \left(1+z\right)^3 + \Omega_\mathrm{\Lambda,0}}.
\end{align}
The PDF of the observed projected proper length RV $L_\mathrm{p,obs}$ becomes
\begin{align}
    f_{L_\mathrm{p,obs}}\left(l_\mathrm{p}\right) = \frac{C\left(l_\mathrm{p}\right)f_{L_\mathrm{p}}\left(l_\mathrm{p}\right)}{\int_0^\infty C\left(l_\mathrm{p}'\right) f_{L_\mathrm{p}}\left(l_\mathrm{p}'\right)\ \mathrm{d}l_\mathrm{p}'},
\label{eq:PDFObservedProjectedProperLength}
\end{align}
where we suppress the $z_\mathrm{max}$-dependence for succinctness.
We note that multiplying $p_\mathrm{obs}(l_\mathrm{p}, z)$ with an $l_\mathrm{p}$- and $z$-independent factor affects the completeness $C(l_\mathrm{p}, z_\mathrm{max})$, but cancels in Eq.~\ref{eq:PDFObservedProjectedProperLength}; $f_{L_\mathrm{p,obs}}$ will be independent of it.
Finally, the PDF of the GRG observed projected proper length RV $L_\mathrm{p,obs}\ \vert\ L_\mathrm{p,obs} > l_\mathrm{p,GRG}$ is
\begin{align}
&f_{L_\mathrm{p,obs}\ \vert\ L_\mathrm{p,obs} > l_\mathrm{p,GRG}}\left(l_\mathrm{p}\right) =\begin{cases}
0 & \text{if } l_\mathrm{p} \leq l_\mathrm{p, GRG}\\
\frac{C\left(l_\mathrm{p}\right)f_{L_\mathrm{p}}\left(l_\mathrm{p}\right)}{\int_{l_\mathrm{p,GRG}}^\infty C\left(l_\mathrm{p}'\right)f_{L_\mathrm{p}}\left(l_\mathrm{p}'\right)\ \mathrm{d}l_\mathrm{p}'} & \text{if } l_\mathrm{p} > l_\mathrm{p, GRG}.
\end{cases}
\label{eq:GRGObservedProjectedProperLength}
\end{align}
We derive these expressions in Appendix~\ref{ap:completenessGeneral}.

\subsubsection{Fuzzy angular length threshold}
\label{sec:fuzzyAngularLengthThreshold}
\begin{figure*}
    \centering
    \begin{subfigure}{.495\textwidth}
    \includegraphics[width=\textwidth]{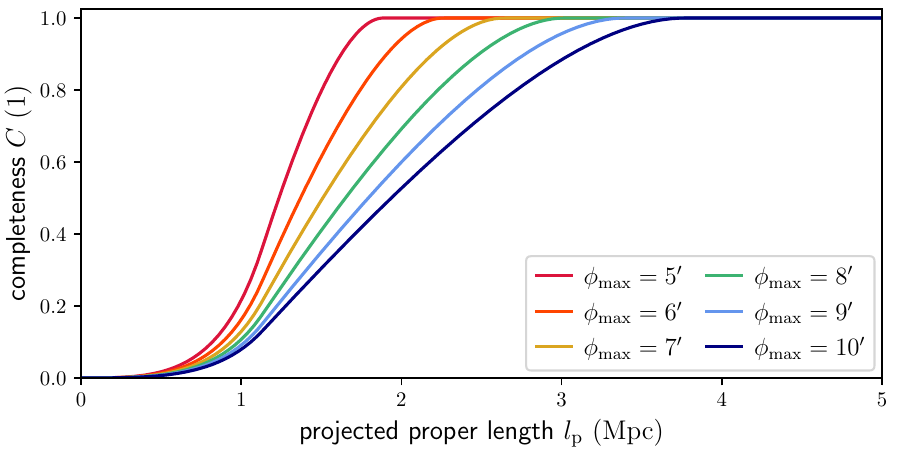}
    \end{subfigure}
    \begin{subfigure}{.495\textwidth}
    \includegraphics[width=\textwidth]{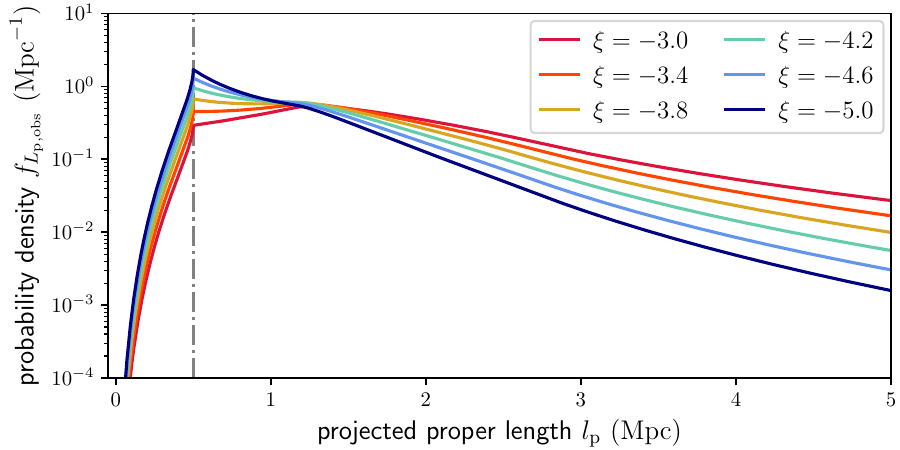}
    \end{subfigure}
    \begin{subfigure}{.495\textwidth}
    \includegraphics[width=\textwidth]{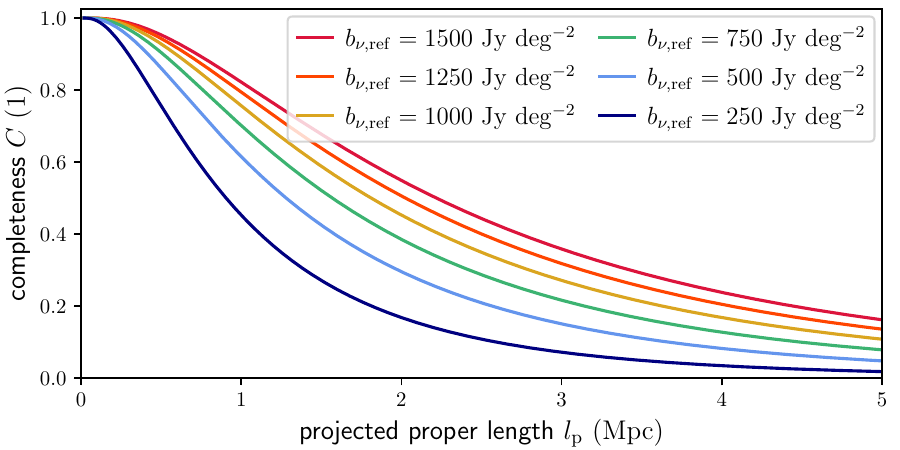}
    \end{subfigure}
    \begin{subfigure}{.495\textwidth}
    \includegraphics[width=\textwidth]{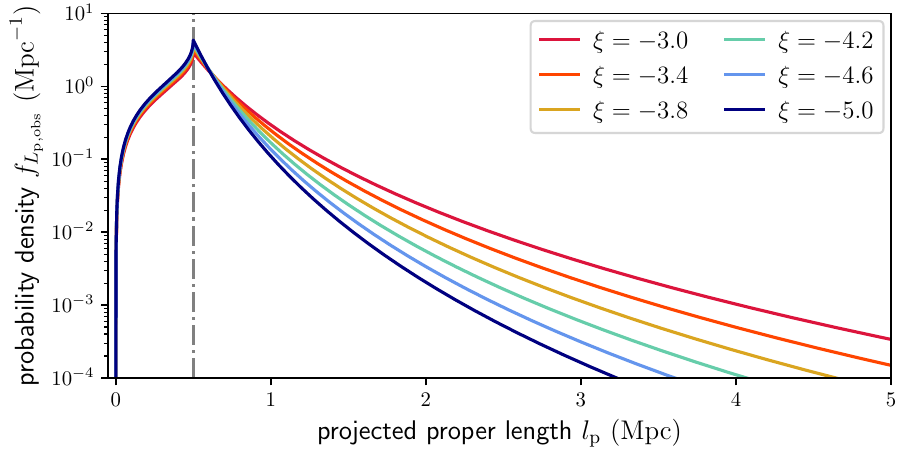}
    \end{subfigure}
    \caption{
    Completeness functions (\textit{left column}) and PDFs of observed projected proper lengths $L_\mathrm{p,obs}$ (\textit{right column}).
    Selection effects leave imprints on the distribution of radio galaxies' $L_\mathrm{p,obs}$.
    In the top row, we show how imposing an angular length threshold in a GRG search campaign leads to incompleteness \textit{(left)}, which causes the PDF $f_{L_\mathrm{p,obs}}$ \textit{(right)} to differ from $f_{L_\mathrm{p}}$.
    We assume RGs with angular length $\phi < \phi_\mathrm{min}$ have probability 0 to be included in a sample, whilst RGs with angular length $\phi > \phi_\mathrm{max}$ have probability 1.
    The inclusion probability is assumed to increase linearly between $\phi_\mathrm{min}$ and $\phi_\mathrm{max}$.
    In the left panel, we fix $\phi_\mathrm{min} = 3'$ and vary $\phi_\mathrm{max}$; in the right panel, we also fix $\phi_\mathrm{max} = 7'$.
    In the bottom row, we show how a survey's surface brightness limitations lead to incompleteness \textit{(left)}, which causes the PDF $f_{L_\mathrm{p,obs}}$ \textit{(right)} to differ from $f_{L_\mathrm{p}}$.
    We assume that lobe surface brightnesses are lognormally distributed; we parametrise the distribution for RGs of intrinsic length $l_\mathrm{ref} = 0.7\ \mathrm{Mpc}$ at $z = 0$ observed at $\nu_\mathrm{obs} = 144\ \mathrm{MHz}$ with a median $b_{\nu,\mathrm{ref}}$ and dispersion parameter $\sigma_\mathrm{ref}$.
    In the left panel, we fix $\sigma_\mathrm{ref} = 1.5$ and vary $b_{\nu,\mathrm{ref}}$; in the right panel, we also fix $b_{\nu,\mathrm{ref}} = 1000\ \mathrm{Jy\ deg^{-2}}$.
    We assume a lobe spectral index $\alpha = -1$, a surface brightness detection threshold $b_{\nu,\mathrm{th}} = 25\ \mathrm{Jy\ deg^{-2}}$, and self-similar growth: $\zeta = -2$.
    For both selection effects, we consider RGs up to cosmological redshift $z_\mathrm{max} = 0.5$ only.
    }
    \label{fig:completeness}
\end{figure*}\noindent
We provide a concrete example of an important observational selection effect in visual searches for GRG candidates in survey images.
To cope with the sheer number of detectable RGs in modern surveys like the LoTSS, a natural criterion is to only add sources to a candidate list if they appear --- by eye --- to have an angular length larger than some threshold.
However, it is hard to precisely assess the angular length of a candidate before actually measuring it; sometimes, a candidate with a smaller angular length than the threshold will feature in the list, while some candidates with a larger angular length than the threshold will not.
This leads to the notion of a `fuzzy angular length threshold', where the probability that an RG with angular length $\phi$ is observed through the visual search increases (e.g. linearly) from 0 to 1 between $\phi_\mathrm{min}$ and $\phi_\mathrm{max}$:
\begin{align}
    p_\mathrm{obs,AL}\left(l_\mathrm{p},z\right) &= \min{\left\{\max{\left\{\frac{\phi\left(l_\mathrm{p},z\right) - \phi_\mathrm{min}}{\phi_\mathrm{max}-\phi_\mathrm{min}}, 0\right\}}, 1\right\}},\label{eq:probabilityObservedAL}\\
    \phi\left(l_\mathrm{p},z\right) &= \frac{l_\mathrm{p}\left(1+z\right)}{r\left(z\right)}.
    \label{eq:angularSize}
\end{align}
See the top row of Fig.~\ref{fig:completeness} for several examples of associated completeness curves $C(l_\mathrm{p})$ and observed projected proper length PDFs.
See Appendix~\ref{ap:completenessAngularLength} for additional information.

\subsubsection{Surface brightness limitations}
\label{sec:surfaceBrightness}
Another important observational selection effect is due to a survey's finite noise level.
The noise determines the surface brightness threshold $b_{\nu,\mathrm{th}}$ (typically comparable to the noise level itself) below which radio galaxy features remain visually undetected.
\paragraph{Fanaroff--Riley class II}
We model the lobe surface brightness RV $B_\nu$ at the central observing frequency $\nu_\mathrm{obs}$ as
\begin{align}
B_\nu = \frac{b_{\nu,\mathrm{ref}} \cdot S}{\left(1+Z\right)^{3-\alpha}}\left(\frac{L}{l_\mathrm{ref}}\right)^\zeta,
\label{eq:surfaceBrightnessRVMainText}
\end{align}
where $b_{\nu,\mathrm{ref}}$ is the median surface brightness of RGs of intrinsic proper length $l_\mathrm{ref}$ at cosmological redshift $z = 0$ and frequency $\nu_\mathrm{obs}$, and $S$ is a lognormally distributed RV with median $1$ and dispersion parameter $\sigma_\mathrm{ref}$ that captures the variability in surface brightness among this population of RGs.
The denominator models the fact that surface brightness is not conserved with distance in a FLRW universe; $Z$ is the cosmological redshift RV up to $z = z_\mathrm{max}$ and $\alpha$ is the typical lobe spectral index.
The exponent $\zeta < 0$ characterises the scaling between intrinsic proper length and surface brightness.
(If RGs are self-similar, so that morphology does not predict length, one finds $\zeta = -2$.)
For this selection effect, the observing probability is
\begin{align}
    p_\mathrm{obs,SB}\left(l_\mathrm{p},z\right) &= \int_{s_\mathrm{min}}^\infty\sqrt{1 - \left(\frac{s_\mathrm{min}}{s}\right)^{-\frac{2}{\zeta}}}f_S\left(s\right) \mathrm{d}s,\label{eq:probabilityObservedSB}\\
    s_\mathrm{min} &= \frac{b_{\nu,\mathrm{th}}}{b_{\nu,\mathrm{ref}}}\left(\frac{l_\mathrm{p}}{l_\mathrm{ref}}\right)^{-\zeta}\left(1+z\right)^{3-\alpha},\\
    f_S\left(s\right) &= \frac{1}{\sqrt{2\pi}\sigma_\mathrm{ref} s}\exp{\left(-\frac{\ln^2s}{2\sigma_\mathrm{ref}^2}\right)}.
\end{align}
We note that $p_\mathrm{obs,SB}$ does not depend on $b_{\nu,\mathrm{th}}$ or $b_{\nu,\mathrm{ref}}$ separately, but on their ratio only.
See the bottom row of Fig.~\ref{fig:completeness} for several examples of associated completeness curves $C(l_\mathrm{p})$ and observed projected proper length PDFs.

\paragraph{Fanaroff--Riley class I}
For FRI RGs, the assumption of a constant surface brightness beyond the core is inaccurate.
The simplest correction in which FRI RGs retain a well-defined notion of length assumes a linearly decreasing surface brightness profile, which peaks at the core and goes to zero at the RG's two endpoints.
(A power-law profile does not work: in such case, the surface brightness only goes asymptotically to zero --- but never actually reaches it.)
In this case, we define RV $B_\nu$ to be the mean surface brightness along an RG's jets, which can be regarded as a typical value for that RG.
As $B_\nu$ again obeys Eq.~\ref{eq:surfaceBrightnessRVMainText}, we find that the formulaic structure of $p_\mathrm{obs, SB}(l_\mathrm{p},z)$ is identical for FRI and FRII giants, except that FRI giants require a change
\begin{align}
    b_{\nu,\mathrm{th}} \to \frac{b_{\nu,\mathrm{th}}}{2\left(1 - \frac{l_\mathrm{p,GRG}}{l_\mathrm{p}}\right)},
\end{align}
which affects $p_\mathrm{obs, SB}(l_\mathrm{p},z)$ through $s_\mathrm{min}$.
There is no change for $l_\mathrm{p} = 2 l_\mathrm{p,GRG}$.
Although the formulaic structure might be the same, the best-fit parameters can differ.
For example, it is possible that $\sigma_\mathrm{ref, FRI} \neq \sigma_\mathrm{ref, FRII}$ or $\zeta_\mathrm{FRI} \neq \zeta_\mathrm{FRII}$.
See Appendix~\ref{ap:completenessSurfaceBrightness} for derivations and numerical implementation considerations.

To include both aforementioned selection effects at the same time, a natural approximation is to assume that the observing probability simply factorises:
\begin{align}
    p_\mathrm{obs}\left(l_\mathrm{p},z\right) \approx p_\mathrm{obs,AL}\left(l_\mathrm{p},z\right) \cdot p_\mathrm{obs,SB}\left(l_\mathrm{p},z\right).
\label{eq:probabilityObserved}
\end{align}

\subsection{GRG number density}
A central question in the field of radio galaxies is: how intrinsically rare are giants?
By counting giants in a search campaign with well-understood selection effects, we can give an answer.
More precisely, one can estimate the comoving number density of giants in the Local Universe, $n_\mathrm{GRG}$, through the number of observed giants up to cosmological redshift $z_\mathrm{max}$ in a uniformly searched region of sky of solid angle $\Omega$, $N_\mathrm{GRG,obs}\left(z_\mathrm{max}, \Omega\right)$.
Then, under the standard assumption $l_\mathrm{p,GRG} > l_\mathrm{min}$, Appendix~\ref{ap:GRGComovingNumberDensity} shows that
\begin{align}
    &n_\mathrm{GRG} = \frac{l_\mathrm{p,GRG}^{\xi+1}}{-(\xi+1)} \cdot \frac{H_0}{c} \cdot\nonumber\\
    &\frac{\frac{4\pi}{\Omega}N_\mathrm{GRG,obs}\left(z_\mathrm{max},\Omega\right)}{\int_{l_\mathrm{p,GRG}}^\infty l_\mathrm{p}^\xi \int_0^{z_\mathrm{max}} p_\mathrm{obs}\left(l_\mathrm{p},z\right) 4\pi r^2\left(z\right)E^{-1}\left(z\right)\ \mathrm{d}z\ \mathrm{d}l_\mathrm{p}}.
\label{eq:GRGComovingNumberDensity}
\end{align}
Although the appropriate $p_\mathrm{obs}(l_\mathrm{p},z)$ varies per search campaign, it is always possible to bound $p_\mathrm{obs}(l_\mathrm{p},z)$ from above --- for example by 1.
In such case, Eq.~\ref{eq:GRGComovingNumberDensity} bounds $n_\mathrm{GRG}$ from below.

\subsection{GRG lobe volume-filling fraction}
Because giants attain cosmological lengths, they might contribute to the energisation and magnetisation of Cosmic Web filaments in regions that smaller radio galaxies cannot reach.
A key statistic that measures the enrichment of the Cosmic Web by giants is the volume-filling fraction (VFF) of their lobes.
Assuming that lobes do not grow along with the expansion of the Universe, the proper VFF changes over cosmic time: $\mathrm{VFF}_\mathrm{GRG}(z) = \mathrm{VFF}_\mathrm{GRG}(z = 0) \cdot (1+z)^3$, where
\begin{align}
    \mathrm{VFF}_\mathrm{GRG}(z = 0) \coloneqq&\ n_\mathrm{GRG} \cdot \mathbb{E}[V\ \vert\ L_\mathrm{p} > l_\mathrm{p,GRG}]\nonumber\\
    =&\ n_\mathrm{GRG} \cdot \mathbb{E}[\Upsilon \cdot L^3\ \vert\ L_\mathrm{p} > l_\mathrm{p,GRG}],
\end{align}
where $V$ is the combined volume of the lobes and $\Upsilon \coloneqq \frac{V}{L^3}$ is a dimensionless RV that captures the diversity in radio galaxy lobe shapes.
We find under self-similar growth
\begin{align}
\mathrm{VFF}_\mathrm{GRG}(z = 0) = n_\mathrm{GRG} \cdot \mathbb{E}[\Upsilon] \cdot \mathbb{E}[L^3\ \vert\ L_\mathrm{p} > l_\mathrm{p,GRG}].
\end{align}
$\mathbb{E}[\Upsilon]$ can be estimated from observations, but one must be wary of selection effects.
Unfortunately, $\mathbb{E}[L^3\ \vert\ L_\mathrm{p} > l_\mathrm{p,GRG}]$ does not exist for $\xi \geq -4$, which is the regime supported by observations.
A useful lower bound then is
\begin{align}
\mathrm{VFF}_\mathrm{GRG}(z = 0) > n_\mathrm{GRG} \cdot \mathbb{E}[\Upsilon] \cdot \mathbb{E}^3[L\ \vert\ L_\mathrm{p} > l_\mathrm{p,GRG}],
\label{eq:VFFLowerBound}
\end{align}
where
\begin{align}
    \mathbb{E}[L\ \vert\ L_\mathrm{p}>l_\mathrm{p,GRG}] = l_\mathrm{p,GRG} \frac{\Gamma\left(-\frac{\xi}{2}-1\right)\Gamma\left(-\frac{\xi}{2}+1\right)}{\Gamma\left(-\frac{\xi}{2}-\frac{1}{2}\right)\Gamma\left(-\frac{\xi}{2}+\frac{1}{2}\right)}.
\end{align}
This is the mean intrinsic proper length of giants, which is only defined when $\xi < -2$.\footnote{
For example, when $\xi = -3$, $\mathbb{E}[L\ \vert\ L_\mathrm{p}>l_\mathrm{p,GRG}] = \frac{3\pi}{4} l_\mathrm{p,GRG}$, which becomes $1.65\ \mathrm{Mpc}$ for $l_\mathrm{p,GRG} \coloneqq 0.7\ \mathrm{Mpc}$ and $2.36\ \mathrm{Mpc}$ for $l_\mathrm{p,GRG} \coloneqq 1\ \mathrm{Mpc}$.
When $\xi = -4$, $\mathbb{E}[L\ \vert\ L_\mathrm{p}>l_\mathrm{p,GRG}] = \frac{16}{3\pi} l_\mathrm{p,GRG}$, which becomes $1.19\ \mathrm{Mpc}$ for $l_\mathrm{p,GRG} \coloneqq 0.7\ \mathrm{Mpc}$ and $1.70\ \mathrm{Mpc}$ for $l_\mathrm{p,GRG} \coloneqq 1\ \mathrm{Mpc}$.}

An alternative is to deviate slightly from our Pareto ansatz and truncate the GRG projected proper length distribution at some $l_\mathrm{p,max}$; then
\begin{align}
    \mathrm{VFF}_\mathrm{GRG}(z = 0) = n_\mathrm{GRG} \cdot \mathbb{E}[\Upsilon_\mathrm{p}] \cdot \frac{\xi + 1}{\xi + 4} \cdot \frac{l_\mathrm{p,max}^{\xi + 4} - l_\mathrm{p,GRG}^{\xi + 4}}{l_\mathrm{p,max}^{\xi + 1} - l_\mathrm{p,GRG}^{\xi + 1}},
\label{eq:VFFTruncated}
\end{align}
where $\Upsilon_\mathrm{p} \coloneqq \frac{V}{L_\mathrm{p}^3}$.
See Appendix~\ref{ap:VFF} for a derivation and further details.

\subsection{Unification model constraints from quasar and non-quasar giants}
\label{sec:unificationModelTheory}
The unification model and its extensions \citep[e.g.][]{Hardcastle12020} form an elegant family of hypotheses that aim to explain the observational diversity of active galaxies.
It posits that active galaxies with quasars differ from those without quasars primarily because of differences in orientation of the dusty tori surrounding SMBHs.
In particular, the central idea is that a quasar appears brighter to the observer than a non-quasar AGN because the axis of its dusty torus happens to be virtually parallel to the line of sight.
As such, only quasars would offer an unobscured view of the luminous accretion disk surrounding the SMBH, whilst also beaming relativistic jet emission towards the observer.
Using our statistical framework, we predict the general ramifications of the basic unification model on a GRG sample.

The basic unification model suggests to divide the radio galaxy population in two, distinguishing between RGs generated by AGN with quasar appearance (quasar RGs) and RGs generated by AGN without quasar appearance (non-quasar RGs).
We assume that quasar RGs have inclination angles $\theta \leq \theta_\mathrm{max}$ or $\theta \geq 180\degree - \theta_\mathrm{max}$ whilst non-quasar RGs have $\theta_\mathrm{max} < \theta < 180\degree - \theta_\mathrm{max}$.\footnote{Geometrically, $\theta_\mathrm{max}$ represents the opening angle of the two coaxial conical gaps in the dusty torus of the AGN.}
If quasar RGs are more closely aligned with the line of sight than non-quasar RGs but are otherwise similar, fewer of them will satisfy $l_\mathrm{p} \geq l_\mathrm{p,GRG}$ and thus be classified as giants.
Therefore, the quasar GRG fraction $f_\mathrm{Q}$ --- the fraction of quasar giants in an actual GRG sample --- constrains $\theta_\mathrm{max}$.
We model $f_\mathrm{Q}$ as an RV:
\begin{align}
    f_\mathrm{Q} \coloneqq \frac{N_\mathrm{Q}}{N};\ \ \ N_\mathrm{Q} \sim \mathrm{Binom}\left(N, p_\mathrm{Q}\right),
\end{align}
where the RV $N_\mathrm{Q}$ is the number of quasar giants in the sample, the constant $N$ is the total number of giants in the sample and the parameter $p_\mathrm{Q}$ is the quasar GRG probability.
Our framework predicts
\begin{align}
    p_\mathrm{Q} \coloneqq& \frac{\mathbb{P}\left(L_\mathrm{p,obs} \geq l_\mathrm{p,GRG},\ \sin{\Theta} \leq \sin{\theta_\mathrm{max}}\right)}{\mathbb{P}\left(L_\mathrm{p,obs} \geq l_\mathrm{p,GRG}\right)}\nonumber\\
    =& \frac{4 \Gamma\left(-\frac{\xi}{2}+1\right)}{-\left(\xi+1\right)\sqrt{\pi}\ \Gamma\left(-\frac{\xi}{2}-\frac{1}{2}\right)}\int_0^{\sin{\theta_\mathrm{max}}}\frac{x^{-\xi}\ \mathrm{d}x}{\sqrt{1-x^2}}.
\label{eq:quasarGRGProbability}
\end{align}
See Appendix~\ref{ap:unificationModel} for a derivation.
Interestingly, as long as quasar giants and non-quasar giants are subject to the same selection effects, these selection effects do not affect $p_\mathrm{Q}$.
\begin{figure}
    \centering
    \includegraphics[width=\columnwidth]{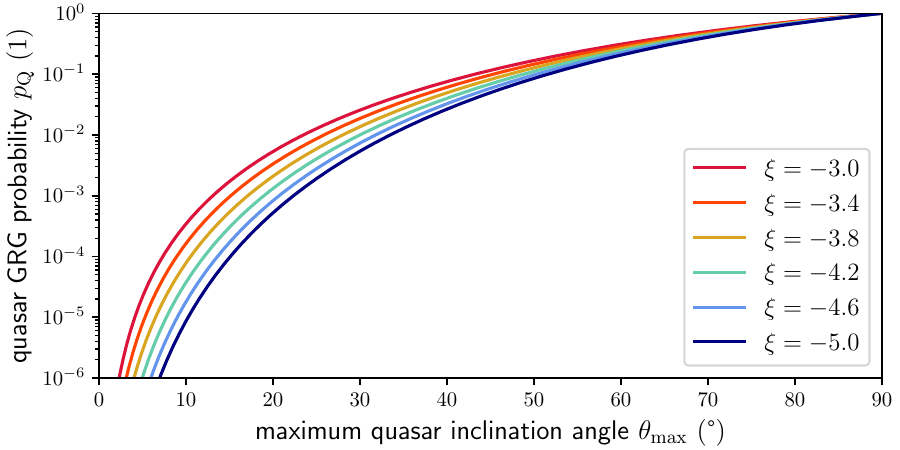}
    \caption{
    Probability $p_\mathrm{Q}$ that an observed giant is a quasar giant under the unification model.
    Under this model, giants generated by AGN with quasar appearance (quasar giants) have inclination angles $\theta \leq \theta_\mathrm{max}$ or $\theta \geq 180\degree - \theta_\mathrm{max}$ and giants generated by AGN without quasar appearance (non-quasar giants) have intermediate $\theta$.
    As long as quasar giants and non-quasar giants are subject to the same selection effects, these selection effects do not affect $p_\mathrm{Q}$.
    Instead, in such case, $p_\mathrm{Q}$ only depends on the maximum quasar inclination angle $\theta_\mathrm{max}$ and the tail index $\xi$.
    }
    \label{fig:quasarGRGProbability}
\end{figure}\noindent
Figure~\ref{fig:quasarGRGProbability} shows that, for all relevant $\xi$, $p_\mathrm{Q}$ is a steeply and monotonically increasing function of $\theta_\mathrm{max}$.
Thus, knowing $\xi$, one can use an empirical $f_\mathrm{Q}$ to determine $p_\mathrm{Q}$ and in turn $\theta_\mathrm{max}$.

Does one expect quasar giants to have a different distribution for $L_\mathrm{p,obs}\ \vert\ L_\mathrm{p,obs} \geq l_\mathrm{p,GRG}$ than non-quasar giants?
Interestingly, our framework allows us to prove that the inclination angle differences between the two classes affect their relative rarity, but not their observed projected proper length distribution.
Under the basic unification model, quasar giants and non-quasar giants obey the same $L_\mathrm{p,obs}\ \vert\ L_\mathrm{p,obs} \geq l_\mathrm{p,GRG}$ if they are subject to the same selection effects.

\subsection{Extreme giants in a sample}
\begin{figure}
    \centering
    \begin{subfigure}{\columnwidth}
    \includegraphics[width=\columnwidth]{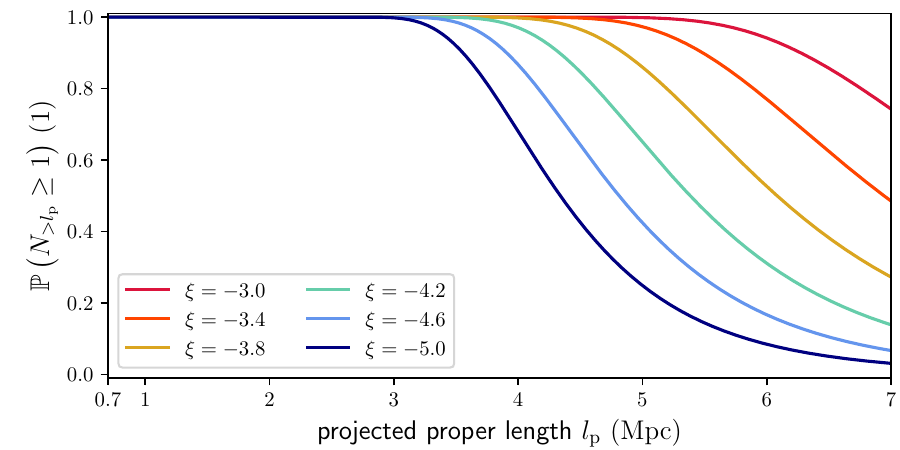}
    \end{subfigure}
    \begin{subfigure}{\columnwidth}
    \includegraphics[width=\columnwidth]{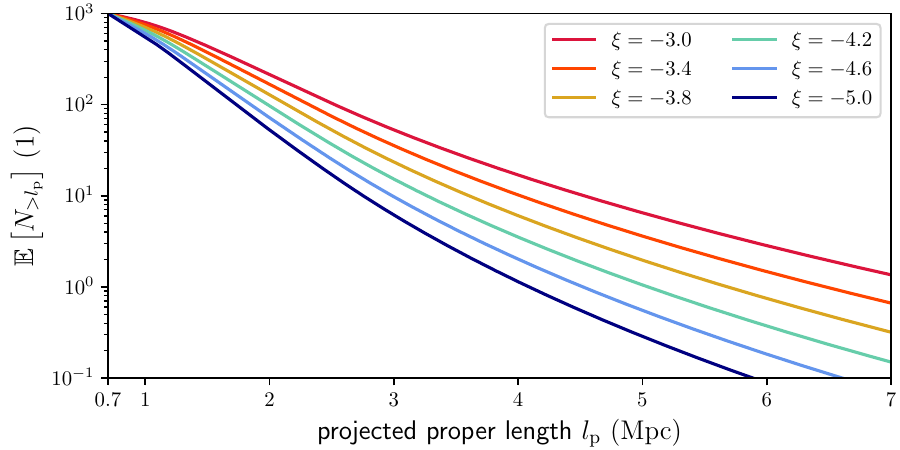}
    \end{subfigure}
    \caption{
    Predictions of the existence and expected number of giants that exceed projected length $l_\mathrm{p}$ in a sample of cardinality $N$, as functions of $l_\mathrm{p}$.
    Both tail index $\xi$ and selection effect parameters affect these predictions.
    We consider a sample of $N = 1000$ giants with redshifts below $z_\mathrm{max} = 0.5$, use $\phi_\mathrm{min} = 3'$, $\phi_\mathrm{max} = 7'$, $b_{\nu,\mathrm{ref}} = 1000\ \mathrm{Jy\ deg^{-2}}$, and $\sigma_\mathrm{ref} = 1.5$, and keep other parameters as in Fig.~\ref{fig:completeness}.
    \textit{Top:} the probability that at least one observed giant has a projected length of at least $l_\mathrm{p}$.
    \textit{Bottom:} the expected number of observed giants with a projected length of at least $l_\mathrm{p}$.
    }
    \label{fig:modelPredictions}
\end{figure}\noindent
The model can predict the occurrence of giants with extreme projected proper lengths in a sample of cardinality $N$.
The probability that an observed GRG has a projected proper length exceeding $l_\mathrm{p}$ is
\begin{align}
    p_{>l_\mathrm{p}} \coloneqq \mathbb{P}\left(L_\mathrm{p,obs} > l_\mathrm{p}\ \vert\ L_\mathrm{p,obs} > l_\mathrm{p,GRG}\right) = \frac{1 - F_{L_\mathrm{p,obs}}\left(l_\mathrm{p}\right)}{1 - F_{L_\mathrm{p,obs}}\left(l_\mathrm{p,GRG}\right)},
\end{align}
so that the number of observed giants with a projected proper length exceeding $l_\mathrm{p}$ is $N_{>l_\mathrm{p}} \sim \mathrm{Binom}(N, p_{>l_\mathrm{p}})$.
Its expectation is $\mathbb{E}[N_{>l_\mathrm{p}}] = N \cdot p_{>l_\mathrm{p}}$.
Furthermore, the probability that the sample contains at least one giant with projected proper length $l_\mathrm{p}$ or higher is
\begin{align}
    \mathbb{P}\left(N_{>l_\mathrm{p}} \geq 1\right) = 1 - \left(1 - p_{>l_\mathrm{p}}\right)^N.
\end{align}
See Appendix~\ref{ap:extremeGRGs} for details.
Figure~\ref{fig:modelPredictions} shows $\mathbb{E}[N_{>l_\mathrm{p}}]$ and $\mathbb{P}(N_{>l_\mathrm{p}} \geq 1)$ for various $\xi$.
As an example, the case $\xi = -3.0$ predicts that a sample of $N = 1000$ giants with redshifts below $z_\mathrm{max} = 0.5$ should contain almost ten giants of $l_\mathrm{p} > 5\ \mathrm{Mpc}$, and still several of $l_\mathrm{p} > 6\ \mathrm{Mpc}$.
Such predictions are useful as they can be directly compared to elementary sample statistics.

\begin{figure*}[p]
    \centering
    \begin{subfigure}{\columnwidth}
    \includegraphics[width=\columnwidth]{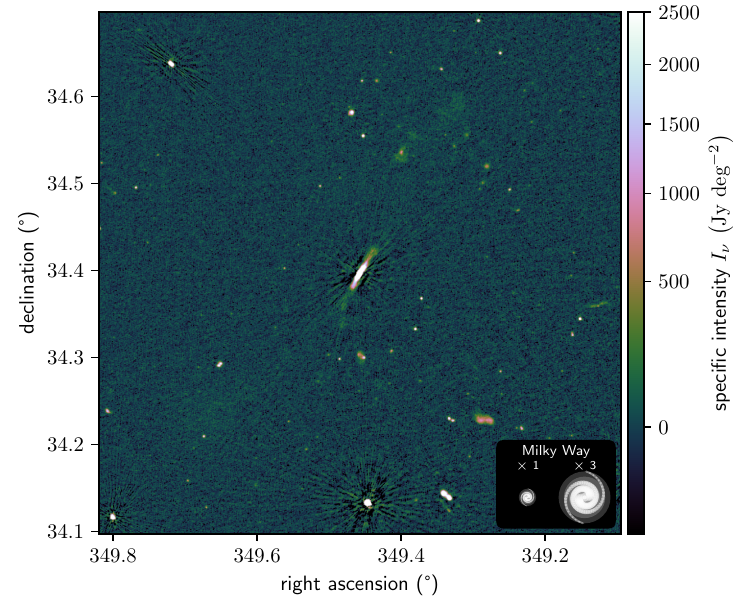}
    \end{subfigure}
    \begin{subfigure}{\columnwidth}
    \includegraphics[width=\columnwidth]{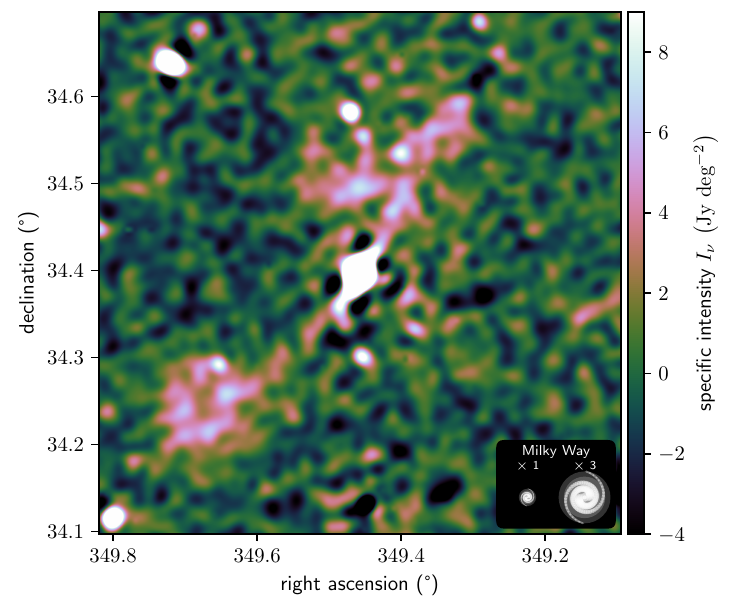}
    \end{subfigure}
    \begin{subfigure}{\columnwidth}
    \includegraphics[width=\columnwidth]{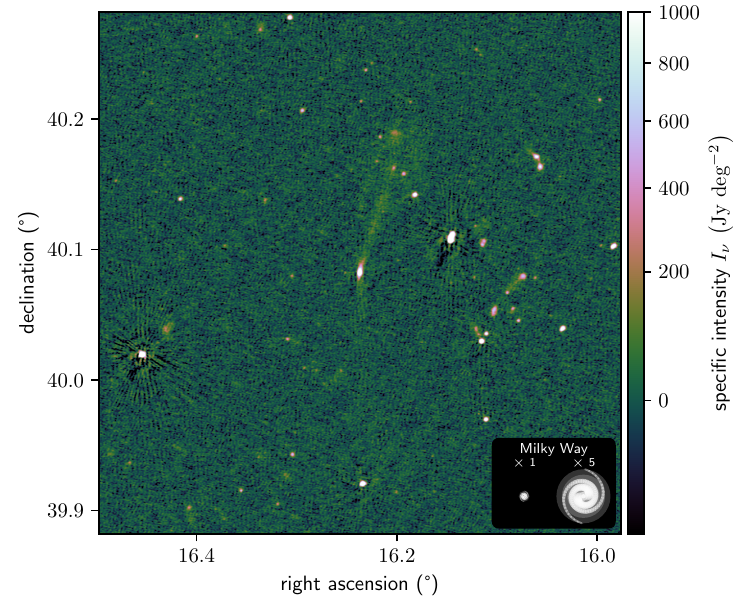}
    \end{subfigure}
    \begin{subfigure}{\columnwidth}
    \includegraphics[width=\columnwidth]{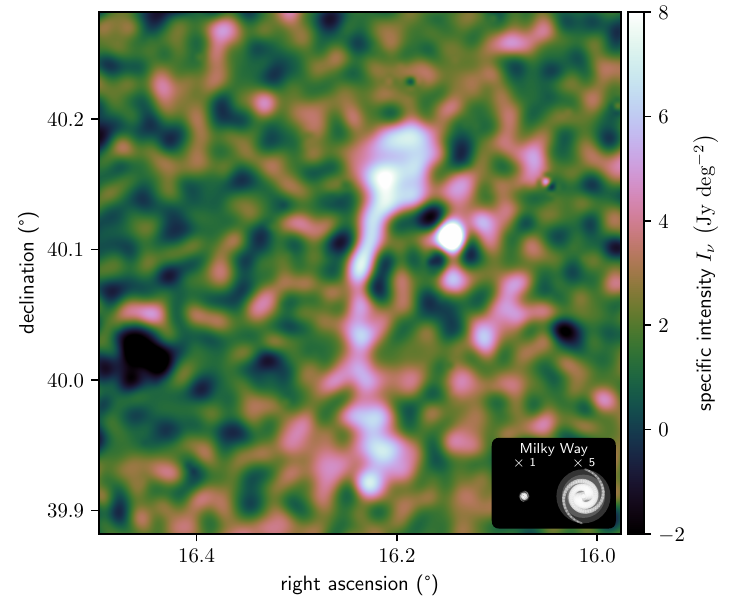}
    \end{subfigure}
    \begin{subfigure}{\columnwidth}
    \includegraphics[width=\columnwidth]{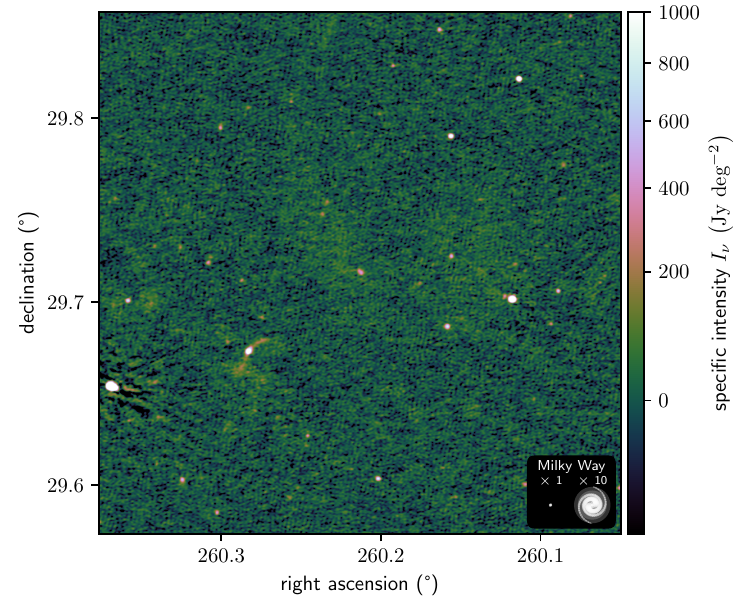}
    \end{subfigure}
    \begin{subfigure}{\columnwidth}
    \includegraphics[width=\columnwidth]{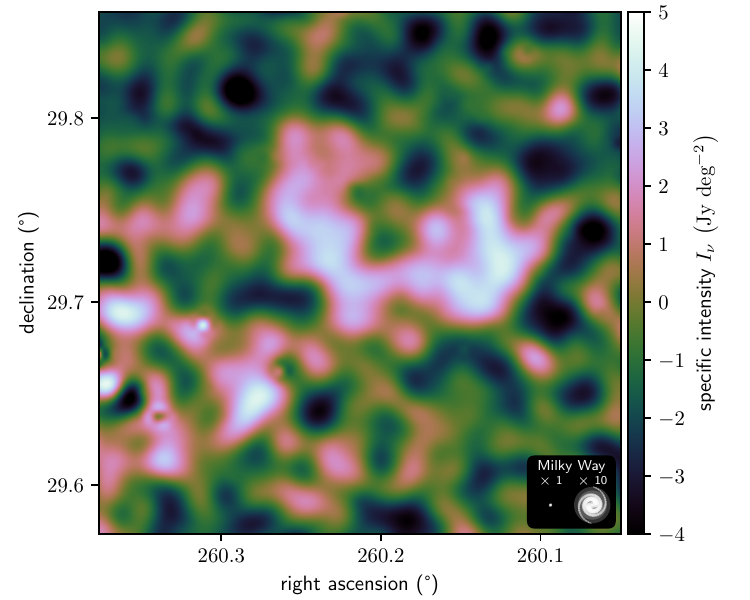}
    \end{subfigure}
    \caption{
    LoTSS DR2 cutouts of three newly discovered giants at $6''$ \textit{(left column)} and $60''$ \textit{(right column)}.
    By subtracting compact sources from calibrated $144\ \mathrm{MHz}$ visibility data and imaging at low resolution ($60''$ and $90''$), we reveal otherwise speculative giant radio galaxies at the unexplored ${\sim}1\ \mathrm{Jy\ deg^{-2}}$ surface brightness level.
    The claimed host galaxy is in the image centre.
    \textit{Top:} a GRG of projected proper length $l_\mathrm{p} = 1.4 \pm 0.3\ \mathrm{Mpc}$, whose angular length $\phi = 32.3 \pm 0.2'$ is larger than that of the full Moon.
    %
    %
    %
    %
    \textit{Middle:} a GRG of $l_\mathrm{p} = 1.6 \pm 0.6\ \mathrm{Mpc}$ and $\phi = 16.4 \pm 0.2'$.
    \textit{Bottom:} a GRG of $l_\mathrm{p} = 3.6 \pm 0.1\ \mathrm{Mpc}$ and $\phi = 8.5 \pm 0.2'$.
    For scale, we show the stellar Milky Way disk (with a diameter of 50 kpc) generated using the \citet{Ringermacher12009} formula, alongside a 3, 5, or 10 times inflated version.
    }
    \label{fig:comparison6Vs90Arcsec}
\end{figure*}\noindent
\pagebreak

\begin{figure}
    \centering
    \includegraphics[width=\columnwidth]{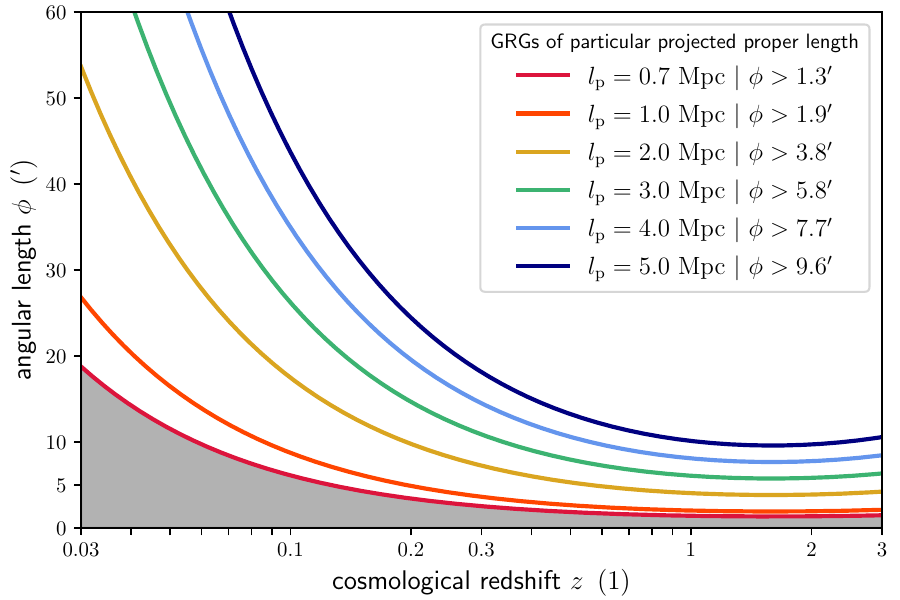}
    \caption{
    Relations between cosmological redshift $z$ and angular length $\phi$ for six giants of different projected lengths $l_\mathrm{p}$.
    Due to the expansion of the Universe, there is a minimum angular length for each $l_\mathrm{p}$.
    If one defines giants as RGs with $l_\mathrm{p} \geq l_\mathrm{p,GRG} = 0.7\ \mathrm{Mpc}$, all giants have an angular length of $1.3'$ or above.
    If one instead defines giants as RGs with $l_\mathrm{p} \geq l_\mathrm{p,GRG} = 1\ \mathrm{Mpc}$, all giants have an angular length of $1.9'$ or above.
    }
    \label{fig:angularLengthRedshift}
\end{figure}\noindent

\begin{figure}
    \centering
    \begin{subfigure}{\columnwidth}
    \includegraphics[width=\columnwidth]{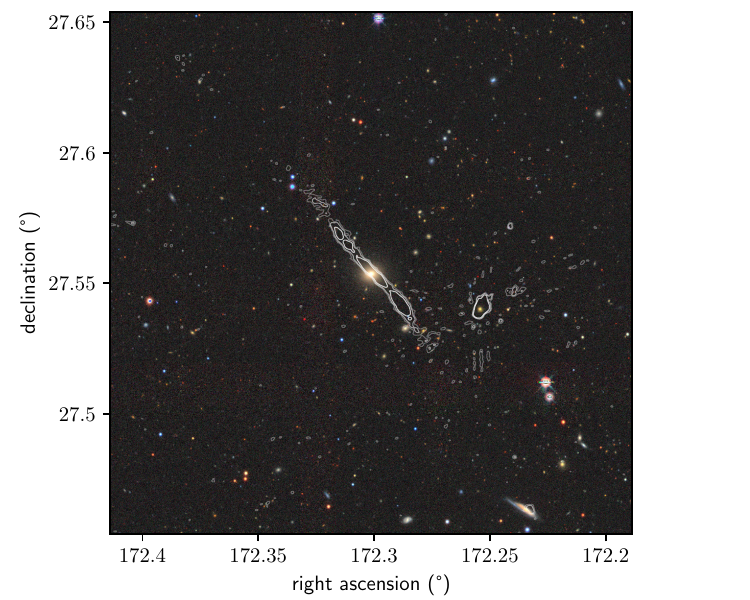}
    \end{subfigure}
    \begin{subfigure}{\columnwidth}
    \includegraphics[width=\columnwidth]{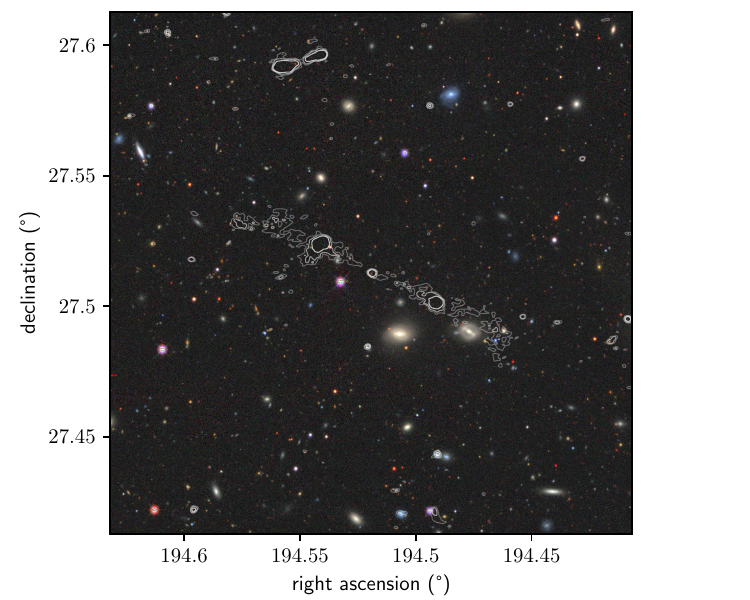}
    \end{subfigure}
    \begin{subfigure}{\columnwidth}
    \includegraphics[width=\columnwidth]{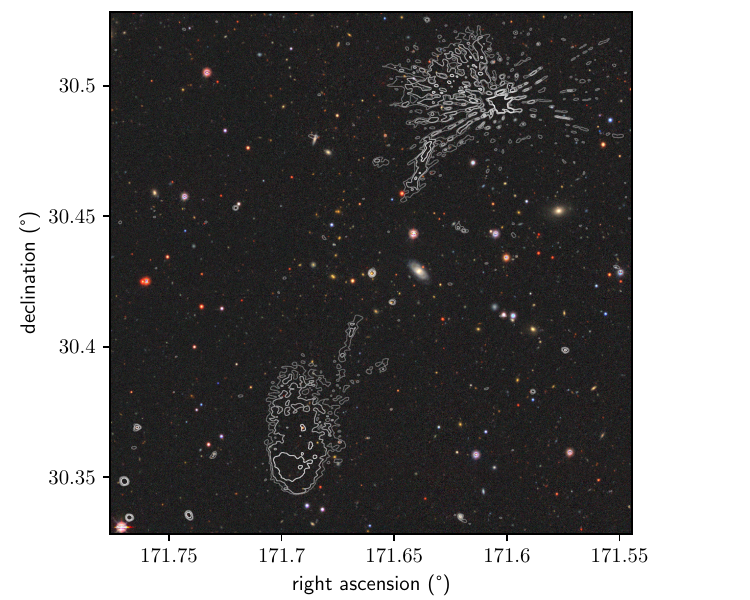}
    \end{subfigure}
    \caption{
    $12' \times 12'$ DESI Legacy Imaging Surveys DR9 $(g,r,z)$-details with LoTSS DR2 $6''$ contours ($3\sigma$, $5\sigma$, $10\sigma$) overlaid.
    At $6''$ resolution, LoTSS images allow for more accurate host galaxy identification in SDSS, Pan-STARRS, and Legacy Survey images than was possible before.
    \textit{Top:} the jet of the giant at rank 33 of Table~\ref{tab:GRGProperties} and shown in the middle-left panel of Fig.~\ref{fig:LoTSSDR2GRGs2}.
    \textit{Middle:} the giant at rank 37 of Table~\ref{tab:GRGProperties}.
    \textit{Bottom:} the giant at rank 43 of Table~\ref{tab:GRGProperties}.
    }
    \label{fig:hostIdentification}
\end{figure}

\section{Sample compilation and properties}
\label{sec:LoTSSDR2GRGSearch}
To measure the intrinsic GRG length distribution, we also require a large sample of giants collected from a single survey through a systematic approach.
This ensures approximately homogeneous selection effects, which we can correct for in subsequent analysis using the statistical framework of Sect.~\ref{sec:theory}.

\subsection{LoTSS DR2}
The Low-Frequency Array \citep[LOFAR;][]{vanHaarlem12013} is a powerful, Pan-European radio interferometer that features both (sub)arcsecond-scale resolution and sensitivity to degree-scale structures.
\citet{Dabhade12020March} have already demonstrated that this combination is ideal for detecting giants: these authors found a record 225 new specimina in the LoTSS DR1, the first data release of the LOFAR's Northern Sky survey at central frequency $\nu_\mathrm{obs} = 144\ \mathrm{MHz}$.\footnote{As in the current work, the authors defined giants using the projected proper length threshold $l_\mathrm{p,GRG} = 0.7\ \mathrm{Mpc}$.}
Excitingly, the recent LoTSS DR2 \citep{Shimwell12022} improves the data calibration and increases the survey footprint from $424\ \mathrm{deg}^2$ to $5635\ \mathrm{deg}^2$ --- that is by more than a factor 13.
By default, the LoTSS features imagery at $6''$ and $20''$ resolutions.
To further facilitate the discovery of giants (among other goals), we reprocessed the LoTSS by subtracting compact sources and imaging at $60''$ and $90''$ resolution; more details are given in \citet{Oei12022Alcyoneus} and \citet{Oei12022MilkyWay}.
This $60''$ and $90''$ imagery has turned out to be effective in highlighting jets and lobes of RGs of large angular and physical extent, whose surface brightnesses are low and which therefore have remained undetected in shallower surveys, and even in the LoTSS DR2 at higher resolutions.
We demonstrate this fact in Fig.~\ref{fig:comparison6Vs90Arcsec} by comparing the LoTSS DR2 at $6''$ and $60''$ for three giants whose discovery has relied on the lower-resolution images.
After the serendipitous discovery of several such hitherto unknown giants in the $60''$ and $90''$ images, we decided to initiate a systematic, multi-resolution, visual GRG search through the area covered by LoTSS DR2 pipeline products as of September 2022.\footnote{In this process, we have skipped the enclosed LoTSS DR1 area, which has already been analysed by \citet{Dabhade12020March}.}
This search comprised of a hundreds-of-hours-long inspection of the LoTSS maps at $6''$ and $60''$, alongside Pan-STARRS DR1 \citep{Chambers12016} and SDSS DR9 \citep{Ahn12012} maps, in \textit{Aladin Desktop 11.0} \citep{Bonnarel12000}.

Reliable automated search strategies do not yet exist for several reasons.
Giants showcase a rich morphological variety (see Sect.~\ref{sec:LoTSSDR2GRGSample}) and are easily confused with other types of astrophysical sources (see Sects.~\ref{sec:GRGCandidatesRadio} and \ref{sec:GRGCandidatesOptical}).
Moreover, the known population is too small to effectively apply supervised learning techniques.
However, it appears possible to find giants in morphological outlier lists of unsupervised learning techniques such as self-organising maps \citep[SOMs;][]{Mostert12021}.
The efficacy of such techniques in GRG searches has not yet been quantified.

\subsection{Angular length threshold}
\label{sec:angularLengthThreshold}
To limit the amount of manual work, we decided to search only for GRG candidates whose angular length exceeds some threshold.
In Fig.~\ref{fig:angularLengthRedshift}, we show how the angular length $\phi$ of giants with six different projected proper lengths $l_\mathrm{p}$ varies as a function of cosmological redshift $z$.
Because of the expansion of the Universe, giants have an arcminute-scale minimum angular length: if $l_\mathrm{p,GRG} = 0.7\ \mathrm{Mpc}$, all giants obey $\phi > 1.3'$.\footnote{If $l_\mathrm{p,GRG} = 1\ \mathrm{Mpc}$, all giants obey $\phi > 1.9'$.}
For the purpose of finding a GRG, it is therefore never useful to inspect a source with an angular length less than $1.3'$.
Our highest priority has been to find giants with $z < 0.2$, which lie in a volume for which the total matter density field is known or will be known in the coming years through the combined power of deep spectroscopic surveys and Bayesian inference frameworks, such as the Bayesian Origin Reconstruction from Galaxies \citep[BORG;][]{Jasche12013, Jasche12015, Jasche12019}.
In an upcoming publication, we combine a sample of low-redshift giants with the BORG to analyse the large-scale environments of giants \citep{Oei12022GiantsCosmicWeb}.
Figure~\ref{fig:angularLengthRedshift} shows that all giants with $l_\mathrm{p} > 1\ \mathrm{Mpc}$ at $z < 0.2$ have an angular length $\phi > 5'$.
For this reason, we have chosen $5'$ as the angular length threshold of our search campaign.
This choice has kept the visual inspection duration to order ${\sim}10^2\ \mathrm{h}$, while still enabling us to target all Mpc-exceeding giants in the Local Universe ($z < 0.2$).
In practice, this threshold is `fuzzy': it is hard to accurately estimate angular lengths by eye before performing an actual measurement, so that our list of GRG candidates does contain some with a smaller angular length than the specified threshold.
Inversely, it will presumably lack some GRG candidates with an angular length exceeding the threshold.

\begin{table*}[h]
\centering
\scriptsize
\begin{tabular}{l l l l l l l l l l}
\hline
rank & host name & host coordinates & redshift & redshift & angular & projected proper & host stellar & host SMBH & host\\
& SDSS DR12 & J2000 ($\degree$) & (1) & type & length ($'$) & length (Mpc) & mass ($10^{11} M_\odot$) & mass ($10^9 M_\odot$) & quasar\\
\hline\arrayrulecolor{lightgray}
$1$ & J081956.41+323537.6 & $124.9851,\ 32.5938$ & $0.749$ $\pm$ $0.073$ & \textit{p} & $11.2$ & $5.07 \pm 0.20$ & $\ldots$ & $\ldots$ & $\ldots$\\
\hline
$2$ & J081421.68+522410.0 & $123.5904,\ 52.4028$ & $0.2467$ $\pm$ $6 \cdot 10^{-5}$ & \textit{s} & $20.8$ & $4.99 \pm 0.04$ & $2.4 \pm 0.4$ & $0.4 \pm 0.2$ & \textit{n}\\
\hline
$3$ & J142910.70+311245.0 & $217.2946,\ 31.2125$ & $0.5921$ $\pm$ $0.0001$ & \textit{s} & $11.7$ & $4.80 \pm 0.06$ & $\ldots$ & $2.3 \pm 2.0$ & $\textit{n}$\\
\hline
$4$ & J131823.42+262622.8 & $199.5976,\ 26.4397$ & $0.6230$ $\pm$ $5 \cdot 10^{-5}$ & \textit{s} & $11.0$ & $4.62 \pm 0.06$ & $\ldots$ & $\ldots$ & \textit{y}\\
\hline
$5$ & J152634.77+262003.2 & $231.6449,\ 26.3342$ & $0.1507$ $\pm$ $2 \cdot 10^{-5}$ & \textit{s} & $28.0$ & $4.56 \pm 0.03$ & $3.7 \pm 0.6$ & $1.4 \pm 0.3$ & \textit{n}\\
\hline
$6$ & J121815.66+382407.5 & $184.5653,\ 38.4021$ & $0.634$ $\pm$ $0.064$ & \textit{p} & $10.6$ & $4.49 \pm 0.21$ & $\ldots$ & $\ldots$ & $\ldots$\\
\hline
$7$ & J175735.88+405154.2 & $269.3995,\ 40.8651$ & $0.585$ $\pm$ $0.036$ & \textit{p} & $10.5$ & $4.29 \pm 0.14$ & $\ldots$ & $\ldots$ & $\ldots$\\
\hline
$8$ & J161622.52+111135.7 & $244.0939,\ 11.1933$ & $0.3574$ $\pm$ $7 \cdot 10^{-5}$ & \textit{s} & $13.4$ & $4.15 \pm 0.05$ & $9.5 \pm 1.8$ & $5.7 \pm 3.1$ & \textit{n}\\
\hline
$9$ & J154709.22+353846.1 & $236.7884,\ 35.6462$ & $0.0794$ $\pm$ $1 \cdot 10^{-5}$ & \textit{s} & $43.8$ & $4.08 \pm 0.01$ & $4.6 \pm 0.1$ & $3.9 \pm 0.9$ & \textit{n}\\
\hline
$10$ & J013406.32+301537.2 & $23.5264,\ 30.2604$ & $0.884$ $\pm$ $0.138$ & \textit{p} & $8.5$ & $4.06 \pm 0.22$ & $\ldots$ & $\ldots$ & $\ldots$\\
\hline
$11$ & J082747.88+662813.6 & $126.9495,\ 66.4705$ & $0.968$ $\pm$ $0.160$ & \textit{p} & $8.2$ & $4.02 \pm 0.20$ & $\ldots$ & $\ldots$ & $\ldots$\\
\hline
$12$ & J012440.54+194003.9 & $21.1689,\ 19.6678$ & $0.578$ $\pm$ $0.162$ & \textit{p} & $9.6$ & $3.90 \pm 0.58$ & $\ldots$ & $\ldots$ & $\ldots$\\
\hline
$13$ & $\ldots$ & $238.4466,\ 28.4763$ & $1.094$ $\pm$ $0.122$ & \textit{p} & $7.7$ & $3.88 \pm 0.12$ & $\ldots$ & $\ldots$ & $\ldots$\\
\hline
$14$ & $\ldots$ & $275.3624,\ 26.6599$ & $0.0850$ $\pm$ $0.0001$ & \textit{s} & $39.0$ & $3.86 \pm 0.02$ & $\ldots$ & $\ldots$ & $\ldots$\\
\hline
$15$ & J162656.58+543421.3 & $246.7358,\ 54.5726$ & $0.4887$ $\pm$ $0.0001$ & \textit{s} & $10.3$ & $3.84 \pm 0.06$ & $\ldots$ & $1.0 \pm 1.6$ & \textit{n}\\
\hline
$16$ & J220605.67+275100.3 & $331.5237,\ 27.8501$ & $0.317$ $\pm$ $0.116$ & \textit{p} & $13.2$ & $3.78 \pm 0.99$ & $\ldots$ & $\ldots$ & $\ldots$\\
\hline
$17$ & J023544.96+310447.5 & $38.9373,\ 31.0799$ & $0.541$ $\pm$ $0.063$ & \textit{p} & $9.6$ & $3.77 \pm 0.23$ & $\ldots$ & $\ldots$ & $\ldots$\\
\hline
$18$ & $\ldots$ & $136.9661,\ 67.1071$ & $0.754$ $\pm$ $0.081$ & \textit{p} & $8.3$ & $3.77 \pm 0.17$ & $\ldots$ & $\ldots$ & $\ldots$\\
\hline
$19$ & J180117.72+510722.4 & $270.3239,\ 51.1229$ & $0.448$ $\pm$ $0.086$ & \textit{p} & $10.3$ & $3.66 \pm 0.42$ & $\ldots$ & $\ldots$ & $\ldots$\\
\hline
$20$ & J123900.69+360924.5 & $189.7529,\ 36.1568$ & $0.5935$ $\pm$ $6 \cdot 10^{-5}$ & \textit{s} & $8.8$ & $3.62 \pm 0.06$ & $\ldots$ & $\ldots$ & \textit{y}\\
\hline
$21$ & J102430.93+381842.8 & $156.1289,\ 38.3119$ & $0.411$ $\pm$ $0.028$ & \textit{p} & $10.7$ & $3.62 \pm 0.16$ & $\ldots$ & $\ldots$ & $\ldots$\\
\hline
$22$ & \textcolor{lightgray}{J172051.08+294256.8} & \textcolor{lightgray}{$260.2129,\ 29.7158$} & $\geq0.620$ $\pm$ $0.044$ & \textit{p} & $8.5$ & $\geq3.56 \pm 0.13$ & \textcolor{lightgray}{$\ldots$} & \textcolor{lightgray}{$\ldots$} & \textcolor{lightgray}{$\ldots$}\\
\hline
$23$ & J090534.54+563052.0 & $136.3939,\ 56.5145$ & $0.898$ $\pm$ $0.056$ & \textit{p} & $7.4$ & $3.55 \pm 0.10$ & $\ldots$ & $\ldots$ & $\ldots$\\
\hline
$24$ & J133105.80+293435.7 & $202.7742,\ 29.5766$ & $0.734$ $\pm$ $0.034$ & \textit{p} & $7.9$ & $3.55 \pm 0.09$ & $\ldots$ & $\ldots$ & $\ldots$\\
\hline
$25$ & J223649.76+251242.5 & $339.2074,\ 25.2118$ & $0.749$ $\pm$ $0.075$ & \textit{p} & $7.8$ & $3.53 \pm 0.15$ & $\ldots$ & $\ldots$ & $\ldots$\\
\hline
$26$ & J230125.38+240148.2 & $345.3558,\ 24.0301$ & $0.450$ $\pm$ $0.089$ & \textit{p} & $9.9$ & $3.53 \pm 0.41$ & $\ldots$ & $\ldots$ & $\ldots$\\
\hline
$27$ & \textcolor{lightgray}{J004848.01+021003.1} & \textcolor{lightgray}{$12.2001,\ 2.1675$} & $\geq0.3604$ $\pm$ $9 \cdot 10^{-5}$ & \textit{s} & $11.2$ & $\geq3.49 \pm 0.05$ & \textcolor{lightgray}{$\ldots$} & \textcolor{lightgray}{$3.0 \pm 1.5$} & \textcolor{lightgray}{\textit{n}}\\
\hline
$28$ & J220239.13+070656.7 & $330.6631,\ 7.1158$ & $0.4649$ $\pm$ $7 \cdot 10^{-5}$ & \textit{s} & $9.6$ & $3.48 \pm 0.05$ & $\ldots$ & $0.9 \pm 0.6$ & \textit{n}\\
\hline
$29$ & J165113.78+320943.4 & $252.8074,\ 32.1621$ & $0.744$ $\pm$ $0.045$ & \textit{p} & $7.7$ & $3.48 \pm 0.10$ & $\ldots$ & $\ldots$ & $\ldots$\\
\hline
$30$ & $\ldots$ & $102.0173,\ 70.8276$ & $0.714$ $\pm$ $0.049$ & \textit{p} & $7.8$ & $3.47 \pm 0.12$ & $\ldots$ & $\ldots$ & $\ldots$\\
\hline
$31$ & $\ldots$ & $11.2869,\ 28.7951$ & $0.668$ $\pm$ $0.042$ & \textit{p} & $8.0$ & $3.46 \pm 0.11$ & $\ldots$ & $\ldots$ & $\ldots$\\
\hline
$32$ & $\ldots$ & $127.9215,\ 67.1934$ & $0.451$ $\pm$ $0.027$ & \textit{p} & $9.5$ & $3.39 \pm 0.12$ & $\ldots$ & $\ldots$ & $\ldots$\\
\hline
$33$ & J112912.14+273313.9 & $172.3006,\ 27.5539$ & $0.0732$ $\pm$ $1 \cdot 10^{-5}$ & \textit{s} & $38.6$ & $3.34 \pm 0.01$ & $3.7 \pm 0.1$ & $3.6 \pm 0.8$ & \textit{n}\\
\hline
$34$ & J163659.07+541725.4 & $249.2461,\ 54.2904$ & $0.5027$ $\pm$ $5 \cdot 10^{-5}$ & \textit{s} & $8.8$ & $3.33 \pm 0.06$ & $\ldots$ & $53.4 \pm 59.1$ & \textit{n}\\
\hline
$35$ & J092826.93+230448.0 & $142.1122,\ 23.0800$ & $0.491$ $\pm$ $0.045$ & \textit{p} & $8.9$ & $3.33 \pm 0.17$ & $\ldots$ & $\ldots$ & $\ldots$\\
\hline
$36$ & J001152.65+310024.3 & $2.9694,\ 31.0068$ & $0.757$ $\pm$ $0.083$ & \textit{p} & $7.3$ & $3.32 \pm 0.15$ & $\ldots$ & $\ldots$ & $\ldots$\\
\hline
$37$ & J125804.46+273046.0 & $194.5186,\ 27.5128$ & $0.741$ $\pm$ $0.083$ & \textit{p} & $7.3$ & $3.29 \pm 0.16$ & $\ldots$ & $\ldots$ & $\ldots$\\
\hline
$38$ & $\ldots$ & $182.5080,\ 44.0903$ & $1.031$ $\pm$ $0.163$ & \textit{p} & $6.6$ & $3.28 \pm 0.15$ & $\ldots$ & $\ldots$ & $\ldots$\\
\hline
$39$ & J084127.02+554627.1 & $130.3626,\ 55.7742$ & $0.7912$ $\pm$ $3 \cdot 10^{-5}$ & \textit{s} & $7.1$ & $3.28 \pm 0.07$ & $\ldots$ & $25.2 \pm 27.0$ & \textit{n}\\
\hline
$40$ & J143011.92+410404.2 & $217.5497,\ 41.0678$ & $0.5868$ $\pm$ $0.0001$ & \textit{s} & $8.0$ & $3.27 \pm 0.06$ & $\ldots$ & $2.1 \pm 1.4$ & \textit{n}\\
\hline
$41$ & J135119.31+340844.1 & $207.8305,\ 34.1456$ & $0.923$ $\pm$ $0.121$ & \textit{p} & $6.7$ & $3.24 \pm 0.14$ & $\ldots$ & $\ldots$ & $\ldots$\\
\hline
$42$ & J134436.33+291239.6 & $206.1514,\ 29.2110$ & $0.766$ $\pm$ $0.134$ & \textit{p} & $7.1$ & $3.24 \pm 0.23$ & $\ldots$ & $\ldots$ & $\ldots$\\
\hline
$43$ & J112638.34+302541.9 & $171.6598,\ 30.4283$ & $0.3049$ $\pm$ $4 \cdot 10^{-5}$ & \textit{s} & $11.6$ & $3.23 \pm 0.04$ & $\ldots$ & $1.5 \pm 0.6$ & \textit{n}\\
\hline
$44$ & J012342.20+293633.1 & $20.9259,\ 29.6092$ & $2.525$ $\pm$ $0.300$ & \textit{p} & $6.5$ & $3.22 \pm 0.11$ & $\ldots$ & $\ldots$ & $\ldots$\\
\hline
$45$ & J223224.15+285753.3 & $338.1007,\ 28.9648$ & $0.566$ $\pm$ $0.038$ & \textit{p} & $8.0$ & $3.21 \pm 0.12$ & $\ldots$ & $\ldots$ & $\ldots$\\
\hline
$46$ & J103731.47+312948.9 & $159.3811,\ 31.4969$ & $0.5228$ $\pm$ $0.0001$ & \textit{s} & $8.3$ & $3.21 \pm 0.06$ & $\ldots$ & $2.2 \pm 1.4$ & \textit{n}\\
\hline
$47$ & J154742.69+384119.4 & $236.9279,\ 38.6887$ & $0.280$ $\pm$ $0.024$ & \textit{p} & $12.2$ & $3.21 \pm 0.20$ & $\ldots$ & $\ldots$ & $\ldots$\\
\hline
$48$ & J114333.93+425800.5 & $175.8914,\ 42.9668$ & $0.802$ $\pm$ $0.075$ & \textit{p} & $6.9$ & $3.20 \pm 0.12$ & $\ldots$ & $\ldots$ & $\ldots$\\
\hline
$49$ & J122329.86+313116.0 & $185.8744,\ 31.5211$ & $0.666$ $\pm$ $0.042$ & \textit{p} & $7.4$ & $3.20 \pm 0.11$ & $\ldots$ & $\ldots$ & $\ldots$\\
\hline
$50$ & $\ldots$ & $212.1410,\ 67.8036$ & $1.103$ $\pm$ $0.181$ & \textit{p} & $6.3$ & $3.17 \pm 0.14$ & $\ldots$ & $\ldots$ & $\ldots$\\
\end{tabular}
\caption{
Properties of the 50 projectively longest giants out of a total of $\numberOfGRGsOei$ discovered during our LoTSS DR2 search campaign.\protect\footnotemark
}
\label{tab:GRGProperties}
\end{table*}
\footnotetext{The giants are ranked by projected proper length.
The column `rank' thus denotes each giant's projected proper length rank within this new sample, not within the total known population.
Lying outside of the coverage, some GRG host galaxies have no SDSS DR12 name.
The column `host coordinates' contains the central right ascension and declination of the host galaxy.
The columns `redshift' and `redshift type' provide cosmological redshift estimates, derived from spectroscopy \textit{s} or photometry \textit{p}.
The column `angular length' denotes the largest great-circle distance between two LoTSS DR2--detectable GRG (end)points.
These angular lengths may increase in future deeper surveys.
We have not measured angular length errors on a case by case basis, but estimate them to be $0.15'$.
The column `projected proper length' propagates both redshift and angular length uncertainty and assumes the \citet{Planck12020} cosmology.
The columns `host stellar mass' and `host SMBH mass', further discussed in Section~\ref{sec:masses} and Appendix~\ref{ap:masses}, provide SDSS-derived estimates of host stellar and supermassive black hole masses.
Finally, the column `host quasar' indicates whether the host's AGN has a quasar appearance: \textit{y} (yes) or \textit{n} (no).
In cases where only a set of candidates containing the host galaxy could be established beyond reasonable doubt, we list the properties of the lowest-redshift candidate.
In this way, the provided projected proper length bounds the actual projected proper length from below.
In such cases, to signify uncertainty, we mark the host name, host coordinates, and physical host properties in grey.
We note that this lowest-redshift candidate is often, but not always, also the most probable host.
For access to these data for all \numberOfGRGsOei newly discovered giants, see Appendix~\ref{sec:supplementaryMaterialData}.
}

\subsection{GRG candidates in the radio}
\label{sec:GRGCandidatesRadio}
We first identified GRG candidates in the LoTSS at $6''$ and $60''$.
These maps serve complementary roles.
The $6''$ images reveal the precise morphology of radio galaxy cores and jets, which are necessary to pinpoint the host galaxy.
Figure~\ref{fig:hostIdentification} provides a representative sense of the host galaxy identification accuracy these data allow for when combined with modern optical surveys such as the DESI Legacy Imaging Surveys \citep{Dey12019} DR9.
In contrast, the $60''$ images have such compact sources removed or highly suppressed, but better highlight diffuse structures, such as RG lobes.
Being similar in morphology, we made sure not to interpret diffuse emission from low-redshift spiral galaxies and their circumgalactic media, or radio halos and relics in galaxy clusters, as RG lobe emission.
We required that all new RGs feature a detection of at least two\footnote{A small fraction of observed RGs are double-double radio galaxies, which show \emph{four} lobes.} lobes, or of at least one lobe and one jet oriented towards the lobe(s), at at least one of the resolutions used in this work ($6''$, $20''$, $60''$, and $90''$).

\subsection{GRG candidates in the optical}
\label{sec:GRGCandidatesOptical}
To confirm that a radio structure really is a radio galaxy, we compared the radio images with optical images of the same sky region.
If a patch of radio emission is indeed due to RG jets or lobes, the patch itself must have \emph{no} codirectional galactic counterpart in the optical.
If the radio emission is due to a low-redshift spiral galaxy or a galaxy cluster instead, a corresponding easily recognisable counterpart \emph{will} exist. 
We also took care not to erroneously associate the lobes of two distinct RGs.
For this reason, we were more cautious to associate a pair of lobes to a suspected host galaxy when, in the optical, one could discern other galaxies in the angular vicinity of the lobes that could have generated them instead.

The Pan-STARRS and SDSS images used for these purposes complement each other, as they differ in quality throughout the sky --- and in particular around sources of high optical flux density.
Neither consistently outperforms the other.
Only Pan-STARRS covers the full Northern Sky and could thus always be relied upon.

\subsection{Host galaxy identification}
We also used the Pan-STARRS and SDSS maps for the identification of host galaxies.
We collected host redshifts from the SDSS DR12 \citep{Alam12015} and \textit{Gaia} \citep{GaiaCollaboration12016} DR3 \citep{GaiaCollaboration12021} through automated \textit{VizieR} queries, from the Galaxy List for the Advanced Detector Era (GLADE) 2.4 \citep{Dalya12018}, and from the DESI DR9 photometric redshift catalogue \citep{Zou12022}.
If redshifts from multiple sources were available, we favoured SDSS over GLADE data, GLADE over \textit{Gaia} data, and \textit{Gaia} over DESI data.
Similarly, we only adopted photometric redshifts if spectroscopic ones were not available.

For a small subset of RGs, a definite host galaxy could not be established beyond reasonable doubt, but a set of candidates containing the host galaxy could.
In these cases, the lowest-redshift candidate provides a lower bound to the RG's projected proper length.\footnote{This is true as long as all candidates have a redshift below that of the angular diameter distance maximum: $z = 1.59$ for the cosmology adopted.}
If this lower bound exceeds $l_\mathrm{p,GRG} = 0.7\ \mathrm{Mpc}$, then the actual projected proper length certainly does so too, and the RG can be classified as a GRG --- despite persisting uncertainty concerning the identity of the host galaxy.

\subsection{LoTSS DR2 GRG sample}
\label{sec:LoTSSDR2GRGSample}
Our search campaign has led to the identification of \numberOfGRGsOei hitherto unknown giants.
To establish novelty, we assembled a literature catalogue with all known giants as of September 2022, combining the catalogue of \citet{Dabhade12020October}\footnote{This catalogue, complete up to and including April 2020, contains the giants found in 40 prior publications (for a list, see \citet{Dabhade12020October}'s Sect.~1: Introduction), alongside their own discoveries.} with the giants discovered in \citet{Galvin12020},
\citet{IshwaraChandra12020}, \citet{Tang12020}, \citet{Bassani12021}, \citet{Bruggen12021}, \citet{Delhaize12021},
\citet{Masini12021}, \citet{Kuzmicz12021}, \citet{Andernach12021}, \citet{Mahato12022}, \citet{Gurkan12022}, and \citet{Simonte12022}.
Fusing our sample with this literature catalogue, we obtain a final catalogue with $N = \numberOfGRGsTotal$ giants.

\begin{figure*}
    \centering
    \includegraphics[width=\textwidth]{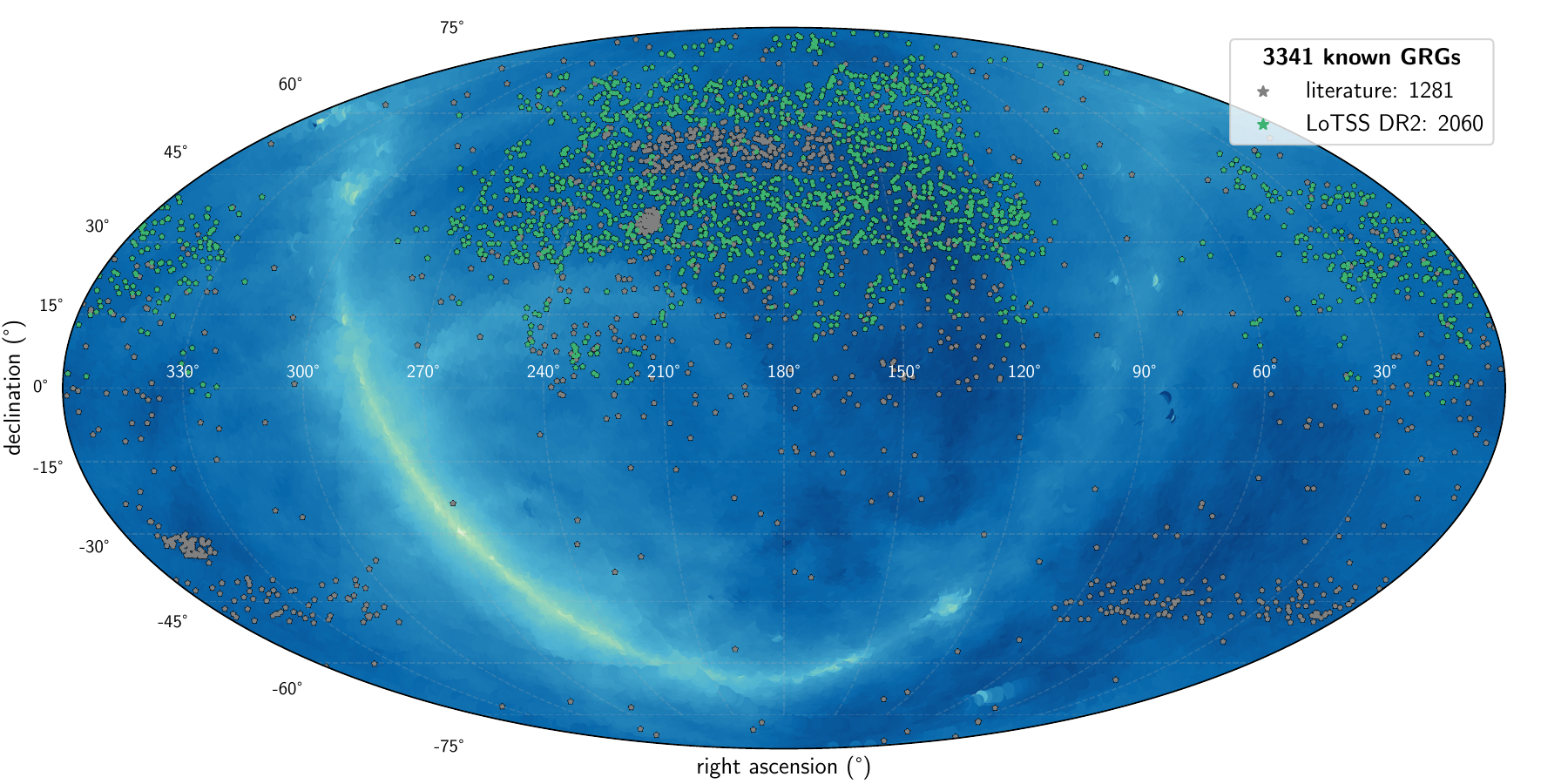}
    \caption{Mollweide view of the sky showing locations of all known giants, of which $62\%$ are discoveries presented in this work.
    In the background, we show the specific intensity function of the Milky Way at $\nu_\mathrm{obs} = 150\ \mathrm{MHz}$ \citep{Zheng12017}.
    The LoTSS DR2 has avoided the Galactic Plane, where extended emission complicates calibration and deconvolution.
    Our search footprint encloses a grey spherical rectangle, which represents the LoTSS DR1 search by \citet{Dabhade12020March}, and a grey spherical cap, which represents the Bo\"otes LOFAR Deep Field search by \citet{Simonte12022}.
    }
    \label{fig:GRGSkyMap}
\end{figure*}

Figure~\ref{fig:GRGSkyMap} shows the locations of all known giants in the sky.
We list basic properties of the 50 projectively largest new discoveries in Table~\ref{tab:GRGProperties}, and refer to Appendix~\ref{sec:supplementaryMaterialData} for access to these data for our entire sample.
In Fig.~\ref{fig:LoTSSDR2GRGs1} and Figs.~\ref{fig:LoTSSDR2GRGs2}--\ref{fig:LoTSSDR2GRGsMix2}, we present images for discoveries with projected proper lengths $l_\mathrm{p}$ in the ranges $5.1$--$4\ \mathrm{Mpc}$, $4$--$3\ \mathrm{Mpc}$, $3$--$2\ \mathrm{Mpc}$, and $2$--$0.7\ \mathrm{Mpc}$, respectively.
For each giant, we use the LoTSS resolution $\theta_\mathrm{FWHM} \in \{6'',\ 20'',\ 60'',\ 90''\}$ that most clearly conveys the morphology through a single image.
The selection reflects our sample's diversity in shapes and sizes and provides a sense of the data quality.

\subsubsection{Angular lengths}
The angular length distribution of the newly found giants is as follows: the smallest $\phi = 1.5'$, the median $\phi = 5.2'$, the largest $\phi = 2.2\degree$, and 80\% of angular lengths fall within $[3.4', 9.8']$.
Thirteen of our discoveries --- listed in Table~\ref{tab:GRGsLargeAngularLength} --- are larger than the Moon in the sky (whose angular diameter varies over time, but here taken to be $\phi = 30'$).
Our search more than doubles the known number of such spectacular giants --- from 10 to 23.

The GRG associated with NGC 2300 (see the middle-left panel of Fig.~\ref{fig:LoTSSDR2GRGsLargerThanLuna}) is the giant with the largest angular length ever found, and the radio galaxy with the largest angular length in the Northern Sky.\footnote{At $\phi = 8\degree$, the Southern Sky's Centaurus A \citep{Cooper11965} is the radio galaxy with the largest angular length overall \citep[e.g.][]{McKinley12022}; despite this, at $l_\mathrm{p} = 0.48\ \mathrm{Mpc}$, it is not a giant.}
It remains possible that the GRG has been generated by spiral galaxy NGC 2276 instead, with which elliptical galaxy NGC 2300 is interacting.
However, this scenario seems unlikely, as only a fraction ${\sim}10^{-3}$ of known giants are hosted by spirals.
Its discovery emphasises that low-frequency interferometers like the LOFAR and the MWA, which are sensitive to degree-scale angular scales, are important to complete a low-redshift census of giant radio galaxies.
Sky-wide, a simple extrapolation of our findings suggests that several (${\sim}10^1$) degree-scale angular length giants similar to the GRG of NGC 2300 still await discovery at the LoTSS DR2 depth.
\begin{center}
\footnotesize
\captionof{table}{
(Extended) Table~\ref{tab:GRGProperties} ranks, host names, angular lengths, and LoTSS DR2 image references for all 13 newly discovered giants that appear larger in the sky than the Moon ($\phi > 30'$).
}
\begin{tabular}{l l l l}
\\
\hline
rank & host name & angular & LoTSS DR2\\
Table~\ref{tab:GRGProperties} & NGC or LEDA & length $(')$ & image reference\\
\hline\arrayrulecolor{lightgray}
$1441$ & NGC 2300 & $2.2\degree$ & Fig.~\ref{fig:LoTSSDR2GRGsLargerThanLuna}, middle left\\
\hline
$141$ & NGC 6185 & $1.0\degree$ & Fig.~\ref{fig:LoTSSDR2GRGsLargerThanLuna}, top left\\
\hline
$9$ & LEDA 56028 & $43.8'$ & Fig.~\ref{fig:LoTSSDR2GRGs1}, bottom left\\
\hline
$144$ & LEDA 54794 & $40.0'$ & -\\
\hline
$14$ & LEDA 5060619 & $39.0'$ & Fig.~\ref{fig:LoTSSDR2GRGs2}, top left\\
\hline
$33$ & LEDA 1811497 & $38.6'$ & Fig.~\ref{fig:LoTSSDR2GRGs2}, middle left\\
\hline
$1692$ & NGC 2789 & $34.5'$ & Fig.~\ref{fig:LoTSSDR2GRGsLargerThanLuna}, bottom right\\
\hline
$1643$ & LEDA 38523 & $33.8'$ & Fig.~\ref{fig:LoTSSDR2GRGsLargerThanLuna}, middle right\\
\hline
$1770$ & NGC 1044 & $32.7'$ & Fig.~\ref{fig:LoTSSDR2GRGsLargerThanLuna}, bottom left\\
\hline
$178$ & LEDA 2048533 & $32.3'$ & Fig.~\ref{fig:comparison6Vs90Arcsec}, top row\\
\hline
$1486$ & NGC 7385 & $32.0'$ & -\\
\hline
$326$ & LEDA 3090801 & $31.0'$ & Fig.~\ref{fig:LoTSSDR2GRGsMix1}, bottom left\\
\hline
$465$ & LEDA 37801 & $30.1'$ & Fig.~\ref{fig:LoTSSDR2GRGsLargerThanLuna}, top right
\end{tabular}
\label{tab:GRGsLargeAngularLength}
\end{center}
\subsubsection{Redshifts}
The redshift distribution of the newly found giants is as follows: the lowest $z = 0.00635 \pm 6 \cdot 10^{-5}$, the median $z = 0.29$, the highest $z = 2.6394 \pm 6 \cdot 10^{-4}$, and 80\% of redshifts fall within $[0.12, 0.68]$.
Because of our focus on giants of large angular length (see Sect.~\ref{sec:angularLengthThreshold}), we have found only 36 giants beyond $z > 1$.
One of these, the GRG at rank 13 of Table~\ref{tab:GRGProperties}, is the largest secure giant found beyond redshift 1.
It lies at $z = 1.1 \pm 0.1$ and spans $l_\mathrm{p} = 3.9 \pm 0.1\ \mathrm{Mpc}$.
Its host galaxy does not appear to contain a quasar.

\subsubsection{Projected proper lengths}
\label{sec:sampleProjectedProperLengths}
With $l_\mathrm{p} = 5.1 \pm 0.2\ \mathrm{Mpc}$ and $l_\mathrm{p} = 4.99 \pm 0.04\ \mathrm{Mpc}$, our LoTSS DR2 sample contains the first two $5\ \mathrm{Mpc}$--scale giants.
We have presented a dedicated analysis of the latter GRG in \citet{Oei12022Alcyoneus}.
11 discoveries have $l_\mathrm{p} \geq 4\ \mathrm{Mpc}$, 53 have $3 \leq l_\mathrm{p} < 4\ \mathrm{Mpc}$, 291 have $2 \leq l_\mathrm{p} < 3\ \mathrm{Mpc}$, 1215 have $1 \leq l_\mathrm{p} < 2\ \mathrm{Mpc}$, and 490 have $0.7 \leq l_\mathrm{p} < 1\ \mathrm{Mpc}$.
The median $l_\mathrm{p} = 1.35\ \mathrm{Mpc}$, and 80\% of projected proper lengths fall within $[0.82\ \mathrm{Mpc}, 2.29\ \mathrm{Mpc}]$.

\begin{figure*}[p]
    \centering
    \begin{subfigure}{\columnwidth}
    \includegraphics[width=\columnwidth]{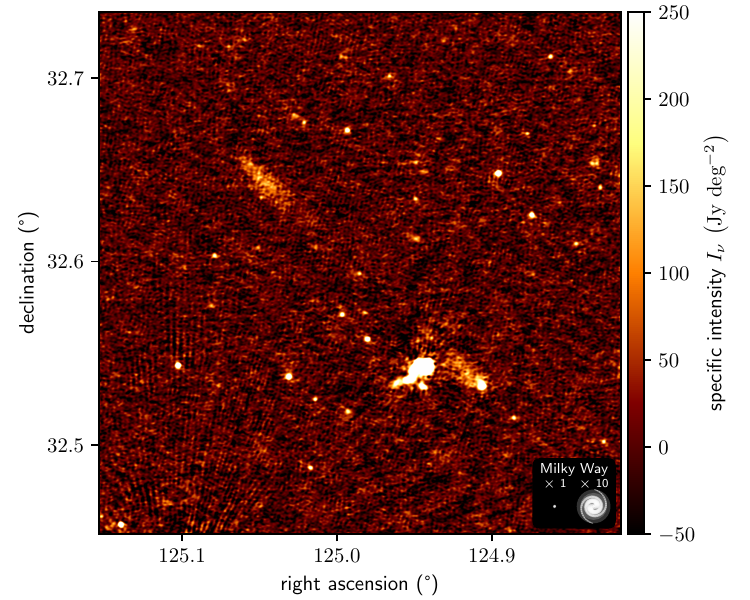}
    \end{subfigure}
    \begin{subfigure}{\columnwidth}
    \includegraphics[width=\columnwidth]{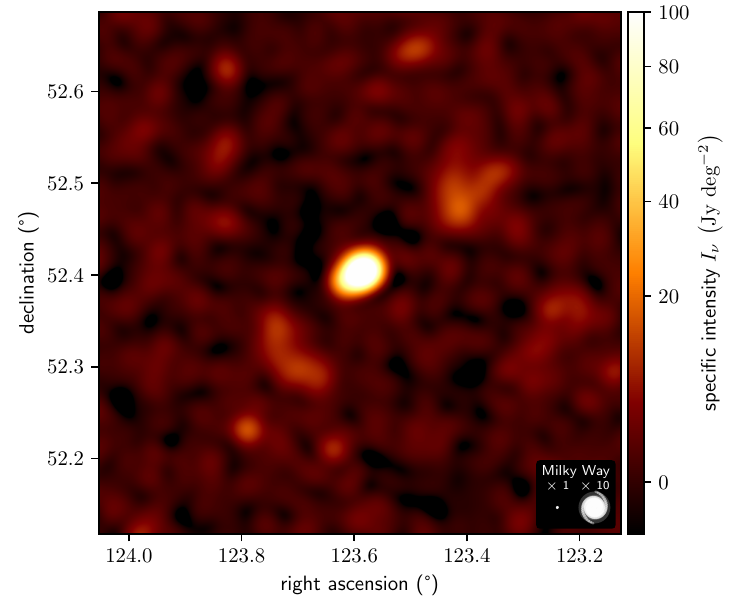}
    \end{subfigure}
    \begin{subfigure}{\columnwidth}
    \includegraphics[width=\columnwidth]{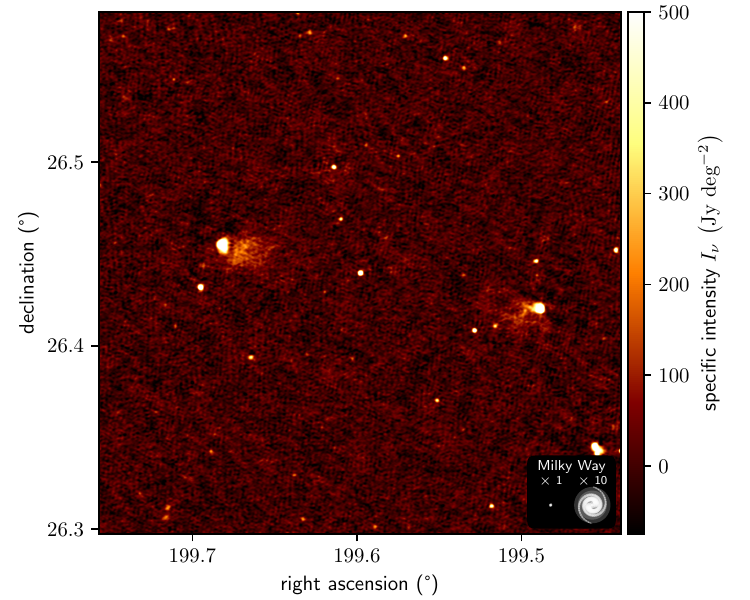}
    \end{subfigure}
    \begin{subfigure}{\columnwidth}
    \includegraphics[width=\columnwidth]{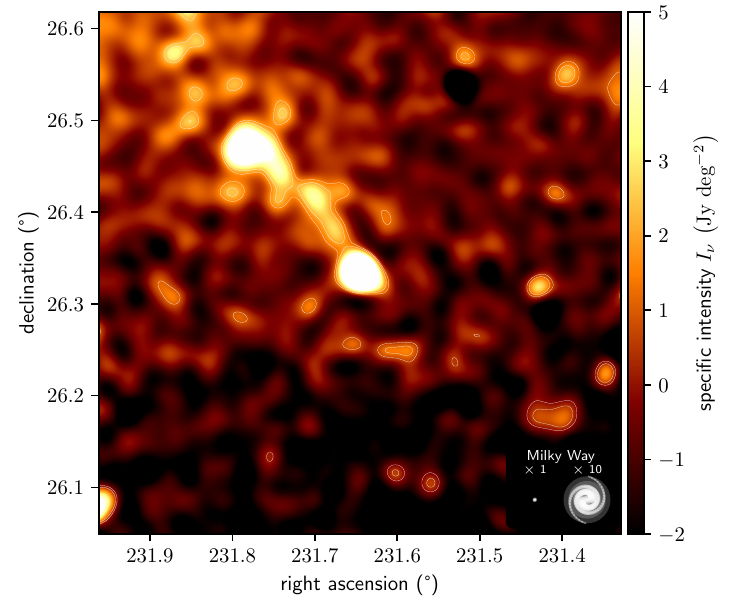}
    \end{subfigure}
    \begin{subfigure}{\columnwidth}
    \includegraphics[width=\columnwidth]{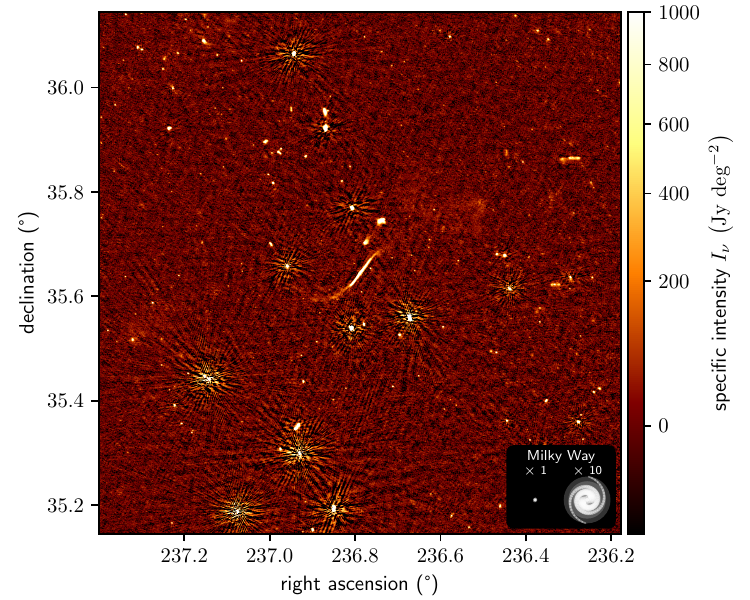}
    \end{subfigure}
    \begin{subfigure}{\columnwidth}
    \includegraphics[width=\columnwidth]{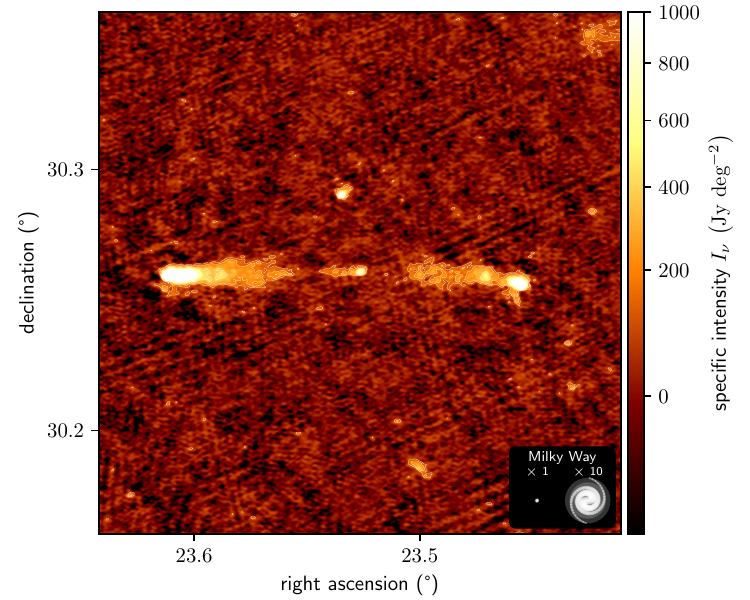}
    \end{subfigure}
    \caption{
    Details of the LoTSS DR2--estimated specific intensity function $I_\nu\left(\hat{r}\right)$ at central observing frequency $\nu_\mathrm{obs} = 144\ \mathrm{MHz}$ and resolutions $\theta_\mathrm{FWHM} \in \{6'',\ 90''\}$, centred around the hosts of newly discovered giants.
    Row-wise from left to right, from top to bottom, the projected proper length $l_\mathrm{p}$ is $5.1\ \mathrm{Mpc}$, $5.0\ \mathrm{Mpc}$ \citep{Oei12022Alcyoneus}, $4.6\ \mathrm{Mpc}$, $4.6\ \mathrm{Mpc}$, $4.1\ \mathrm{Mpc}$, and $4.1\ \mathrm{Mpc}$; in the same order, $\theta_\mathrm{FWHM}$ is $6''$, $90''$, $6''$, $90''$, $6''$, and $6''$.
    The GRG in the bottom-left panel appears larger in the sky than the Moon.
    In the middle-right panel, contours signify 2.5 and 3.5 sigma-clipped standard deviations (SDs) above the sigma-clipped median; in the bottom-right panel, they signify 3, 5, and 10 such SDs.
    For scale, we show the stellar Milky Way disk (with a diameter of 50 kpc) generated using the \citet{Ringermacher12009} formula, alongside a 10 times inflated version.
    }
    \label{fig:LoTSSDR2GRGs1}
\end{figure*}\noindent
\begin{figure*}[p]
    \centering
    \begin{subfigure}{\columnwidth}
    \includegraphics[width=\columnwidth]{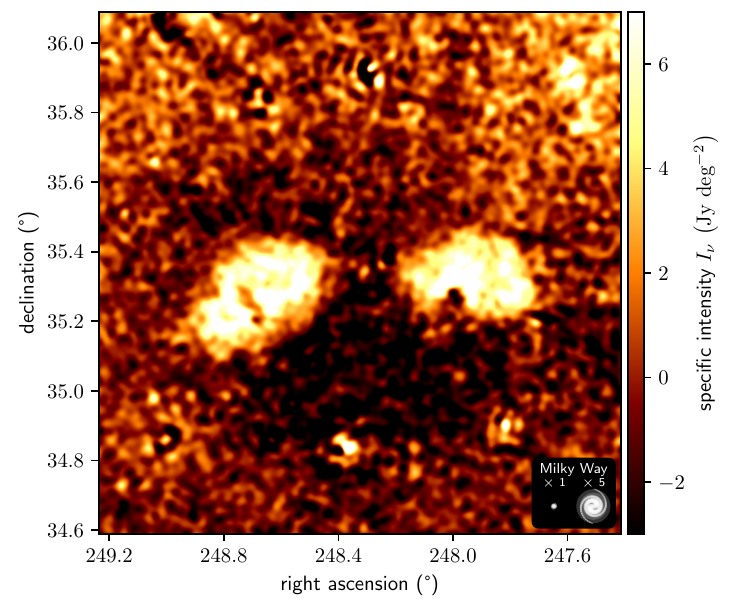}
    \end{subfigure}
    \begin{subfigure}{\columnwidth}
    \includegraphics[width=\columnwidth]{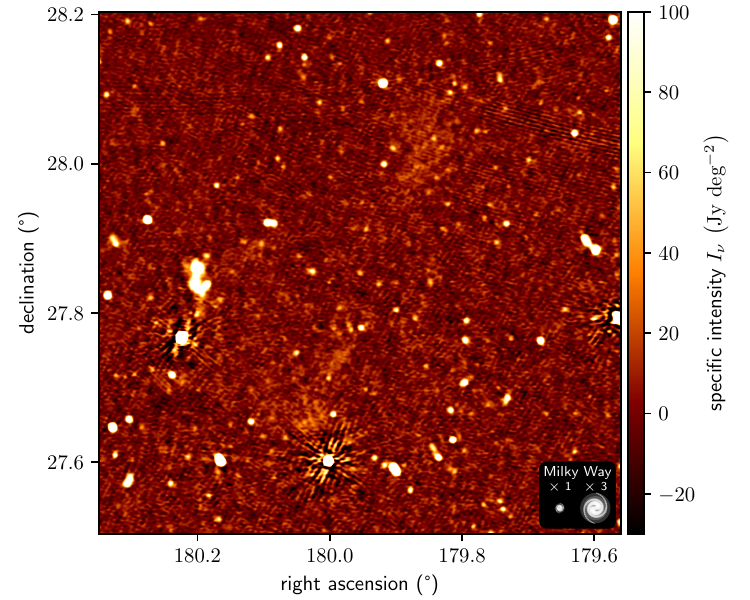}
    \end{subfigure}
    \begin{subfigure}{\columnwidth}
    \includegraphics[width=\columnwidth]{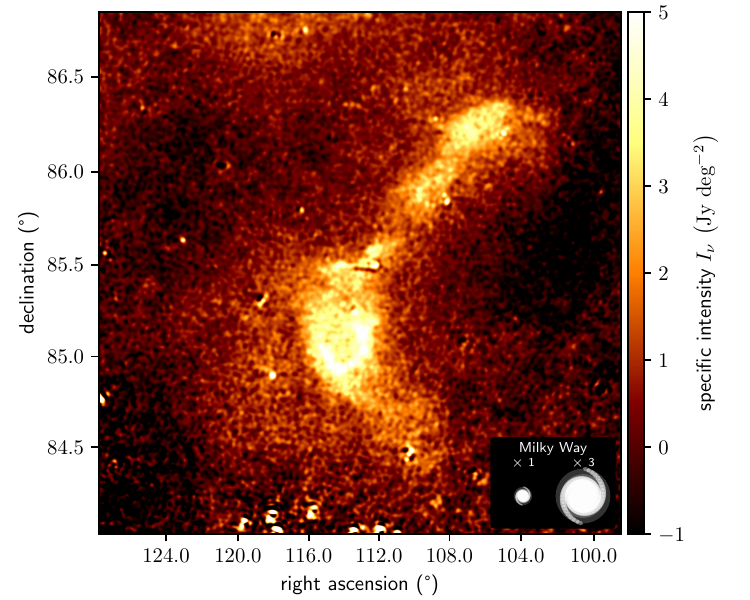}
    \end{subfigure}
    \begin{subfigure}{\columnwidth}
    \includegraphics[width=\columnwidth]{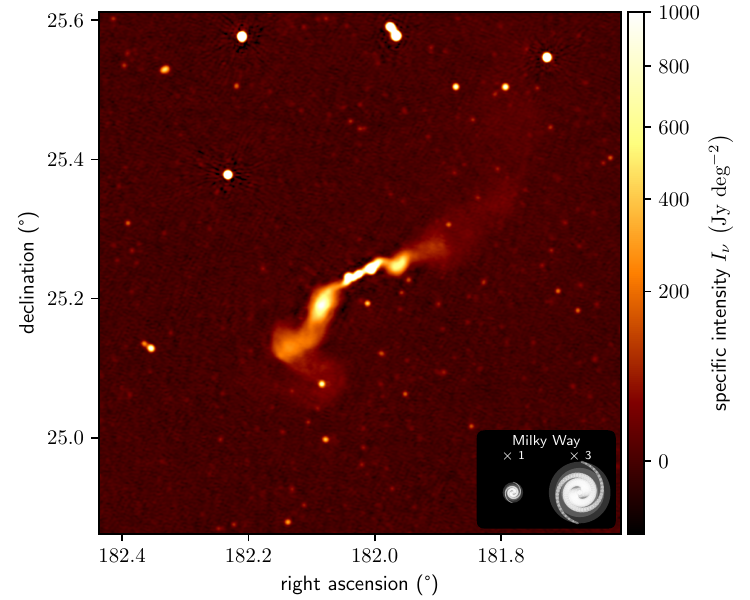}
    \end{subfigure}
    \begin{subfigure}{\columnwidth}
    \includegraphics[width=\columnwidth]{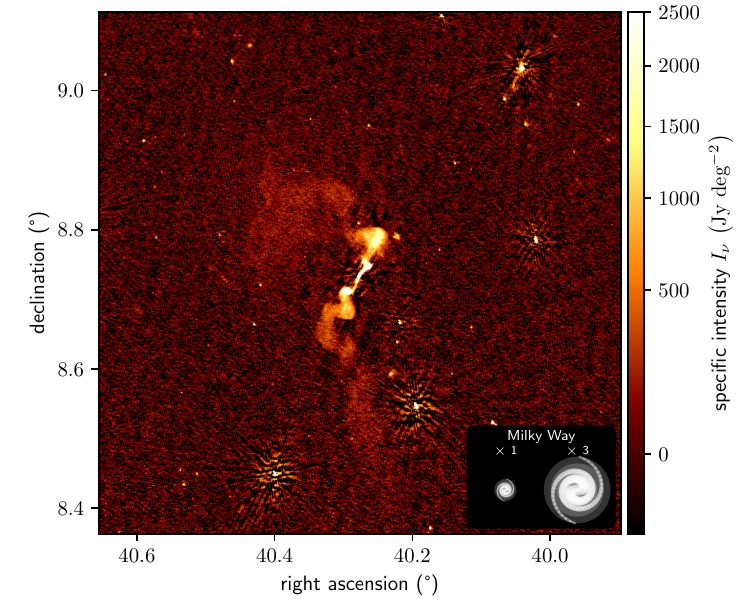}
    \end{subfigure}
    \begin{subfigure}{\columnwidth}
    \includegraphics[width=\columnwidth]{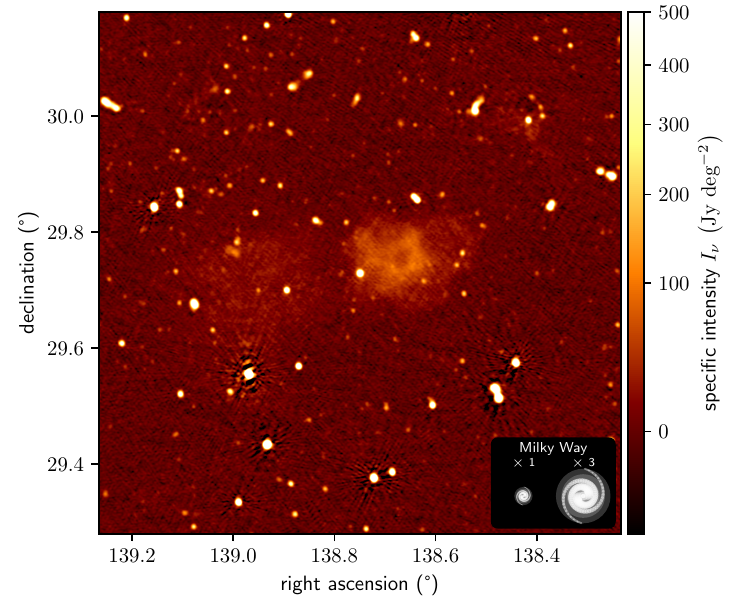}
    \end{subfigure}
    \caption{
    Details of the LoTSS DR2--estimated specific intensity function $I_\nu\left(\hat{r}\right)$ at central observing frequency $\nu_\mathrm{obs} = 144\ \mathrm{MHz}$ and resolutions $\theta_\mathrm{FWHM} \in \{6'',\ 20'', 90''\}$, centred around the hosts of newly discovered giants.
    Row-wise from left to right, from top to bottom, the projected proper length $l_\mathrm{p}$ is $2.5\ \mathrm{Mpc}$, $1.9\ \mathrm{Mpc}$, $1.1\ \mathrm{Mpc}$, $1.0\ \mathrm{Mpc}$, $0.9\ \mathrm{Mpc}$, and $0.9\ \mathrm{Mpc}$; in the same order, $\theta_\mathrm{FWHM}$ is $90''$, $20''$, $90''$, $20''$, $6''$, and $20''$.
    All appear larger in the sky than the Moon.
    The top-left panel shows the giant of NGC 6185, a \emph{spiral} galaxy.
    This is the first spiral galaxy--hosted giant known with $l_\mathrm{p} > 2\ \mathrm{Mpc}$.
    The middle-left panel shows a structure we interpret as a radio galaxy belonging to the elliptical galaxy NGC 2300.
    At $\phi = 2.2\degree$, this giant has the largest angular length of all uncovered thus far.
    }
    \label{fig:LoTSSDR2GRGsLargerThanLuna}
\end{figure*}\noindent

\subsubsection{Stellar and supermassive black hole masses}
\label{sec:masses}
Following \citet{Oei12022Alcyoneus}, we collected host stellar masses $M_\star$ from \citet{Chang12015} and \citet{Salim12018}, and estimated host SMBH masses $M_\bullet$ via SDSS DR12 stellar velocity dispersions \citep{Alam12015} and the M-sigma relation of \citet{Kormendy12013}'s Eq. (7).
From all \numberOfGRGsTotal giants in our final catalogue, only 732 (22\%) could be assigned a stellar mass in this way, and only 1115 (33\%) an SMBH mass; for both quantities, our LoTSS DR2 sample accounts for four-fifths of the resulting subpopulation. 
Fig.~\ref{fig:stellarMass} shows both $M_\star$ and $M_\bullet$ in relation to projected proper length $l_\mathrm{p}$.

The median $M_\star = 3.4 \cdot 10^{11}\ M_\odot$, and 80\% of stellar masses fall within $[1.8 \cdot 10^{11}\ M_\odot,\ 5.3 \cdot 10^{11}\ M_\odot]$.
We discover two giants whose hosts, J150329.07+374850.3 and J073505.24+415827.5, are the least massive known: both have a stellar mass $M_\star = 4.8 \cdot 10^{10}\ M_\odot$.
These are small giants, with $l_\mathrm{p} = 0.8\ \mathrm{Mpc}$ and $l_\mathrm{p} = 0.7\ \mathrm{Mpc}$, respectively.
The top panel of Fig.~\ref{fig:stellarMass} hints at a weak positive correlation between $M_\star$ and $l_\mathrm{p}$, which future work should confirm or reject.

The median $M_\bullet = 1.0 \cdot 10^9\ M_\odot$, and 80\% of SMBH masses fall within $[0.4 \cdot 10^9\ M_\odot,\ 2.2 \cdot 10^9\ M_\odot]$.
The SMBH masses of J123703.24+275819.5, J163659.07+541725.4, and J103129.54+502959.1 are the highest estimated yet, with $M_\bullet = 2 \cdot 10^{11}\ M_\odot$, $M_\bullet = 5 \cdot 10^{10}\ M_\odot$, and $M_\bullet = 5 \cdot 10^{10}\ M_\odot$, respectively.
The latter masses equal the theoretical maximum mass of accreting black holes of typical spin \citep{King12016}.
Curiously, although J103129.54+502959.1's $M_\bullet$ is among the highest estimated SMBH mass of any GRG host, the GRG itself is relatively small: $l_\mathrm{p} = 0.81 \pm 0.06\ \mathrm{Mpc}$.
Conversely, the bottom panel of Fig.~\ref{fig:stellarMass} shows that multi-Mpc giants can have hosts with SMBH masses that are two orders of magnitude lower than J103129.54+502959.1's.

\subsubsection{Spiral or lenticular host galaxies}
Remarkably, although NGC 6185 is a spiral galaxy of Hubble--de Vaucouleurs class SAa \citep{Jansen12000}, it appears to have generated the giant shown in the top-left panel of Fig.~\ref{fig:LoTSSDR2GRGsLargerThanLuna}.
Such systems are exceedingly rare: not only are few RGs giants, but also virtually all giants have an elliptical galaxy as their host.
With $l_\mathrm{p} = 2.54 \pm 0.01\ \mathrm{Mpc}$, this is the largest known spiral galaxy--hosted radio galaxy.
Hitherto, the largest spiral galaxy--hosted giant in the literature has been J2345-0449 \citep{Bagchi12014}, with $l_\mathrm{p} = 1.6\ \mathrm{Mpc}$.
Given the favourably low redshift of $z = 0.03430 \pm 7 \cdot 10^{-5}$, NGC 6185 and its enigmatic giant solicit a dedicated analysis \citep{Oei12022NGC6185}.
Besides NGC 6185, spiral or lenticular galaxies J080403.40+404809.3 and J091459.66+294348.8 (known alternatively as NGC 2789; see the bottom-right panel of Fig.~\ref{fig:LoTSSDR2GRGsLargerThanLuna}) have also generated giants; these have projected proper lengths $l_\mathrm{p} = 1.1\ \mathrm{Mpc}$ and $l_\mathrm{p} = 0.9\ \mathrm{Mpc}$, respectively.
Morphological host classification through SDSS, Pan-STARRS, and DESI imagery is reliable only up to $z \sim 0.1$--$0.2$, depending on viewing angle and various other factors.
Our LoTSS DR2 sample contains 342 giants with definite hosts at $z = 0.15$ or below, among which are all 3 giants with spiral or lenticular hosts discussed here.
It thus appears that the fraction of GRG hosts that is of such non-elliptical nature is ${\sim}1\%$.
A more detailed morphological characterisation of the sample appears possible, for example using data from \citet{Hart12016}, but this is beyond the scope of the current work.

\begin{figure*}
    \centering
    \begin{subfigure}{\columnwidth}
    \includegraphics[width=\linewidth]{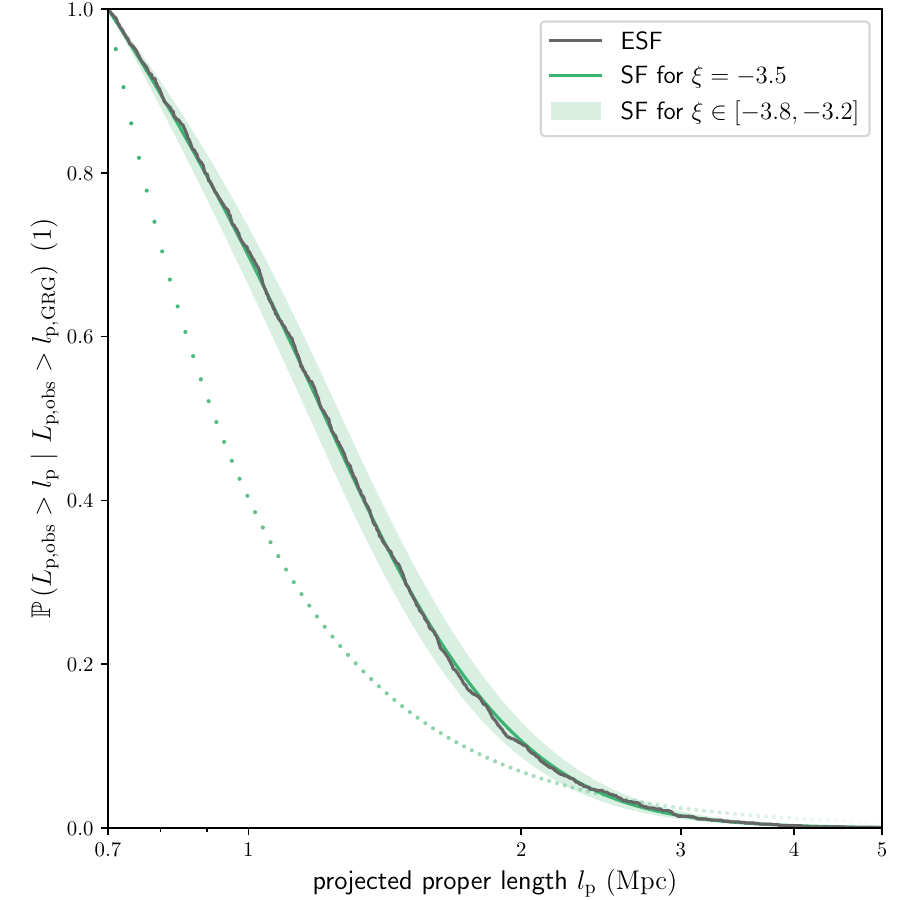}
    \end{subfigure}
    \begin{subfigure}{\columnwidth}
    \includegraphics[width=\linewidth]{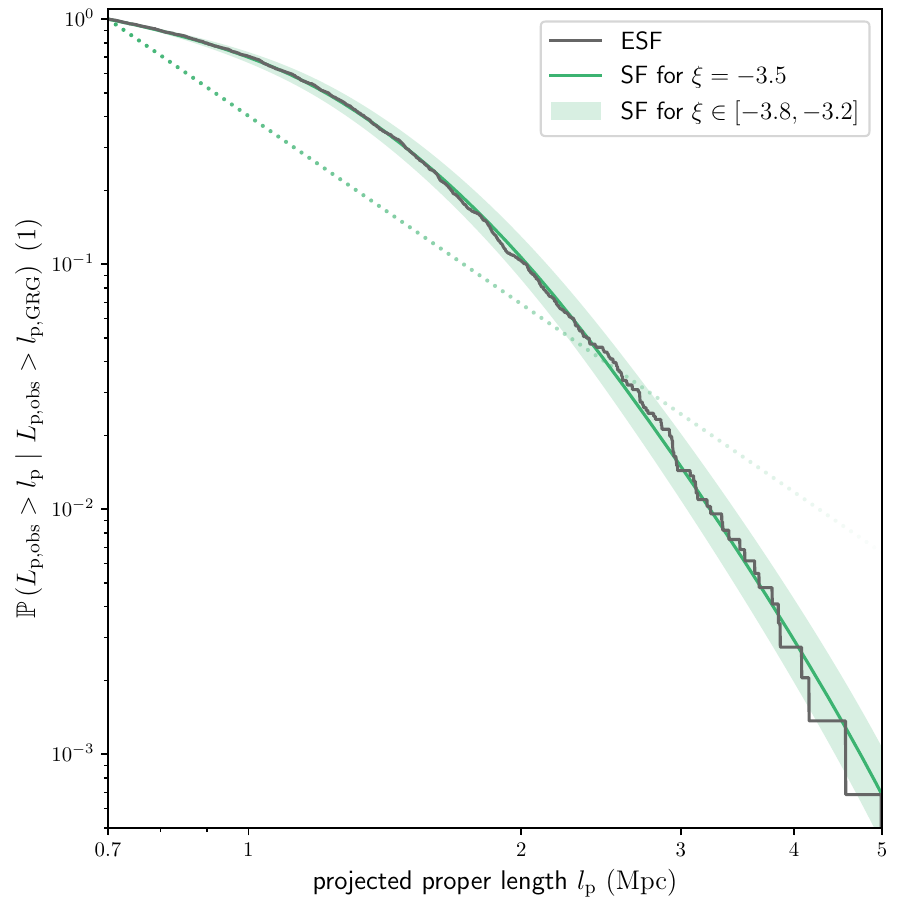}
    \end{subfigure}
    \caption{
    Empirical survival function of the observed giant radio galaxy projected proper length RV (ESF; dark grey) and the corresponding survival function $1 - F_{L_\mathrm{p,obs}\ \vert\ L_\mathrm{p,obs} > l_\mathrm{p,GRG}}$ (SF; green curve) using the maximum a posteriori probability parameters (MAP; see Table~\ref{tab:posterior}).
    Observed GRG projected lengths are well described by a Pareto distribution modified to include selection effects.
    Keeping the selection effect parameters fixed, we show how models vary with tail index $\xi$ (green range).
    We also show the selection effect--free SF $1 - F_{L_\mathrm{p}\ \vert\ L_\mathrm{p} > l_\mathrm{p,GRG}}$ using the MAP $\xi$ (green dots).
    We included all LoTSS DR2 search campaign giants up to $z_\mathrm{max} = 0.5$.
    \textit{Left:} logarithmic horizontal axis and linear vertical axis.
    \textit{Right:} logarithmic horizontal axis and logarithmic vertical axis.
    }
    \label{fig:ECDFCompleteness}
\end{figure*}\noindent

\section{Results}
\label{sec:results}
After first building a statistical framework and then collecting a large sample of giants from a single survey through a systematic approach, we were ready to infer the intrinsic GRG length distribution.
In particular, we aimed to establish whether the RG intrinsic proper length RV $L$ is well described by a Pareto distribution, and if so, what its tail index $\xi$ is.
Subsequently, we inferred derived quantities, such as the comoving GRG number density in the Local Universe.

\subsection{Giant radio galaxy length distribution}
\subsubsection{Empirical survival function}
From our LoTSS DR2 GRG sample, we computed the empirical cumulative distribution function (ECDF) of the GRG observed projected proper length RV $L_\mathrm{p,obs}\ \vert\ L_\mathrm{p,obs} > l_\mathrm{p,GRG}$ (see Eq.~\ref{eq:GRGObservedProjectedProperLength}).
We only included giants for which $l_\mathrm{p}$ itself --- rather than a lower bound --- is known, and set either $z_\mathrm{max} = 0.5$ or $z_\mathrm{max} = 0.25$; this retained 1473 or 811 giants for analysis, respectively.\footnote{We explored two choices for $z_\mathrm{max}$ as our model assumes that $\xi$ remains constant between $z \in [0, z_\mathrm{max}]$. We further discuss this assumption in Section~\ref{sec:XiEvolution}.}
The empirical survival function (ESF), which equals one minus the ECDF, is shown in dark grey in both panels of Fig.~\ref{fig:ECDFCompleteness}.
Just as the ECDF approximates the CDF, the ESF approximates the survival function (SF).
In this case, for any $l_\mathrm{p}$, the ESF provides the probability that a randomly drawn LoTSS DR2 GRG will have a projected proper length exceeding $l_\mathrm{p}$.
If $L_\mathrm{p,obs}\ \vert\ L_\mathrm{p,obs} > l_\mathrm{p,GRG}$ were Paretian, its ESF (hereafter: `the' ESF) would resemble a straight line in Fig.~\ref{fig:ECDFCompleteness}'s right panel --- both axes have logarithmic scaling.
However, the ESF clearly displays curvature.

\subsubsection{Selection effects}
\label{sec:selectionEffects}
Under the ansatz that $L$ is Paretian, as proposed in Sect.~\ref{sec:intrinsicProperLength}, the aforementioned ESF's curvature implies a significant role for observational selection effects.
The reason is the following.
The ansatz implies that the GRG projected proper length RV $L_\mathrm{p}\ \vert\ L_\mathrm{p} > l_\mathrm{p,GRG}$ is also Paretian (see Eq.~\ref{eq:projectedProperLengthPDF}), as is $L_\mathrm{p,obs}\ \vert\ L_\mathrm{p,obs} > l_\mathrm{p,GRG}$ when selection effects are negligible (see Sect.~\ref{sec:generalConsiderations} and set $C(l_\mathrm{p}) = 1$).
But in our case $L_\mathrm{p,obs}\ \vert\ L_\mathrm{p,obs} > l_\mathrm{p,GRG}$ is not Paretian: its ESF is curved.
To avoid contradiction, we must relax at least one assumption.
If the Pareto ansatz is held, then selection effects must be at play.

When selection effects are non-negligible, we must devise a procedure to disentangle them from the data if our $\xi$ estimate is to be uncontaminated.
To this end, we performed joint Bayesian inference with a model that includes both $\xi$ \emph{and} parameters that describe the selection effects.

In particular, we considered the roles of the fuzzy angular length threshold and surface brightness selection effects, introduced in Sects.~\ref{sec:fuzzyAngularLengthThreshold} and \ref{sec:surfaceBrightness}, respectively.
In Sect.~\ref{sec:angularLengthThreshold}, we explain that we have attempted to maintain a $5'$ angular length threshold during our LoTSS DR2 GRG search.
If we want to use Sect.~\ref{sec:fuzzyAngularLengthThreshold} to correct for the bias against faraway and physically small giants that this threshold has imprinted onto our sample, we must estimate parameters $\phi_\mathrm{min}$ and $\phi_\mathrm{max}$.
A natural choice is to assume that they lie symmetrically around the intended angular length threshold of $5'$ --- but at what distance from it?
We propose to consider this distance, $\frac{1}{2}(\phi_\mathrm{max} - \phi_\mathrm{min})$, as a yet unknown model parameter that we fitted to the data.
Similarly, if we want to use Sect.~\ref{sec:surfaceBrightness} to correct for the bias against physically large giants that the limited depth of the LoTSS DR2 imprints onto our sample, we must estimate parameters $b_{\nu,\mathrm{ref}}$ and $\sigma_\mathrm{ref}$.
Our approach was to, again, fit these parameters --- possibly making use of any available prior knowledge.

The meaning of $b_{\nu,\mathrm{ref}}$ depends on the choice of $l_\mathrm{ref}$ and $\nu_\mathrm{obs}$; we defined $l_\mathrm{ref} \coloneqq 0.7\ \mathrm{Mpc}$ and used $\nu_\mathrm{obs} = 144\ \mathrm{MHz}$.
We furthermore assumed $b_{\nu,\mathrm{th}} = 1 \cdot \sigma_{I_\nu}$, with $\sigma_{I_\nu} = 25\ \mathrm{Jy\ deg^{-2}}$ being the typical LoTSS DR2 $6''$ RMS noise \citep{Shimwell12022}.

\subsubsection{Prior}
\begin{figure}[h!]
    \centering
    \begin{subfigure}{\columnwidth}
    \includegraphics[width=\columnwidth]{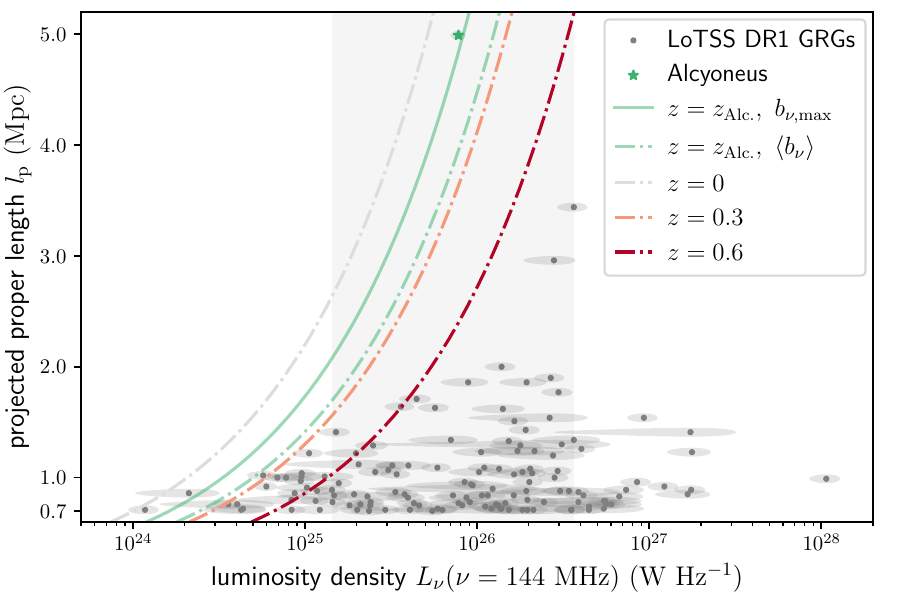}
    \end{subfigure}
    \begin{subfigure}{\columnwidth}
    \includegraphics[width=\columnwidth]{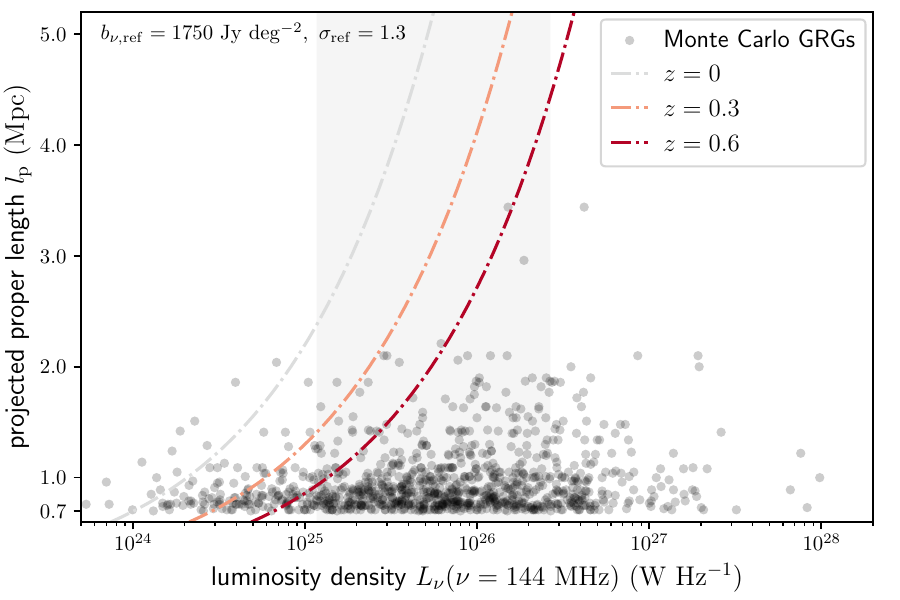}
    \end{subfigure}
    \begin{subfigure}{\columnwidth}
    \includegraphics[width=\columnwidth]{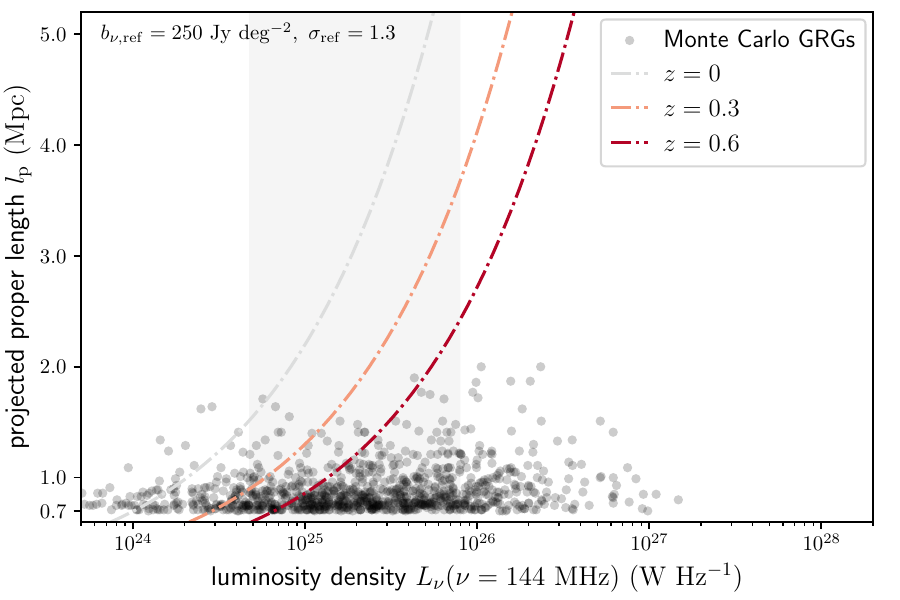}
    \end{subfigure}
    \caption{
    Comparison between luminosity density--projected proper length relations for observed and simulated giants.
    Each dash-dotted curve denotes a family of giants at a given redshift, assuming $f_{L_\nu} = 0.3$ and $f_l = 0.3$, whose mean lobe surface brightnesses equal the LoTSS DR2 noise level and who are thus borderline-detectable.
    The grey band denotes the median-centred luminosity density range that contains 68\% of giants.
    \textit{Top:} 139 LoTSS DR1 giants \citep{Dabhade12020March}, alongside Alcyoneus, with redshift $z_\mathrm{Alc.} = 0.25$ \citep{Oei12022Alcyoneus}.
    The solid green curve is similar to the dash-dotted green curve, but represents maximum instead of mean lobe surface brightness.
    \textit{Middle:} 1000 simulated giants, assuming $b_{\nu,\mathrm{ref}} = 1750\ \mathrm{Jy\ deg^{-2}}$ and $\sigma_\mathrm{ref} = 1.3$.
    \textit{Bottom:} 1000 simulated giants, assuming $b_{\nu,\mathrm{ref}} = 250\ \mathrm{Jy\ deg^{-2}}$ and $\sigma_\mathrm{ref} = 1.3$.
    }
    \label{fig:surfaceBrightness}
\end{figure}\noindent
We exploited two sources of prior knowledge.
Firstly, we attempted to directly estimate $b_{\nu,\mathrm{ref}}$ and $\sigma_\mathrm{ref}$ by selecting from all LoTSS DR2 giants with $l_\mathrm{p} \leq 1\ \mathrm{Mpc}$ a random subset of size 50 (10\%).
For these giants, we estimated the mean surface brightnesses of both lobes from the LoTSS DR2 $6''$ imagery, differentiating between the brighter and the fainter lobe.
Because our goal was to estimate $b_{\nu,\mathrm{ref}}$, we attempted to undo cosmological and growth-induced surface brightness dimming assuming a universal lobe spectral index $\alpha = -1$ and self-similar growth: $\zeta = -2$.
The resulting surface brightnesses correspond to $z = 0$, and to the epoch in each giant's life when $l = l_\mathrm{ref}$.
The median of the corrected bright lobe mean surface brightnesses is $b_{\nu,\mathrm{ref}} = 1.3 \cdot 10^3\ \mathrm{Jy\ deg^{-2}}$, whilst the median of the corrected faint lobe mean surface brightnesses is $b_{\nu,\mathrm{ref}} = 0.7 \cdot 10^3\ \mathrm{Jy\ deg^{-2}}$.
All lobes taken together, the median becomes $b_{\nu,\mathrm{ref}} = 1.0 \cdot 10^3\ \mathrm{Jy\ deg^{-2}}$.
We performed maximum likelihood estimation assuming the surface brightness distribution is lognormal and found $\sigma_\mathrm{ref} = 1.3$, again using all lobes.
Note, however, that we have only used observed giants here, whilst $b_{\nu,\mathrm{ref}}$ and $\sigma_\mathrm{ref}$ should correspond to the entire population of giants.
As fainter giants will have preferentially fallen out, we might have overestimated $b_{\nu,\mathrm{ref}}$ and underestimated $\sigma_\mathrm{ref}$.

To probe whether we had overestimated $b_{\nu,\mathrm{ref}}$ and underestimated $\sigma_\mathrm{ref}$, we used data from \citet{Dabhade12020March} and a Monte Carlo approach.
First, for their sample of 239 LoTSS DR1 giants, we computed total luminosity densities $L_\nu$ at rest-frame frequency $\nu = 144\ \mathrm{MHz}$.
(Given that LoTSS DR1 and DR2 noise levels are similar, this population is also representative of the LoTSS DR2.)
In the top panel of Fig.~\ref{fig:surfaceBrightness}, we show $L_\nu$ versus $l_\mathrm{p}$ for all 139 for which $z < 0.6$.
To increase the range of lengths covered, we additionally show data on Alcyoneus \citep{Oei12022Alcyoneus}.
Next, under the same assumptions of a constant spectral index and self-similar growth, we derived a simple luminosity density--surface brightness relationship that allows for back-and-forth conversion between the two --- at least, given projected lengths and redshifts.
The mean lobe surface brightness $\langle b_{\nu} \rangle$ is proportional to the total luminosity density $L_\nu$, and assuming a pair of spherical lobes
\begin{align}
    \langle b_{\nu} \rangle = \frac{2 f_{L_\nu} \cdot L_\nu}{\pi^2 \cdot \mathbb{E}[D](\eta(f_l)) \cdot f_l^2 \cdot l_\mathrm{p}^2 \cdot (1 + z)^{3-\alpha}}.
\label{eq:surfaceBrightnessLobeMean}
\end{align}
Here $f_{L_\nu}$ is the fraction of the total luminosity density that belongs to the lobes, $f_l$ is the fraction of the RG's axis length that lies inside the lobes, and $\mathbb{E}[D](\eta(f_l))$ is the mean deprojection factor as given by Eq.~\ref{eq:deprojectionFactorLobesExpectation}.
The peak surface brightness $b_{\nu,\mathrm{max}}$ relates to $\langle b_\nu \rangle$ as $b_{\nu,\mathrm{max}} = \frac{3}{2}\langle b_\nu \rangle$.
Section~\ref{ap:surfaceBrightnessPrior} contains derivations for both these results.
For Alcyoneus, at $z_\mathrm{Alc.} = 0.25$, $f_{L_\nu} = 0.3$ and $f_l = 0.3$ \citep{Oei12022Alcyoneus}.
Assuming these parameter values, again in the top panel of Fig.~\ref{fig:surfaceBrightness}, we show luminosity density--projected length pairs of RGs at $z = z_\mathrm{Alc.}$ whose peak (solid green curve) or mean (dash-dotted green curve) surface brightness equals the LoTSS DR2 noise level.
Thus, each curve represents a family of borderline-detectable giants at Alcyoneus's redshift.
The other dash-dotted curves indicate similar barely detectable families, but for other redshifts.
Without optimising any free parameters, the curves correctly predict that Alcyoneus's lobes have surface brightnesses comparable to the $6''$ LoTSS DR2 noise level and explain the absence of observations in the top-left corner of the figure.
We conclude that Eq.~\ref{eq:surfaceBrightnessLobeMean} appears reasonable, but note that RGs may significantly differ in their values of $f_{L_\nu}$ and $f_l$.\footnote{This is also the reason that some giants in Fig.~\ref{fig:surfaceBrightness} cross their redshift's dash-dotted curve, which represents $f_{L_\nu} = 0.3$ and $f_l = 0.3$ only.}

Bolstered, we made use of Eq.~\ref{eq:surfaceBrightnessLobeMean} to Monte Carlo simulate --- for particular values of $b_{\nu,\mathrm{ref}}$ and $\sigma_\mathrm{ref}$ --- luminosity density--projected length relationships as they appear to observers.
The simulated giants have projected lengths adopted from the observed giants, randomly sampled redshifts up to $z = 0.6$ assuming a spatially constant GRG number density, and randomly sampled reference surface brightnesses (i.e. those for RGs at $z = 0$ that have intrinsic length $l = l_\mathrm{ref}$) whose distribution is determined by $b_{\nu,\mathrm{ref}}$ and $\sigma_\mathrm{ref}$.
We then used Eq.~\ref{eq:surfaceBrightnessRVMainText} to compute surface brightnesses as they would be observed, and retained only those giants whose surface brightness exceeds the LoTSS DR2 noise level.
For these simulated detectable giants, we finally generated luminosity densities using Eq.~\ref{eq:surfaceBrightnessLobeMean}, assuming wide uniform distributions $f_{L_\nu} \sim f_l \sim \mathrm{Uniform}(0.1, 0.9)$.
The middle and bottom panels of Fig.~\ref{fig:surfaceBrightness} show results for $b_{\nu,\mathrm{ref}} = 1750\ \mathrm{Jy\ deg^{-2}}$ and $b_{\nu,\mathrm{ref}} = 250\ \mathrm{Jy\ deg^{-2}}$, respectively; we adopted $\sigma_\mathrm{ref} = 1.3$ from our LoTSS DR2 GRG surface brightness measurements.
The median luminosity density of the $z < 0.6$ LoTSS DR1 giants is $L_\nu = 1.1 \cdot 10^{26}\ \mathrm{W\ Hz^{-1}}$, that of the $b_{\nu,\mathrm{ref}} = 1750\ \mathrm{Jy\ deg^{-2}}$ simulated giants is $L_\nu = 0.6 \cdot 10^{26}\ \mathrm{W\ Hz^{-1}}$, and that of the $b_{\nu,\mathrm{ref}} = 250\ \mathrm{Jy\ deg^{-2}}$ simulated giants is $L_\nu = 0.2 \cdot 10^{26}\ \mathrm{W\ Hz^{-1}}$.\footnote{
For another interesting case, $b_{\nu,\mathrm{ref}} = 500\ \mathrm{Jy\ deg^{-2}}$ (not shown), the median $L_\nu = 0.3 \cdot 10^{26}\ \mathrm{W\ Hz^{-1}}$.}
Interestingly, the higher reference surface brightness median provides a better fit to the data.
Even in case our Monte Carlo approach would predict luminosity densities that are biased low by a factor two, the higher median remains favoured.

In conclusion, it seems reasonable to suppose that our measurement $b_{\nu,\mathrm{ref}} = 1.0 \cdot 10^3\ \mathrm{Jy\ deg^{-2}}$ is not, or only mildly, biased high by selection effects.
Still, we take a conservative approach and in setting priors we assume a 75\% error on our measurement of $b_{\nu,\mathrm{ref}}$ and a 50\% error on our measurement of $\sigma_\mathrm{ref}$.
Thus, the priors for $b_{\nu,\mathrm{ref}}$ and $\sigma_\mathrm{ref}$ --- which we choose to be Gaussian --- have 68\% credible intervals $[250\ \mathrm{Jy\ deg^{-2}}, 1750\ \mathrm{Jy\ deg^{-2}}]$ and $[0.65, 1.95]$.
We retain flat priors for $\xi$ and $\frac{1}{2}(\phi_\mathrm{max} - \phi_\mathrm{min})$.
\begin{figure*}
    \centering
    \includegraphics[width=\textwidth]{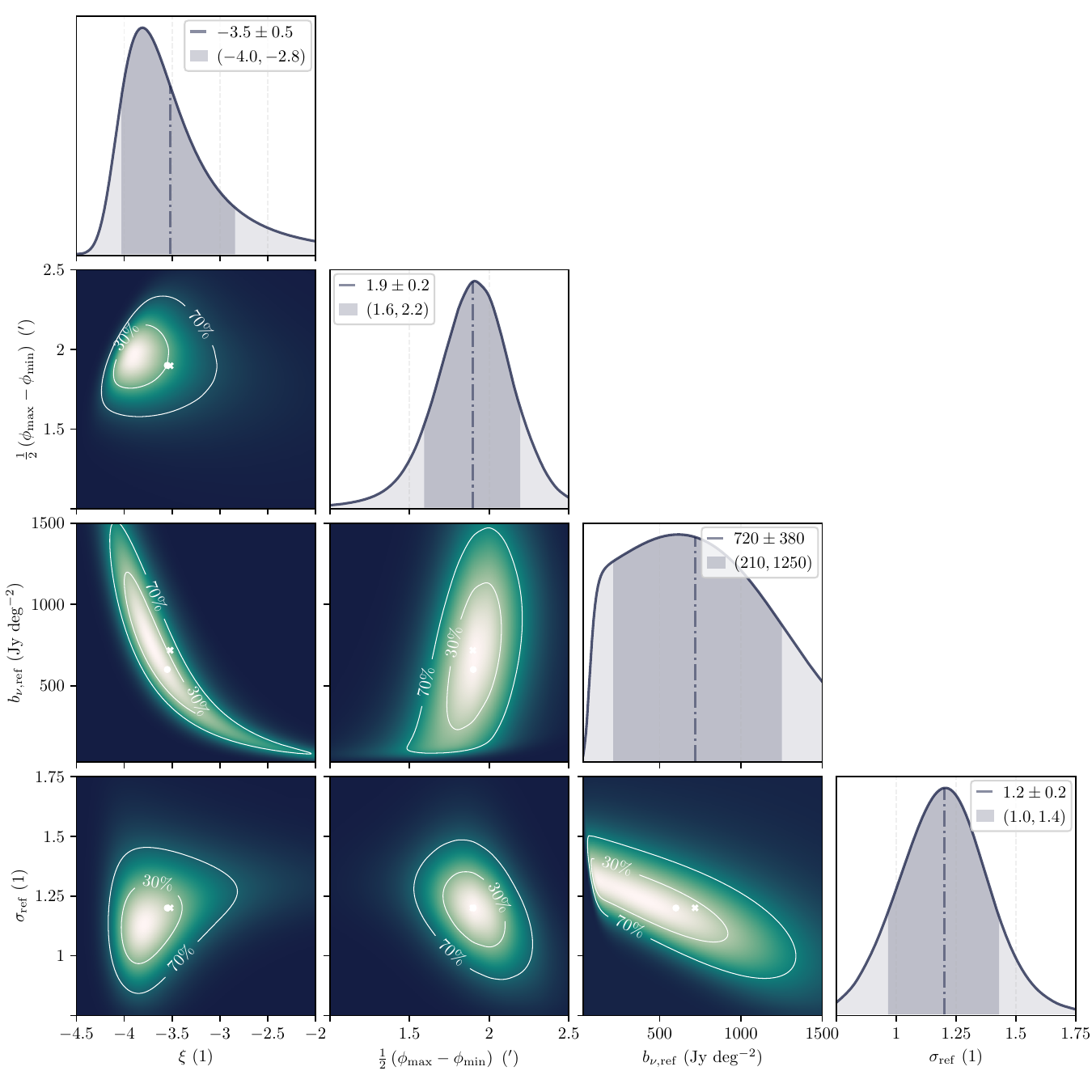}
    \caption{
    Joint posterior distribution over $\xi$ --- the parameter of interest --- and $\frac{1}{2}(\phi_\mathrm{max} - \phi_\mathrm{min})$, $b_{\nu,\mathrm{ref}}$ and $\sigma_\mathrm{ref}$ --- the selection effect parameters, based on 1473 projected lengths of LoTSS DR2 giants up to $z_\mathrm{max} = 0.5$.
    We show all two-parameter marginals of the posterior, with contours enclosing $30\%$ and $70\%$ of total probability.
    We mark the maximum a posteriori probability parameters (white circle) and the posterior mean parameters (white cross).
    The single-parameter marginals again show the estimated posterior mean, now marked by a vertical line, alongside shaded median-centred $80\%$ credible intervals.
    To compare the posterior to the likelihood function, which is also the posterior for a uniform prior, see Fig.~\ref{fig:likelihood}.
    }
    \label{fig:posterior}
\end{figure*}\noindent
\subsubsection{Inference}
\label{sec:inference}
To compute the posterior distribution for $\xi$, $\frac{1}{2}(\phi_\mathrm{max} - \phi_\mathrm{min})$, $b_{\nu,\mathrm{ref}}$ and $\sigma_\mathrm{ref}$, we first brute-force evaluated the likelihood function over a regular grid that covers a total of 3.3 million parameter combinations.\footnote{The computation took a few thousand CPU hours to complete, but can be trivially distributed among nodes, and within a node among CPUs.
Model extensions that introduce additional parameters shall necessitate more efficient inference techniques, such as Markov chain Monte Carlo.}
For each proposed parameter quartet, we computed the PDF of $L_\mathrm{p,obs}\ \vert\ L_\mathrm{p,obs} > l_\mathrm{p,GRG}$, and obtained the likelihood assuming that the LoTSS DR2 GRG projected proper lengths are IID draws from it.
To obtain the PDF, we successively evaluated Eqs.~\ref{eq:probabilityObservedAL}, \ref{eq:probabilityObservedSB}, \ref{eq:probabilityObserved}, \ref{eq:completeness}, \ref{eq:PDFProjectedProperLength}, and \ref{eq:GRGObservedProjectedProperLength}, alongside their direct dependencies.
This required the numerical evaluation of five integrals.
Compared with using Riemann sums, we achieved substantial accuracy improvements at virtually no added numerical cost by approximating these integrals with the trapezoid rule and the composite Simpson's rule.

We summarise the likelihood function in Table~\ref{tab:likelihood} and Fig.~\ref{fig:likelihood}.
To obtain the posterior, we simply multiplied the likelihood function by the prior and normalised the result.
\begin{center}
\captionof{table}{
Maximum a posteriori probability (MAP) and posterior mean and standard deviation (SD) estimates of the free parameters in intrinsic GRG length distribution inference.\protect\footnotemark
}
$z_\mathrm{max} = 0.5$:
\vspace{1mm}\\
\begin{tabular}{c c c}
\hline
parameter & MAP & posterior mean and SD\\
 [3pt] \hline\arrayrulecolor{lightgray}
$\xi$ & $-3.55$ & $-3.5 \pm 0.5$\\
\hline
$\frac{1}{2}(\phi_\mathrm{max} - \phi_\mathrm{min})$ & $1.9'$ & $1.9\pm 0.2'$\\
\hline
$b_{\nu,\mathrm{ref}}$ & $600\ \mathrm{Jy\ deg^{-2}}$ & $720 \pm 380\ \mathrm{Jy\ deg^{-2}}$\\
\hline
$\sigma_\mathrm{ref}$ & $1.2$ & $1.2 \pm 0.2$
\end{tabular}
\vspace{2mm}\\
$z_\mathrm{max} = 0.25$:
\vspace{1mm}\\
\begin{tabular}{c c c}
\hline
parameter & MAP & posterior mean and SD\\
 [3pt] \hline\arrayrulecolor{lightgray}
$\xi$ & $-3.5$ & $-3.5 \pm 0.4$\\
\hline
$\frac{1}{2}(\phi_\mathrm{max} - \phi_\mathrm{min})$ & $1.85'$ & $1.7\pm 0.3'$\\
\hline
$b_{\nu,\mathrm{ref}}$ & $900\ \mathrm{Jy\ deg^{-2}}$ & $1020 \pm 490\ \mathrm{Jy\ deg^{-2}}$\\
\hline
$\sigma_\mathrm{ref}$ & $1.15$ & $1.3 \pm 0.4$
\end{tabular}
\label{tab:posterior}
\end{center}
\footnotetext{The model assumes $\xi$ is constant for $z \in [0, z_\mathrm{max}]$.
We determined the posterior twice: for $z_\mathrm{max} = 0.5$, using 1473 giants, and for $z_\mathrm{max} = 0.25$, using 811 giants.}
In Table~\ref{tab:posterior}, for each parameter, we list the maximum a posteriori probability (MAP) estimate, alongside estimates for the posterior mean and standard deviation.
In Fig.~\ref{fig:posterior}, we visualise all one- and two-dimensional posterior marginals, in which we mark the MAP (white circle) and the posterior mean (white cross).
The joint marginal for $\xi$ and $b_{\nu,\mathrm{ref}}$ shows that these parameters have a strong negative correlation, indicating that with current data, the steep slope of the ESF at high $l_\mathrm{p}$ can equally be described with a steep intrinsic slope and mild surface brightness selection (i.e. $\xi$ low and $b_{\nu,\mathrm{ref}}$ high), or by a shallow intrinsic slope and strong surface brightness selection (i.e. $\xi$ high and $b_{\nu,\mathrm{ref}}$ low).
We leave it up to future studies to break this degeneracy, either by using larger samples, by measuring $b_{\nu,\mathrm{ref}}$ directly, or by improving survey sensitivities so that surface brightness selection effect modelling becomes superfluous altogether.
\subsubsection{Goodness of fit}
In both panels of Fig.~\ref{fig:ECDFCompleteness}, we compare the ESF and SF of $L_\mathrm{p,obs}\ \vert\ L_\mathrm{p,obs} > l_\mathrm{p,GRG}$ for $z_\mathrm{max} = 0.5$, using the MAP parameters for the latter.
The model appears able to produce a tight fit to the data.
The mean and standard deviation of the ESF--SF residuals are $0.01\%$ and $0.3\%$, whilst the mean and standard deviation of the absolute ESF--SF residuals are both $0.2\%$.
Using a Kolmogorov--Smirnov test, we formally verified that our \emph{best} parameters are indeed \emph{good} parameters --- in the sense that they represent a plausible model underlying the data.
The Kolmogorov--Smirnov statistic is the maximum deviation between the ESF and SF, and equals $1\%$ in our case.
The p-value --- the probability that an ESF--SF discrepancy of at least this magnitude would occur if the SF represents the true underlying distribution --- is $p = 99\%$.
For any reasonable significance level, we do not reject the null hypothesis.
The model, given our best parameters, indeed represents a possible description of the data.
We conclude that the distribution of GRG intrinsic proper lengths, after correcting for selection effects, is consistent with a single Pareto distribution with tail index $\xi = -3.5$.
We show the SF of this distribution in both panels of Fig.~\ref{fig:ECDFCompleteness} (fading green dots).
For low $l_\mathrm{p}$, the observed slope is shallower (due to angular length selection), whilst for high $l_\mathrm{p}$, the observed slope is steeper (due to surface brightness selection).

A quasi-Pareto distribution can arise naturally as the tail of a lognormal distribution \citep[e.g.][]{Malevergne12011}, and there are reasons to believe that the entire radio galaxy length distribution is indeed approximately lognormal \citep{Oei12022GiantsCosmicWeb}.
This provides an explanation of the approximately Paretian nature of the giant radio galaxy length distribution found in this section.
The specific value of the tail index $\xi$ is set by both the physics of radio galaxy growth and the distribution of radio galaxies over large-scale environments, the latter of which we measure in \citet{Oei12022GiantsCosmicWeb}.
Our result $\xi = -3.5 \pm 0.5$ is a new constraint for dynamical models such as those of \citet{Turner12015} and \citet{Hardcastle12018}.
\begin{figure}
    \centering
    \includegraphics[width=\columnwidth]{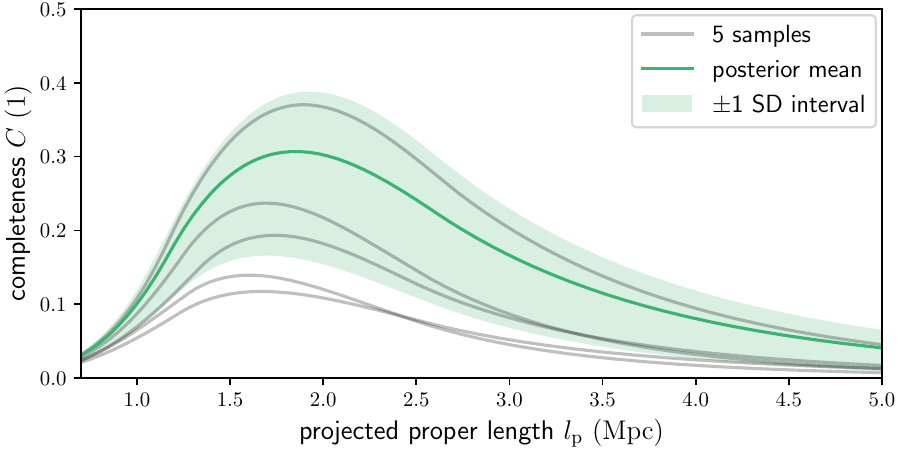}
    \caption{
    Completeness $C$ of a sample of giant radio galaxies up to cosmological redshift $z_\mathrm{max}$ as a function of projected proper length $l_\mathrm{p}$.
    From samples of the posterior distribution, we infer the LoTSS DR2 GRG search campaign completeness up to $z_\mathrm{max} = 0.5$.
    We show completeness curves for five randomly selected samples (grey) and for the posterior mean (dark green).
    We also show an interval around the completeness mean with the completeness standard deviation (SD) as the half-width (light green).
    The completeness peaks around $l_\mathrm{p} = 2\ \mathrm{Mpc}$.
    }
    \label{fig:completenessPosterior}
\end{figure}

\subsection{Giant radio galaxy number density}
\label{sec:numberDensity}
\begin{figure}
    \centering
    \includegraphics[width=\columnwidth]{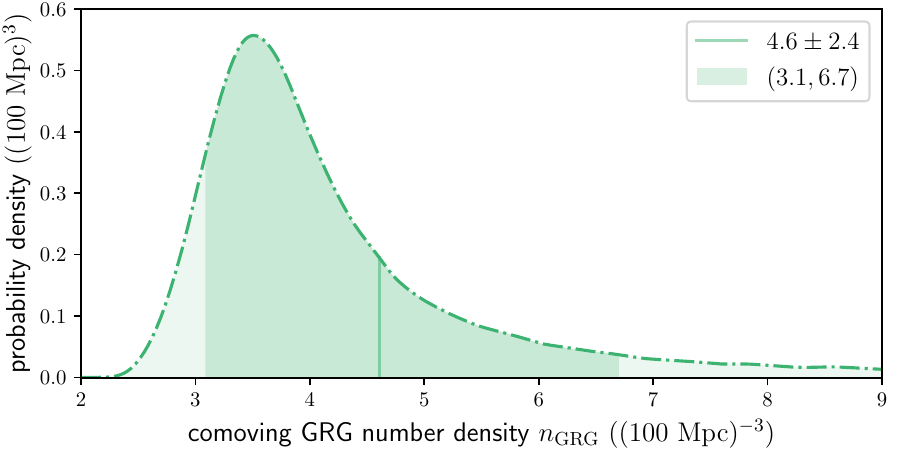}
    \caption{
    PDF of the comoving GRG number density $n_\mathrm{GRG}$.
    We mark the mean (vertical line) and the median-centred $80\%$ credible interval (darker range).
    In the Local Universe, the average number of giants per comoving cube with $100\ \mathrm{Mpc}$ sides is $4.6 \pm 2.4$.
    We define giants through $l_\mathrm{p,GRG} = 0.7\ \mathrm{Mpc}$, and define the Local Universe to be up to cosmological redshift $z_\mathrm{max} = 0.5$.
    }
    \label{fig:numberDensityOfGRGs}
\end{figure}
If in addition to our discoveries, we know how many giants our search campaign has missed, then we can infer the true comoving GRG number density in the Local Universe.
The posterior distribution over selection effect parameters $\frac{1}{2}(\phi_\mathrm{max} - \phi_\mathrm{min})$, $b_{\nu,\mathrm{ref}}$ and $\sigma_\mathrm{ref}$ induces a probability distribution over the search completeness function $C(l_\mathrm{p})$.
$C(l_\mathrm{p})$ denotes the probability that a giant of projected proper length $l_\mathrm{p}$ in comoving space up to $z = z_\mathrm{max}$ is detected through the search.
We first generated parameter samples from our posterior using rejection sampling, and then used each to calculate a $C(l_\mathrm{p})$ sample.
We show the distribution over $C(l_\mathrm{p})$ for $z_\mathrm{max} = 0.5$ in Fig.~\ref{fig:completenessPosterior}.
For small $l_\mathrm{p}$, $C$ is low as many giants drop out due to angular length selection; for large $l_\mathrm{p}$, $C$ is low as many giants drop out due to surface brightness selection.
The completeness peaks around $l_\mathrm{p} \sim 2\ \mathrm{Mpc}$; however, even there the majority of giants remains undetected.

We inferred a probability distribution over the true comoving GRG number density $n_\mathrm{GRG}$ by combining Eqs.~\ref{eq:completeness} and \ref{eq:GRGComovingNumberDensity} with the LoTSS DR2 GRG catalogue and samples from Sect.~\ref{sec:inference}'s posterior.
The resulting skewed distribution, with mean and SD $n_\mathrm{GRG} = 4.6 \pm 2.4\ (100\ \mathrm{Mpc})^{-3}$ and 80\% credible interval $3.1$--$6.7\ (100\ \mathrm{Mpc})^{-3}$, is shown in Fig.~\ref{fig:numberDensityOfGRGs}.
We note that, although the uncertainty in $b_{\nu,\mathrm{ref}}$ induces a large uncertainty in $C$ from $l_\mathrm{p} \sim 1.5\ \mathrm{Mpc}$ onwards, the completeness uncertainty at large projected lengths does not substantially contribute to the uncertainty in $n_\mathrm{GRG}$.
This is because the GRG population is dominated by smaller giants, for which the completeness appears better constrained.

What picture arises regarding the abundance of giant radio galaxies in the Local Universe's large-scale structure?
If we model the Cosmic Web through comoving cubic unit cells \citep{Oei12022MASSCW} with $50\ \mathrm{Mpc}$ sides, and each cubic unit cell contributes one cluster and three filaments, then a cube with $100\ \mathrm{Mpc}$ sides features a total of eight clusters and $24$ filaments.
For comparison, in a $(100\ \mathrm{Mpc})^3$ volume up to $z_\mathrm{max} = 0.5$, the SDSS-III cluster catalogue of \citet{Wen12012} contains on average $11.2$ clusters of any mass, and $4.5$ clusters of mass $M_{200} > 10^{14}\ M_\odot$.
Since clusters contain ${\sim}20\%$ of giants \citep{Oei12022GiantsCosmicWeb}, we find the average number of giants per cluster to be ${\sim}10^{-1}$.
If one assumes that filaments contain the remaining ${\sim}80\%$ of giants, and uses the fact that the average number of filaments per cluster is of order unity, it follows that the average number of giants per filament is also ${\sim}10^{-1}$.
In all likelihood, most clusters and filaments do not currently contain a giant.

\subsection{Giant radio galaxy lobe volume-filling fraction}
Because giant radio galaxies enrich the IGM with hot plasma and magnetic fields far beyond the circumgalactic media of their hosts, they may provide a meaningful contribution to the heating and magnetisation of --- in particular --- the most rarefied parts of the filament IGM.
By combining the GRG number density and the GRG jet power distribution \citep[e.g.][]{Dabhade12020October}, one could estimate the instantaneous heating and magnetisation contributions directly.
We recommend such analysis for future research.

We evaluated Eq.~\ref{eq:VFFTruncated} to obtain an estimate of the fraction of the Local Universe's proper volume that GRG lobes occupy.
We used Alcyoneus as a reference giant, for which $V = 2.5 \pm 0.3\ \mathrm{Mpc^3}$ and $l_\mathrm{p} = 4.99 \pm 0.04\ \mathrm{Mpc}$ \citep{Oei12022Alcyoneus}; this suggests $\mathbb{E}[\Upsilon_\mathrm{p}] \approx 2\%$.
Future work should determine whether Alcyoneus's case is typical, as observations, such as those shown in Figs.~\ref{fig:LoTSSDR2GRGs1}--\ref{fig:LoTSSDR2GRGsLargerThanLuna}, suggest that giants exhibit a large variety of shapes --- and thus total lobe volume--cubed length ratios.
Interestingly, simulations by \citet{Krause12012} have found that these shapes also depend on environmental parameters such as ambient pressure and density.
Truncating the GRG projected length distribution at $l_\mathrm{p,max} = 7\ \mathrm{Mpc}$, so that its support is exactly an order of magnitude, Eq.~\ref{eq:VFFTruncated} predicts $\mathrm{VFF}_\mathrm{GRG}(z = 0) = 3\substack{+4\\-1} \cdot 10^{-7}$.\footnote{
We found the weaker constraints $\mathrm{VFF}_\mathrm{GRG}(z = 0) > 13\substack{+21\\-4} \cdot 10^{-8}$, $\mathrm{VFF}_\mathrm{GRG}(z = 0) > 11\substack{+16\\-3} \cdot 10^{-8}$, and $\mathrm{VFF}_\mathrm{GRG}(z = 0) > 6\substack{+4\\-1} \cdot 10^{-8}$ using Eqs.~\ref{eq:VFFLowerBound}, \ref{eq:VFF1}, and \ref{eq:VFF2}, respectively.}

Whether this result is sensitive to changes in $l_\mathrm{p,max}$ depends on $\xi$, with $\xi = -3$ being a special value under self-similar growth.
In that case, small and large giants contribute equally to $\mathrm{VFF}_\mathrm{GRG}$: although large giants are rarer ($f_{L_\mathrm{p}} \propto l_\mathrm{p}^{-3}$), their larger lobe volumes ($V \propto l_\mathrm{p}^3$) exactly compensate.
For $\xi < -3$, small giants provide the dominant contribution to $\mathrm{VFF}_\mathrm{GRG}$ and the choice of $l_\mathrm{p,max}$ can be irrelevant; for $\xi > -3$, large giants dominate and the choice of $l_\mathrm{p,max}$ always matters.

If we assume that giants occur in clusters and filaments only and use the fact that clusters and filaments comprise about 5\% of the Local Universe's volume \citep{Forero-Romero12009}, then the GRG lobe VFF within clusters and filaments specifically is $\mathrm{VFF}_\mathrm{GRG}(z = 0) = 5\substack{+8\\-2} \cdot 10^{-6}$.
We conclude that, at each given moment, GRG lobes occupy just a small fraction of the WHIM and ICM.
If the enrichment of the IGM by giants is to affect the WHIM and ICM on a large scale, mixing processes in the IGM are necessary and many galaxies must be able to form giants at some point in their evolution.

\subsection{Unification model constraints from quasar and non-quasar giants}
\label{sec:unificationModel}
\begin{figure}
    \centering
    \includegraphics[width=\columnwidth]{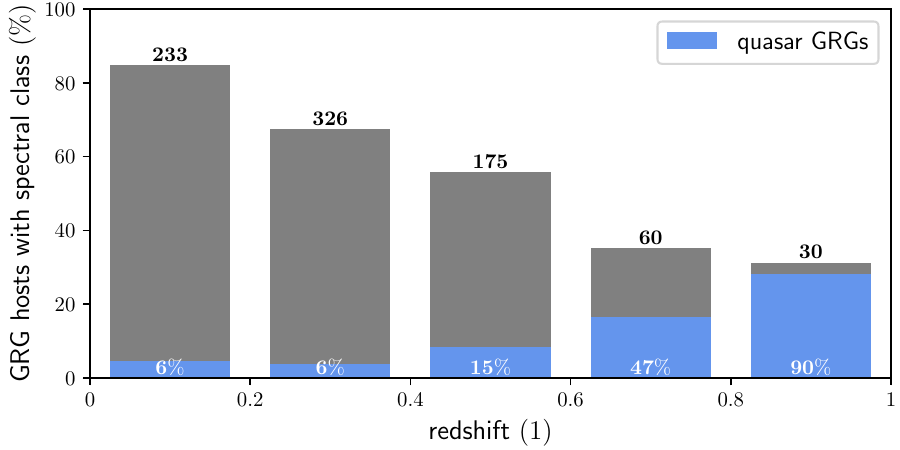}
    \caption{
    Binary classification into quasar and non-quasar giants based on SDSS DR12 host spectra, for 5 redshift intervals.
    We selected all giants with definite hosts within the SDSS-covered sky patch bounded by right ascensions $120\degree$ and $250\degree$, and declinations $27\degree$ and $62\degree$.
    As redshift increases, the fraction of hosts with a known spectral class decreases (grey bars, with the absolute number of such hosts in black), whilst the fraction of quasar identifications increases (blue bars and white percentages).
    }
    \label{fig:spectralClass}
\end{figure}
\begin{figure}
    \centering
    \begin{subfigure}{\columnwidth}
    \includegraphics[width=\columnwidth]{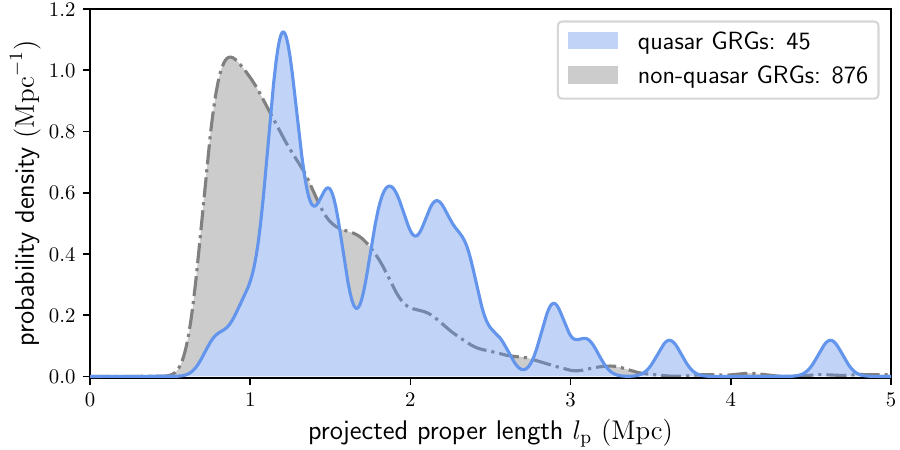}
    \end{subfigure}
    \begin{subfigure}{\columnwidth}
    \includegraphics[width=\columnwidth]{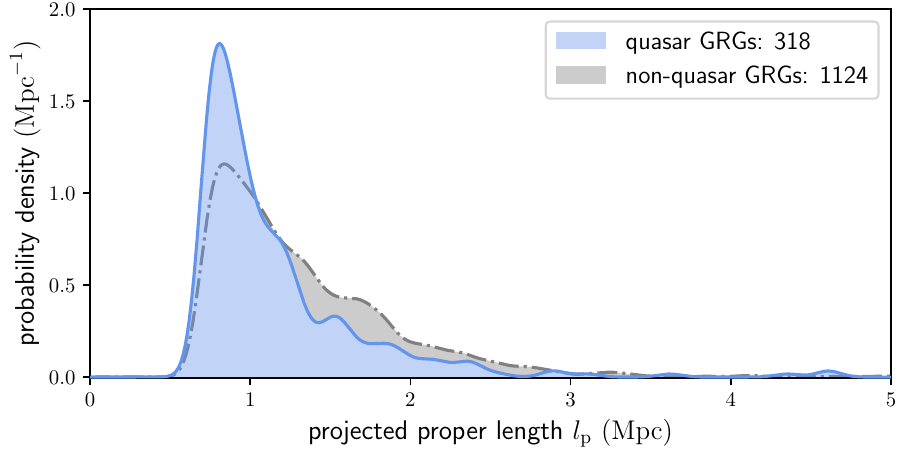}
    \end{subfigure}
    \caption{
    Observed projected proper length PDFs for SDSS-classified quasar and non-quasar giants, obtained through kernel density estimation.
    We used a Gaussian kernel with $\sigma_\mathrm{KDE} = 75\ \mathrm{kpc}$.
    For both panels, two-sample Kolmogorov--Smirnov tests yield $p < 1$\textperthousand.
    However, given the severe impact of selection effects, we could not reject the null hypothesis that quasar giants and non-quasar giants have the same underlying projected proper length distribution.
    \textit{Top:} newly discovered (LoTSS DR2) giants with SDSS spectral class labels.
    \textit{Bottom:} all known giants with SDSS spectral class labels.
    }
    \label{fig:distributionLengthProjectedQSO}
\end{figure}\noindent
In Section~\ref{sec:unificationModelTheory}, we have predicted general ramifications of the basic unification model on a GRG sample.
We constrained and tested this model with our LoTSS DR2 GRG sample.

First, for all 3198 giants with definitively identified hosts, we queried the SDSS DR12 spectral class \texttt{spCl}: a Boolean label indicating whether or not the host contains a quasar.
As many hosts have no SDSS DR12 spectrum, or even fall outside of SDSS DR12 coverage, we retrieved host classifications for just 1442 of these giants (45\%).
Of these classified giants, 318 are quasar giants (22\%) (of which 45 (14\%) are discoveries presented in this work) and 1124 are non-quasar giants (78\%) (of which 876 (78\%) are discoveries presented in this work).
Therefore, the apparent LoTSS DR2 quasar GRG fraction $f_\mathrm{Q} = \frac{45}{45 + 876} = 5\%$.
However, spectral class labels are preferentially available for GRG hosts with higher optical flux densities, such as those at low redshifts or those containing quasars, because they are more probable spectroscopic targets.
Through Fig.~\ref{fig:spectralClass}, we demonstrate that the fraction of observed GRG hosts with spectral class labels indeed decreases with redshift, whilst the fraction of quasar identifications increases.
For each redshift interval, in white, we denote the fraction of quasar giants within the classified population.
If spectral class labels would have been available for the non-classified observed populations as well, the quasar GRG fractions would probably have been lower.
Assuming that all observed giants whose hosts have an unknown spectral class are non-quasar giants, we find quasar GRG fractions $f_\mathrm{Q} = 5\%$, $4\%$, $8\%$, $16\%$, and $28\%$, for redshift intervals $0$--$0.2$, $0.2$--$0.4$, $0.4$--$0.6$, $0.6$--$0.8$, and $0.8$--$1$, respectively.
The true quasar GRG fraction for a given redshift interval might still differ from the aforementioned \emph{observed} quasar GRG fraction: namely, if selection effects make a given quasar GRG easier (or harder) to detect than a given non-quasar GRG.
At higher redshifts, quasar giants certainly appear easier to detect than non-quasar giants, as hosts without quasars often become too faint to optically identify.
For the lowest redshift intervals, this problem does not exist, and we therefore consider the observed quasar GRG fraction $f_\mathrm{Q}=5\%$ to be closest to the true one.
A thorough analysis of the impact of selection effects on $f_\mathrm{Q}$ is a topic for future research.

Using Eq.~\ref{eq:quasarGRGProbability}, and assuming a quasar GRG probability $p_\mathrm{Q} = 5\%$, we found a maximum inclination angle $\theta_\mathrm{max} = 39\substack{+2\\-3}\degree$.
For $p_\mathrm{Q} \sim \mathrm{Uniform}(4\%, 6\%)$, the result remained the same.
In conclusion, if the basic unification model considered in Sect.~\ref{sec:unificationModelTheory} is correct, then observations of giants predict that quasars are AGN seen along lines-of-sight that make an angle of at most $\theta_\mathrm{max} = 39\substack{+2\\-3}\degree$ with the black hole rotation axis.

Finally, we tested whether the RV $L_\mathrm{p,obs}\ \vert\ L_\mathrm{p,obs} \geq l_\mathrm{p,GRG}$ has the same distribution for quasar and non-quasar giants, as predicted by the unification model.
In the top panel of Fig.~\ref{fig:distributionLengthProjectedQSO}, we show PDFs approximated through kernel density estimation (KDE) for newly discovered (LoTSS DR2) SDSS-classified quasar giants and non-quasar giants up to $z_\mathrm{max} = \infty$.
Despite the small number of quasar giants, $N_\mathrm{Q} = 45$, the two-sample Kolmogorov--Smirnov (KS) test yielded a low p-value $p < 1$\textperthousand: we rejected the null hypothesis that these distributions stem from a single underlying one.
However, this does not mean that the unification model hypothesis should immediately be rejected as well, because the selective availability of SDSS DR12 spectral class labels induces a severe selection effect.\footnote{
In fact, the top panel of Fig.~\ref{fig:distributionLengthProjectedQSO} shows expected behaviour for a GRG search campaign with a (fuzzy) angular length threshold selection effect if the unification model is correct.
At high redshifts, a GRG must have a larger projected length to pass the angular length threshold than at low redshifts.
Thus, sampled high-redshift giants are physically larger than sampled low-redshift giants.
Because the fraction of high-redshift quasar giants that is detectable and spectrally classifiable is higher than the fraction of high-redshift non-quasar giants that is detectable and spectrally classifiable, sampled quasar giants will be physically larger than sampled non-quasar giants.
To draw this latter conclusion, we must also invoke the fact that, under the unification model, the projected length distributions of quasar and non-quasar giants are the same.}
This effect can be tempered by choosing a lower $z_\mathrm{max}$, but for choices such as $z_\mathrm{max} = 0.5$ and $0.25$, just $N_\mathrm{Q} = 16$ and $10$ quasar giants remain --- too few to extract meaningful information.
The bottom panel of Fig.~\ref{fig:distributionLengthProjectedQSO} again shows observed projected proper length KDE PDFs, but now for \emph{all} SDSS-classified quasar giants and non-quasar giants up to $z_\mathrm{max} = \infty$; in this case, $N_\mathrm{Q} = 318$.
Again, the two-sample KS test yields $p < 1$\textperthousand, but this time quasar giants appear smaller than non-quasar giants.
Because we aggregated samples here that have different selection effects imprinted, it becomes hard to draw clear conclusions.
Limiting $z_\mathrm{max}$ reduces the severity of most selection effects, but of course comes at the cost of reducing the sample size.
Interestingly, the corresponding quasar GRG and non-quasar GRG projected length distributions do become more alike; for instance for $z_\mathrm{max} = 0.5$ and $0.25$, the two-sample KS test yields $p = 1\%$ and $4\%$, respectively.
In conclusion, because quasar giants are intrinsically rare and the availability of spectral class labels is biased towards low redshifts and hosts containing quasars, it is challenging to robustly test the unification model with current GRG observations.
We refrain from drawing final conclusions, and recommend a careful future analysis.

\section{Discussion}
\label{sec:discussion}
\subsection{Radio galaxy length definitions}
How large are radio galaxies?
Despite the simplicity of this question and more than half a century of research on radio galaxies, their intrinsic length distribution has not yet been rigorously characterised.
In this work, we have carried out the first precision analysis of the tail of the radio galaxy intrinsic length distribution.
Precision analyses tend to raise questions; firstly, whether the studied observable is well defined, and secondly whether one could conceive of more informative observables --- that is to say those that make it easier to reveal underlying physical mechanisms.
This work's main observable is the radio galaxy projected proper length; we argue that it is neither well defined nor maximally informative.
\subsubsection{The current length definition: survey-dependence}
Contemporary research uses a survey-dependent definition for radio galaxy angular lengths, which then makes projected proper lengths survey-dependent too.

The angular length is canonically defined as the largest possible angular separation between two directions for which the RG's specific intensity function $I_{\nu,\mathrm{RG}}$ exceeds $b_{\nu,\mathrm{th}}$: some specified factor of order unity times the image noise $\sigma_{I_\nu}$.
A complication is that not only $I_{\nu,\mathrm{RG}}$, but also $\sigma_{I_\nu}$ varies with observing frequency; the latter because of observational factors such as $\left(u,v\right)$-coverage, bandpass, radio-frequency interference (RFI), ionospheric weather, the sky density of bright calibrators and the performance of calibration algorithms.
$I_{\nu,\mathrm{RG}}$ additionally depends on resolution, at least for point sources; $\sigma_{I_\nu}$ additionally depends on resolution and integration time.
As a result, both $I_{\nu,\mathrm{RG}}$ and the concrete value of $b_{\nu,\mathrm{th}}$ used in the angular length definition change from image to image.
Each study thus far has therefore implicitly used a different definition for angular length, instead of a shared, absolute one.
As the projected proper length follows from combining the angular length with the host redshift, it suffers from the same problem.

Whether the survey-dependence of the current length definition is problematic, depends on the radio galaxy.
For archetypal FRII RGs, the angular length is roughly equal to the angular distance between the hotspots.
These are an FRII RG's brightest morphological components, and are thus the first to be picked up by a survey.
In contrast, archetypal FRI RGs, which gradually fade with distance from the host, can have significantly larger angular lengths in surveys of higher sensitivity.
The giants in the middle-right and bottom-left panel of Fig.~\ref{fig:LoTSSDR2GRGsLargerThanLuna} are good examples: these radio galaxies were known before the LoTSS DR2, but were not known to be giants; similarly, more sensitive surveys are poised to assign them even larger extents.
If we are to move towards precision science, it therefore makes sense --- at least for FRI RGs --- to more explicitly recognise that the angular and projected proper lengths are functions of the observing frequency $\nu_\mathrm{obs}$ and a surface brightness threshold $b_{\nu,\mathrm{th}}$.
If catalogues would explicitly state for what combination  $\left(\nu_\mathrm{obs}, b_{\nu,\mathrm{th}}\right)$ they provide $\phi = \phi\left(\nu_\mathrm{obs},b_{\nu,\mathrm{th}}\right)$ and $l_\mathrm{p} = l_\mathrm{p}\left(\nu_\mathrm{obs},b_{\nu,\mathrm{th}}\right)$, it is possible to homogenise a collection of data sets by using universal angular and projected proper length definitions for all RGs.\footnote{Our LoTSS DR2 GRG $\phi$ and $l_\mathrm{p}$ correspond to $b_{\nu,\mathrm{th}} \coloneqq 1 \cdot \sigma_{I_\nu}$; on average, $\sigma_{I_\nu}(\nu_\mathrm{obs} = 144\ \mathrm{MHz}, \theta_\mathrm{FWHM} = 6'') = 25\ \mathrm{Jy\ deg^{-2}}$.}
If two angular lengths have been measured for the same RG, for instance $\phi_1$ at $\left(\nu_{\mathrm{obs},1}, b_{\nu,\mathrm{th},1}\right)$ and $\phi_2$ at $\left(\nu_{\mathrm{obs},2}, b_{\nu,\mathrm{th},2}\right)$, then we can estimate $\phi$ for any desired $\left(\nu_\mathrm{obs}, b_{\nu,\mathrm{th}}\right)$ through interpolation.
For example, the interpolation formula for a symmetric radio galaxy with jets or lobes of constant spectral index $\alpha$, and with a specific intensity function contribution which fades to zero linearly with angular distance from the host, is
\begin{align}
\phi = \max \left\{\phi_1 + \frac{b_{\nu,\mathrm{th}} - b_{\nu,\mathrm{th},1}\left(\frac{\nu_\mathrm{obs}}{\nu_{\mathrm{obs},1}}\right)^\alpha}{b_{\nu,\mathrm{th},2}\left(\frac{\nu_\mathrm{obs}}{\nu_{\mathrm{obs},2}}\right)^\alpha - b_{\nu,\mathrm{th},1}\left(\frac{\nu_\mathrm{obs}}{\nu_{\mathrm{obs},1}}\right)^\alpha}\left(\phi_2-\phi_1\right), 0\right\}.
\end{align}
\subsubsection{The ideal length definition: physical relevance}
Alcyoneus, shown in Fig.~\ref{fig:LoTSSDR2GRGs1}'s top-right panel, is a 5 Mpc giant whose ageing lobes are revealed for the first time by the LoTSS DR2 \citep{Oei12022Alcyoneus}.
Future image sensitivity improvements shall reveal more hitherto unseen, fading lobes around known RGs that formed in the aftermath of earlier AGN activity episodes.
Could a large fraction of sufficiently old RGs turn out to be giants, once such sensitivity improvements start providing evidence of earlier and earlier AGN activity episodes?

Up to now, it has been informative to include all visible plasma in the angular length measurement.
In future images, some of the visible plasma might be of such low pressure that it has become physically insignificant, in the sense that it does not affect the thermodynamics of the surrounding IGM anymore; we propose to exclude such plasma from a radio galaxy length.
Practically, one option is to introduce an absolute threshold: for example to include plasma of pressure $P \geq 10^{-17}\ \mathrm{Pa}$ only --- this is the pressure of the warm--hot intergalactic medium (WHIM) in the $\rho_\mathrm{BM} \sim 10\ \Omega_\mathrm{BM,0}\ \rho_\mathrm{c,0}$ and $T \sim 5 \cdot 10^5\ \mathrm{K}$ regime.
Another option is to introduce an environment-dependent pressure threshold; this would mean that a cluster RG sees its length measured against a higher pressure threshold than a filament RG, because its plasma becomes thermodynamically irrelevant sooner.
A problem with (equipartition or minimum energy) pressure--based length definitions is that pressure is a derived quantity: the specific intensity only determines the product of pressure and line-of-sight length through the lobe.
\subsection{Moving beyond line segment projection}
\label{sec:movingBeyondLineSegmentProjection}
\begin{figure}
    \centering
    \includegraphics[width=\columnwidth]{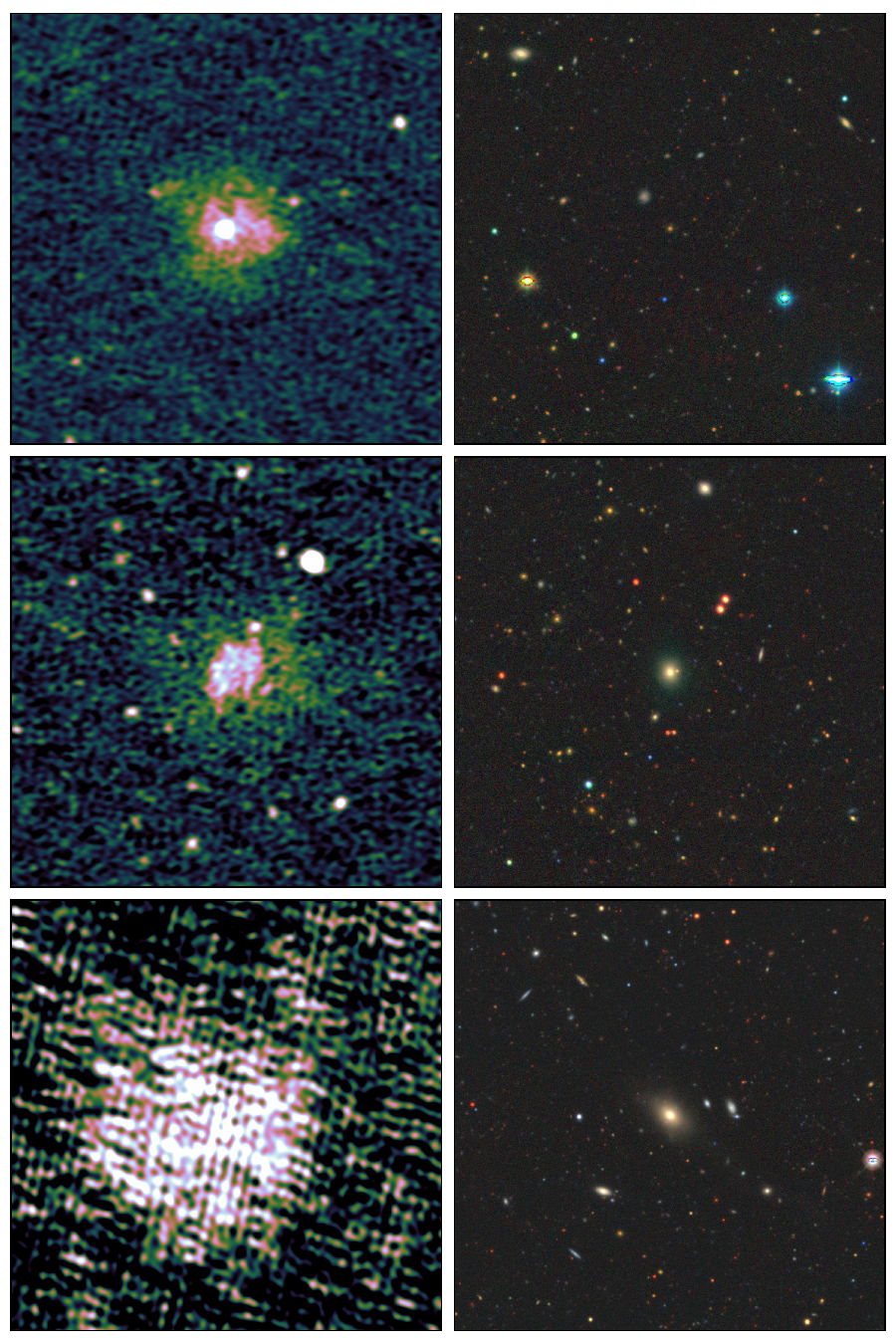}
    \caption{
    Suspected line-of-sight RGs in the radio and optical.
    We show $5' \times 5'$ LoTSS DR2 $6''$ cutouts \textit{(left column)} and corresponding DESI Legacy Imaging Surveys $(g,r,z)$-cutouts \textit{(right column)}.
    Due to its lobes, an RG whose axis aligns closely to the line of sight has a non-zero projected length.
    From top to bottom row, the SDSS host names are J125027.47+642034.3, J092220.72+560234.9, and J023100.61+032922.1.
    }
    \label{fig:RGsAlongLoS}
\end{figure}
In this work, we have adopted a classical approach to treating projection, by modelling radio galaxies as line segments.
When a radio galaxy's inclination angle $\theta$ is close to $0\degree$ or $180\degree$, this approach predicts that the angular length $\phi$ and thus the projected proper length $l_\mathrm{p}$ vanishes.
In reality however, because radio galaxy lobes have non-zero volumes, the RV $L_\mathrm{p}$ will never tend to zero.
Figure~\ref{fig:RGsAlongLoS} illustrates this point through three sources interpreted to be radio galaxies aligned closely with the line of sight.
In each case, $\phi$ remains of arcminute scale.

A more realistic approach, outlined in Appendix~\ref{ap:deprojectionFactor}, moves beyond the simplistic line segment geometry by adding two lobes of radius $R$ to the endpoints of the line segment, whose length is given by RV $L$.
The ratio $\frac{L}{L_\mathrm{p}}$, which was unbounded under the classical approach, is now at most $1 + \frac{1}{\eta}$, where $\eta \coloneqq \frac{2R}{L}$.
(The classical approach simply is the limit $\eta = 0$.)
We suggest a recalculation of this work's theoretical and applied results under this more realistic RG geometry as a direction for future research.
To obtain the applied results, one must either fix $\eta$ as a hyperparameter, or include it as an additional model parameter.
We note that even using conservatively low values of $\eta$, such as $\eta = \frac{1}{10}$, will represent an improvement in realism over $\eta = 0$.

\subsection{Unmodelled selection effects}
In this work, we have modelled both an angular length and a surface brightness selection effect.
Several other plausible selection effects have not been included in the forward model, as we have judged each to be of minor importance.
However, in unison, they could have a non-negligible influence on the distribution of the observed projected proper GRG length RV $L_\mathrm{p,obs}\ \vert\ L_\mathrm{p,obs} > l_\mathrm{p,GRG}$.
Some of their influence might have been absorbed by the parameters of the \emph{included} selection effects --- that is by $\frac{1}{2}(\phi_\mathrm{max} - \phi_\mathrm{min})$, $b_{\nu,\mathrm{ref}}$, and $\sigma_\mathrm{ref}$ --- or, worse still, by $\xi$.

One of these unmodelled selection effects is that RGs whose axes are oriented almost parallel to the line of sight are more likely to be rejected from a sample than RGs whose axes are closer to the plane of the sky, as the former do not always have a characteristic double-lobe appearance.
For instance, perhaps not all readers would regard our identification of the sources in the left column of Fig.~\ref{fig:RGsAlongLoS} as RGs convincing.
Nevertheless, as remarked in Sect.~\ref{sec:unificationModelTheory}, conditioning $L_\mathrm{p,obs}\ \vert\ L_\mathrm{p,obs} > l_\mathrm{p,GRG}$ on inclination angle does not affect its distribution.
As a result, this selection effect does not necessitate forward model modifications.\\
Furthermore, as shown in Sect.~\ref{sec:unificationModel}, there is a selection effect at play that favours the selection of quasar giants over non-quasar giants at high redshift, as host galaxies with quasars are more luminous in the optical and therefore have a better chance to be picked up in optical imagery.
This selection effect will get less severe once deeper photometric surveys become available.

We encountered three more selection effects during our LoTSS DR2 GRG search.
The larger an RG --- and especially an FRII RG --- becomes, the harder it is for an observer to identify its host galaxy, as an increasing number of plausible host candidates can lie interspersed in the strip of sky between the lobes.
The severity of this effect, which diminishes the prevalence of the largest giants in a sample, depends on the balance chosen between avoiding false discoveries and avoiding rejections of true discoveries.
Another selection effect runs against RGs in galaxy clusters.
Such environments can contain multiple adjacent RGs, making it at times unclear which lobe belongs to which RG.
When no confident double-lobe associations can be made, the RGs involved fail to make it into the sample.
If galaxy clusters contain primarily smaller giants, this selection effect induces a bias against smaller giants.
A final unmodelled bias runs against RGs at the end of their life cycle.
Once the AGN stops launching jets for a prolonged period, it becomes hard to identify the host galaxy, which will no longer present as a bright, compact radio source.
This effect preferentially deselects larger giants, which are even more likely to approach the end of their life than smaller giants.
\subsection{Does the choice of prior matter?}
We have taken a conservative approach to constraining the posterior distribution through the prior: we have left tail index $\xi$ and angular length selection half-width $\frac{1}{2}(\phi_\mathrm{max} - \phi_\mathrm{min})$ fully unconstrained, and have adopted wide Gaussian priors for reference surface brightness parameters $b_{\nu,\mathrm{ref}}$ and $\sigma_\mathrm{ref}$, despite measuring them explicitly under assumptions.
We provide summary statistics of the posterior in Table~\ref{tab:posterior} and visualise its one- and two-parameter marginals in Fig.~\ref{fig:posterior}.
Does our choice of prior significantly affect the inferences?
To explore this question, we chose a different reasonable prior and compared results.
One such prior is the fully uniform prior, which equivalises the posterior and the likelihood function.
We provide analogous summary statistics of the likelihood function in Table~\ref{tab:likelihood} and visualise analogous marginals in Fig.~\ref{fig:likelihood}.
Reassuringly, no statistically significant parameter changes occur.
In particular, for $z_\mathrm{max} = 0.5$, $\xi = -3.5 \pm 0.5$ becomes $\xi = -3.4 \pm 0.5$ upon changing to the uniform prior; for $z_\mathrm{max} = 0.25$, $\xi = -3.5 \pm 0.4$ even remains the same.
However, given the strong likelihood degeneracy between $\xi$ and $b_{\nu,\mathrm{ref}}$ apparent in Fig.~\ref{fig:likelihood}, more stringent priors on $b_{\nu,\mathrm{ref}}$ \emph{are} able to meaningfully shift $\xi$'s posterior mean.
Such priors shall be appropriate only after studying the surface brightness properties of large radio galaxies --- and the associated selection effect --- in more detail.

\subsection{Cosmological evolution of the GRG length distribution}
\label{sec:XiEvolution}
In this work we have assumed that the parameter $\xi$, which fully characterises the intrinsic GRG length distribution under the ansatz of Paretianity, remains constant throughout the chosen redshift range $[0, z_\mathrm{max}]$.
To test this assumption, we have analysed our LoTSS DR2 GRG sample in Sect.~\ref{sec:results} up to both $z_\mathrm{max} = 0.5$ and $z_\mathrm{max} = 0.25$.
For $z_\mathrm{max} = 0.5$, we included giants that existed in the last 5 Gyr of the Universe's history, and found $\xi = -3.5 \pm 0.5$.
Meanwhile, for $z_\mathrm{max} = 0.25$, we included giants that existed in the last 3 Gyr of the Universe's history only, and found the very similar $\xi = -3.5 \pm 0.4$.
Thus, our analysis did not produce evidence that $\xi$ evolves over cosmic time.
However, given the large error bars, a modest time evolution cannot be excluded.
Furthermore, the data sets that underlie these inferences are not disjoint: the 811 giants that inform the lower-maximum-redshift analysis make up 55\% of the 1473 giants that inform the higher-maximum-redshift analysis.

Whether a time evolution of $\xi$ is expected is presumably tied to whether giant radio galaxy growth varies with environmental density at a given epoch, because the combined effects of the Universe's expansion and ongoing large-scale structure formation can similarly change environmental density.
Combining our LoTSS DR2 GRG sample with Cosmic Web reconstructions to explore giant growth as a function of environmental density is the topic of a forthcoming work \citep{Oei12022GiantsCosmicWeb}.
Interestingly, the recent exploration by \citet{Lan12021} that compared the environments of giants and non-giants did not find significant differences.

The most straightforward model extension is to again assume that RG lengths are Pareto distributed with tail index $\xi$, but now $\xi = \xi(z)$.
This function's first-degree Maclaurin polynomial, which provides the linearisation at the present day, is
\begin{align}
    \xi(z) \approx \xi(z = 0) + \frac{\mathrm{d}\xi}{\mathrm{d}z}(z = 0)\cdot z.
\end{align}
One would adopt $\xi(z = 0)$ and $\frac{\mathrm{d}\xi}{\mathrm{d}z}(z = 0)$ as the parameters of interest, replacing what used to be a constant $\xi$; the number of model parameters would thus increase by one.
However, an attempt to infer the cosmic evolution of $\xi$ appears promising only once the major selection effects are better constrained.

\section{Conclusions}
\label{sec:conclusions}
In this work, we have performed Bayesian inference on a LoTSS-derived sample of \numberOfGRGsOei giant radio galaxy projected proper lengths, using a one-parameter model that assumes a spatially homogeneous, non-evolving population of radio galaxies with stick-like geometry, Pareto-distributed lengths, and isotropic inclination angles.
Before fitting to data, we extended the forward model with two selection effects typical of contemporary manual GRG search campaigns.
The best-fit survival function tightly reproduces the empirical one, leaving permille-scale absolute residuals.
Having quantified the most important selection effects, we estimated the true comoving giant radio galaxy number density in the Local Universe.
\begin{enumerate}
    \item We developed an analytical model through which statistical questions about radio galaxy (RG) lengths can be rigorously answered.
In the current work, we applied this model to giant radio galaxies.
We adopted the ansatz that the RG intrinsic proper length $L$, as measured in three spatial dimensions, is a random variable (RV) with a Pareto Type I distribution (i.e. a simple power-law distribution) characterised by tail index $\xi$.
Next, by assuming that RGs have no preferential orientation with respect to the observer, we derived the distribution of the RG projected proper length $L_\mathrm{p}$.
By conditioning, one obtains the version relevant for giants, $L_\mathrm{p}\ \vert\ L_\mathrm{p} > l_\mathrm{p,GRG}$ (where we chose $l_\mathrm{p,GRG} \coloneqq 0.7\ \mathrm{Mpc}$).
This RV is again Paretian, with the same tail index $\xi$.
In summary, for giant radio galaxies, projection retains Paretianity.
Finally, observers face selection effects; we modelled the observed projected proper length $L_\mathrm{p,obs}$ by considering an angular length threshold selection effect and a surface brightness selection effect.
The angular length threshold selection effect assumes a linearly increasing angular-length-dependent probability of sample inclusion around a particular pre-defined threshold, meant to emulate manual visual searches that only target RGs of some angular length and above.
The surface brightness selection effect assumes that giants are self-similar, and that their lobes have lognormally distributed surface brightnesses which must be above-noise to secure sample inclusion.
The GRG observed projected proper length $L_\mathrm{p,obs}\ \vert\ L_\mathrm{p,obs} > l_\mathrm{p,GRG}$ again follows through conditioning.
We assumed our data to be realisations of this RV.
\item The model also yielded explicit expressions for the (posterior) distribution of $L\ \vert\ L_\mathrm{p} = l_\mathrm{p}$.
This allows one to deproject RGs in a statistical sense, providing the intrinsic proper length given the projected proper length in the limit of negligible selection effects.
We also present practical expressions for the mean and variance of $L\ \vert\ L_\mathrm{p} = l_\mathrm{p}$.
To unravel the driving factors that allow some RGs to become giants, most authors search for correlations between host or environmental physical parameters and the GRG \emph{projected} length.
However, if there is a causal chain that connects host or environmental parameters to GRG length, the connection will be to the \emph{intrinsic} length; the observer's vantage point does not play a role in the physics.
Therefore, the projection effect merely serves as a multiplicative noise source.
We suggest that future analyses should recognise the projection effect as such, and correlate host or environmental parameters with the intrinsic, rather than projected, proper length --- using statistical deprojection.
\item Through a manual visual search of the LoTSS DR2 pipeline products, the latest version of the LOFAR's Northern Sky survey at 144 MHz, we discovered a population of \numberOfGRGsOei previously unknown giants.
This is the largest single contribution to the literature yet, and increases the community-wide census by a factor 2.6.
We present 11 discoveries with $l_\mathrm{p} \geq 4\ \mathrm{Mpc}$, 53 with $3 \leq l_\mathrm{p} < 4\ \mathrm{Mpc}$, 291 with $2 \leq l_\mathrm{p} < 3\ \mathrm{Mpc}$, 1215 with $1 \leq l_\mathrm{p} < 2\ \mathrm{Mpc}$, and 490 with $0.7 \leq l_\mathrm{p} < 1\ \mathrm{Mpc}$.
Our study extends the known breadth of the giant radio galaxy phenomenon.
Among the findings are both the giant hosted by J081956.41+323537.6 and Alcyoneus \citep{Oei12022Alcyoneus}, at $l_\mathrm{p} = 5.1\ \mathrm{Mpc}$ and $l_\mathrm{p} = 5.0\ \mathrm{Mpc}$ the projectively largest giants ever found.
We discover that multi-Mpc radio galaxies can be generated before redshift 1, despite the Universe's mean density being an order of magnitude higher, and by spiral galaxies, whose stellar masses are typically an order of magnitude lower than those of ellipticals.
We discover giants whose hosts have a record-low stellar mass $M_\star = 4.8 \cdot 10^{10}\ M_\odot$.
We also discover giants whose hosts have a record-high supermassive black hole mass $M_\bullet \gtrsim 5 \cdot 10^{10}\ M_\odot$; interestingly, with $l_\mathrm{p} = 0.8\ \mathrm{Mpc}$, one of these giants is relatively small.
We more than double the number of known giants with angular lengths exceeding that of the Moon; one discovery, at $2.2\degree$, is the angularly largest radio galaxy in the Northern Sky and the angularly largest giant overall.
Excitingly, our LoTSS DR2 search has been far from exhaustive: many thousands of readily identifiable giants still await discovery in this public data set.
\item Using our LoTSS DR2 GRG sample up to $z_\mathrm{max} = 0.5$, we generated a posterior distribution over $\xi$ and three selection effect parameters.
Our model provides an excellent fit to the data, with absolute residuals being on average 2\textperthousand.
We inferred that the intrinsic proper length distribution of the largest radio galaxies resembles a Pareto distribution with tail index $\xi = -3.5 \pm 0.5$.
Our analysis did not yield evidence for an evolving $\xi$ in the last $5\ \mathrm{Gyr}$ of cosmic time.
\item The selection effect parameters estimated through the posterior are far from nuisance parameters, as they allowed us to statistically undo the selection effects imprinted on our LoTSS DR2 GRG data.
As a result, we could for the first time estimate the true comoving giant radio galaxy number density $n_\mathrm{GRG}$ in the Local Universe up to $z_\mathrm{max}$.
We relied on the crucial assumption that the surface brightness distribution of RGs with intrinsic length $l_\mathrm{ref} = 0.7\ \mathrm{Mpc}$ at $z = 0$ and frequency $\nu_\mathrm{obs} = 144\ \mathrm{MHz}$ is unimodal --- and lognormal in particular.
We furthermore assumed a lobe spectral index $\alpha = -1$ and self-similar growth.
We found $n_\mathrm{GRG}(l_\mathrm{p,GRG} = 0.7\ \mathrm{Mpc}, z_\mathrm{max} = 0.5) = 5 \pm 2\ (100\ \mathrm{Mpc})^{-3}$.
The implication is that giant radio galaxies are truly rare --- not only from a current observational perspective, but also from a cosmological one.
Current GRG lobes occupy just a few millionths of the IGM volume.
At any given moment in time, most clusters and filaments --- the building blocks of modern large-scale structure --- do not harbour giants.
\end{enumerate}
Giants embody the most extreme known mechanism by which galaxies can affect the Cosmic Web around them.
Whereas this work has explored the geometric properties of giants, a thorough exploration of their Cosmic Web energisation and magnetisation potential is a future frontier.
Excitingly, the interactions between giants and the ethereal intergalactic medium may also allow for new constraints on the thermodynamics in filaments.

\begin{acknowledgements}
M.S.S.L. Oei warmly thanks Frits Sweijen for coding the very useful \url{https://github.com/tikk3r/legacystamps} and for tireless ICT advice, and Aleksandar Shulevski for comments that have improved the manuscript.
In dear memory of Nicoline. By staying strong despite life's challenges, you showed what it means to be a giant.\\
M.S.S.L. Oei, R.J. van Weeren, and A. Botteon acknowledge support from the VIDI research programme with project number 639.042.729, which is financed by The Netherlands Organisation for Scientific Research (NWO).
A. Drabent acknowledges support by the BMBF Verbundforschung under the grant 05A20STA.\\
The LOFAR is the Low-Frequency Array designed and constructed by ASTRON.
It has observing, data processing, and data storage facilities in several countries, which are owned by various parties (each with their own funding sources), and which are collectively operated by the ILT Foundation under a joint scientific policy.
The ILT resources have benefited from the following recent major funding sources: CNRS--INSU, Observatoire de Paris and Université d'Orléans, France; BMBF, MIWF--NRW, MPG, Germany; Science Foundation Ireland (SFI), Department of Business, Enterprise and Innovation (DBEI), Ireland; NWO, The Netherlands; the Science and Technology Facilities Council, UK; Ministry of Science and Higher Education, Poland; the Istituto Nazionale di Astrofisica (INAF), Italy.\\
This research made use of the Dutch national e-infrastructure with support of the SURF Cooperative (e-infra 180169) and the LOFAR e-infra group.
The J\"ulich LOFAR Long Term Archive and the German LOFAR network are both coordinated and operated by the J\"ulich Supercomputing Centre (JSC), and computing resources on the supercomputer JUWELS at JSC were provided by the Gauss Centre for Supercomputing e.V. (grant CHTB00) through the John von Neumann Institute for Computing (NIC).\\
This research made use of the University of Hertfordshire high-performance computing facility and the LOFAR-UK computing facility located at the University of Hertfordshire and supported by STFC [ST/P000096/1], and of the Italian LOFAR IT computing infrastructure supported and operated by INAF, and by the Physics Department of the University of Turin (under an agreement with Consorzio Interuniversitario per la Fisica Spaziale) at the C3S Supercomputing Centre, Italy.\\
Funding for SDSS-III has been provided by the Alfred P. Sloan Foundation, the Participating Institutions, the National Science Foundation, and the U.S. Department of Energy Office of Science. The SDSS-III web site is \url{http://www.sdss3.org/}.
SDSS-III is managed by the Astrophysical Research Consortium for the Participating Institutions of the SDSS-III Collaboration including the University of Arizona, the Brazilian Participation Group, Brookhaven National Laboratory, Carnegie Mellon University, University of Florida, the French Participation Group, the German Participation Group, Harvard University, the Instituto de Astrof\'isica de Canarias, the Michigan State/Notre Dame/JINA Participation Group, Johns Hopkins University, Lawrence Berkeley National Laboratory, Max Planck Institute for Astrophysics, Max Planck Institute for Extraterrestrial Physics, New Mexico State University, New York University, Ohio State University, Pennsylvania State University, University of Portsmouth, Princeton University, the Spanish Participation Group, University of Tokyo, University of Utah, Vanderbilt University, University of Virginia, University of Washington, and Yale University.\\
The Pan-STARRS1 Surveys (PS1) and the PS1 public science archive have been made possible through contributions by the Institute for Astronomy, the University of Hawaii, the Pan-STARRS Project Office, the Max-Planck Society and its participating institutes, the Max Planck Institute for Astronomy, Heidelberg and the Max Planck Institute for Extraterrestrial Physics, Garching, The Johns Hopkins University, Durham University, the University of Edinburgh, the Queen's University Belfast, the Harvard--Smithsonian Center for Astrophysics, the Las Cumbres Observatory Global Telescope Network Incorporated, the National Central University of Taiwan, the Space Telescope Science Institute, the National Aeronautics and Space Administration under Grant No. NNX08AR22G issued through the Planetary Science Division of the NASA Science Mission Directorate, the National Science Foundation Grant No. AST-1238877, the University of Maryland, E\"otv\"os Lor\'and University (ELTE), the Los Alamos National Laboratory, and the Gordon and Betty Moore Foundation.\\
This work has made use of data from the European Space Agency (ESA) mission
{\it Gaia} (\url{https://www.cosmos.esa.int/gaia}), processed by the {\it Gaia}
Data Processing and Analysis Consortium (DPAC,
\url{https://www.cosmos.esa.int/web/gaia/dpac/consortium}). Funding for the DPAC
has been provided by national institutions, in particular the institutions
participating in the {\it Gaia} Multilateral Agreement.\\
The Legacy Surveys consist of three individual and complementary projects: the Dark Energy Camera Legacy Survey (DECaLS; Proposal ID \#2014B-0404; PIs: David Schlegel and Arjun Dey), the Beijing--Arizona Sky Survey (BASS; NOAO Prop. ID \#2015A-0801; PIs: Zhou Xu and Xiaohui Fan), and the Mayall z-band Legacy Survey (MzLS; Prop. ID \#2016A-0453; PI: Arjun Dey). DECaLS, BASS and MzLS together include data obtained, respectively, at the Blanco telescope, Cerro Tololo Inter-American Observatory, NSF's NOIRLab; the Bok telescope, Steward Observatory, University of Arizona; and the Mayall telescope, Kitt Peak National Observatory, NOIRLab. The Legacy Surveys project is honored to be permitted to conduct astronomical research on Iolkam Du'ag (Kitt Peak), a mountain with particular significance to the Tohono O'odham Nation.
NOIRLab is operated by the Association of Universities for Research in Astronomy (AURA) under a cooperative agreement with the National Science Foundation.
This project used data obtained with the Dark Energy Camera (DECam), which was constructed by the Dark Energy Survey (DES) collaboration. Funding for the DES Projects has been provided by the U.S. Department of Energy, the U.S. National Science Foundation, the Ministry of Science and Education of Spain, the Science and Technology Facilities Council of the United Kingdom, the Higher Education Funding Council for England, the National Center for Supercomputing Applications at the University of Illinois at Urbana-Champaign, the Kavli Institute of Cosmological Physics at the University of Chicago, Center for Cosmology and Astro-Particle Physics at the Ohio State University, the Mitchell Institute for Fundamental Physics and Astronomy at Texas A\&M University, Financiadora de Estudos e Projetos, Fundacao Carlos Chagas Filho de Amparo, Financiadora de Estudos e Projetos, Fundacao Carlos Chagas Filho de Amparo a Pesquisa do Estado do Rio de Janeiro, Conselho Nacional de Desenvolvimento Cientifico e Tecnologico and the Ministerio da Ciencia, Tecnologia e Inovacao, the Deutsche Forschungsgemeinschaft and the Collaborating Institutions in the Dark Energy Survey. The Collaborating Institutions are Argonne National Laboratory, the University of California at Santa Cruz, the University of Cambridge, Centro de Investigaciones Energeticas, Medioambientales y Tecnologicas-Madrid, the University of Chicago, University College London, the DES-Brazil Consortium, the University of Edinburgh, the Eidgen\"ossische Technische Hochschule (ETH) Z\"urich, Fermi National Accelerator Laboratory, the University of Illinois at Urbana-Champaign, the Institut de Ciencies de l'Espai (IEEC/CSIC), the Institut de Fisica d'Altes Energies, Lawrence Berkeley National Laboratory, the Ludwig Maximilians Universit\"at M\"unchen and the associated Excellence Cluster Universe, the University of Michigan, NSF's NOIRLab, the University of Nottingham, the Ohio State University, the University of Pennsylvania, the University of Portsmouth, SLAC National Accelerator Laboratory, Stanford University, the University of Sussex, and Texas A\&M University.
BASS is a key project of the Telescope Access Program (TAP), which has been funded by the National Astronomical Observatories of China, the Chinese Academy of Sciences (the Strategic Priority Research Program “The Emergence of Cosmological Structures” Grant \# XDB09000000), and the Special Fund for Astronomy from the Ministry of Finance. The BASS is also supported by the External Cooperation Program of Chinese Academy of Sciences (Grant \# 114A11KYSB20160057), and Chinese National Natural Science Foundation (Grant \# 11433005).
The Legacy Survey team makes use of data products from the Near-Earth Object Wide-field Infrared Survey Explorer (NEOWISE), which is a project of the Jet Propulsion Laboratory/California Institute of Technology. NEOWISE is funded by the National Aeronautics and Space Administration.
The Legacy Surveys imaging of the DESI footprint is supported by the Director, Office of Science, Office of High Energy Physics of the U.S. Department of Energy under Contract No. DE-AC02-05CH1123, by the National Energy Research Scientific Computing Center, a DOE Office of Science User Facility under the same contract; and by the U.S. National Science Foundation, Division of Astronomical Sciences under Contract No. AST-0950945 to NOAO.
\end{acknowledgements}

{\footnotesize\bibliography{cite}}

\begin{appendix}
\section{Framework derivations and details}
\label{sec:supplementaryMaterialPowerLawModel}
\subsection{Intrinsic proper length}
\label{ap:intrinsicLengthDistribution}
Let $f_L: \mathbb{R} \to \mathbb{R}_{\geq 0}$ be the PDF of the distribution of intrinsic (i.e. 3D) proper (i.e. not comoving) radio galaxy lengths $L$.
We assume that $f_L$ follows a power law between $l_\mathrm{min}$ and $l_\mathrm{max}$:
\begin{align}
f_L\left(l\right) = \begin{cases}
f_L\left(l_\mathrm{ref}\right)\left(\frac{l}{l_\mathrm{ref}}\right)^\xi & l \in [l_\mathrm{min}, l_\mathrm{max}],\\
0 & \text{otherwise}.
\end{cases}
\end{align}
Then, when $\xi \neq -1$,
\begin{align}
1 &= \int_\mathbb{R} f_L\left(l\right)\ \mathrm{d}l
    = \int_{l_\mathrm{min}}^{l_\mathrm{max}} f_L\left(l_\mathrm{ref}\right)\left(\frac{l}{l_\mathrm{ref}}\right)^\xi\ \mathrm{d}l\nonumber\\
    &= \frac{f_L\left(l_\mathrm{ref}\right) l_\mathrm{ref}}{\xi + 1} \left(\left(\frac{l_\mathrm{max}}{l_\mathrm{ref}}\right)^{\xi + 1} - \left(\frac{l_\mathrm{min}}{l_\mathrm{ref}}\right)^{\xi + 1}\right).
\label{eq:normalisation}
\end{align}
This equation provides the normalisation factor $f_L\left(l_\mathrm{ref}\right)$ given $l_\mathrm{min}$, $l_\mathrm{max}$, $\xi$ and an arbitrary choice for $l_\mathrm{ref} \neq 0$.

When $l_\mathrm{max} = \infty$, $L$ has a Pareto Type I distribution and we must have $\xi < -1$ for the integrals of Eq.~\ref{eq:normalisation} to converge.
Throughout the remaining analysis, for the sake of simplicity, we assume $l_\mathrm{max} = \infty$ and $\xi < -1$ and choose $l_\mathrm{ref} = l_\mathrm{min}$.
Then $f_L\left(l_\mathrm{ref}\right) = f_L\left(l_\mathrm{min}\right) = -\frac{\xi+1}{l_\mathrm{min}}$.
The corresponding CDF is
\begin{align}
    F_L(l) = \begin{cases}
    0 & \text{if } l \leq l_\mathrm{min}\\
    1 - \left(\frac{l}{l_\mathrm{min}}\right)^{\xi + 1} & \text{if } l > l_\mathrm{min}.
    \end{cases}
    \label{eq:CDFLengthProperIntrinsicRG}
\end{align}
In reality, radio galaxies cannot become arbitrarily long \citep{Hardcastle12018}: at some distance from the host, all energy carried by the jets will have been radiated away, used to perform work on the IGM, transferred to CMB photons through inverse Compton scattering, or converted into heat.
However, at the moment of writing, the implied maximum length $l_\mathrm{max}$ remains ill-constrained.
As we see throughout Appendix~\ref{sec:supplementaryMaterialPowerLawModel}, a major advantage of simple model assumptions is that explicit and thus insightful analytic expressions can be derived.
Such easy-to-evaluate expressions complement the results of expensive numerical simulations, which aim to maximise realism; our aim to maximise insight is best served by setting $l_\mathrm{max} = \infty$.
From Eq.~\ref{eq:CDFLengthProperIntrinsicRG}, we see that for $l_\mathrm{min} = 0.7\ \mathrm{Mpc}$ and $\xi = -3.5$, the tail of the intrinsic proper length distribution ($l > 5\ \mathrm{Mpc}$) contains less than $1\%$ of probability.
Thus, for most framework applications, our choice $l_\mathrm{max} = \infty$ is unproblematic --- except when higher powers of $L$ are involved.
For example, the volume-filling fraction calculations of Appendix~\ref{ap:VFF} necessitate considering $L^3$; for realistic $\xi$, results exist only for finite $l_\mathrm{max}$.

Upon relabelling $\xi \rightarrow -\alpha - 1$, $l_\mathrm{min} \rightarrow k$, $L \rightarrow X$, and $l \rightarrow x$, one obtains the literature's most common form of the Pareto Type I PDF:
\begin{align}
    f_X\left(x\right) = \begin{cases}
    \frac{\alpha k^\alpha}{x^{\alpha + 1}} & x \geq k,\\
    0 & x < k.
    \end{cases}
\end{align}

\subsection{Projected proper length}
\subsubsection{Distribution for RGs}
\label{ap:projectedLengthDistribution}
Let $f_{L_\mathrm{p}}: \mathbb{R} \to \mathbb{R}_{\geq 0}$ be the PDF of the distribution of projected proper radio galaxy lengths.
The PDF $f_{L_\mathrm{p}}$ follows from the associated CDF $F_{L_\mathrm{p}}: \mathbb{R} \to [0, 1]$ through differentiation; $F_{L_\mathrm{p}}$ and $f_L$ relate through
\begin{align}
    F_{L_\mathrm{p}}\left(l_\mathrm{p}\right) &\coloneqq \mathbb{P}\left(L_\mathrm{p} \leq l_\mathrm{p}\right) = \int_\mathbb{R} \mathbb{P}\left(L\sin{\Theta} \leq l_\mathrm{p}\ \vert\ L = l\right) f_L\left(l\right)\ \mathrm{d}l\nonumber\\
    &= \int_\mathbb{R} \mathbb{P}\left(\sin{\Theta} \leq \frac{l_\mathrm{p}}{l}\right) f_L\left(l\right)\ \mathrm{d}l.
\end{align}
We note that $F_{L_\mathrm{p}}\left(l_\mathrm{p}\right)$ vanishes for $l_\mathrm{p} \leq 0$: $f_L\left(l\right)$ vanishes when $l$ is negative, whilst $\mathbb{P}\left(\sin{\Theta} \leq \frac{l_\mathrm{p}}{l}\right)$ vanishes when $l$ is positive.
Clearly, the interesting case is $l_\mathrm{p} > 0$.
We can differentiate between two cases: $l_\mathrm{p} \leq l_\mathrm{min}$, and $l_\mathrm{p} > l_\mathrm{min}$.
In the first case, considering that $f_L$ has support from $l_\mathrm{min}$ onwards only, we have $\frac{l_\mathrm{p}}{l} \leq 1$, and
\begin{align}
    F_{L_\mathrm{p}}\left(l_\mathrm{p}\right) &= \int_{l_\mathrm{min}}^\infty \left(\mathbb{P}\left(\Theta \leq \arcsin{\frac{l_\mathrm{p}}{l}}\right)\right.\nonumber\\
    &+ \left.\mathbb{P}\left(\Theta \geq \pi - \arcsin{\frac{l_\mathrm{p}}{l}}\right)\right) \frac{-\left(\xi+1\right)}{l_\mathrm{min}}\left(\frac{l}{l_\mathrm{min}}\right)^\xi\ \mathrm{d}l\nonumber\\
    &= -\frac{\xi + 1}{l_\mathrm{min}}\int_{l_\mathrm{min}}^\infty 2 \mathbb{P}\left(\Theta \leq \arcsin{\frac{l_\mathrm{p}}{l}}\right) \left(\frac{l}{l_\mathrm{min}}\right)^\xi\ \mathrm{d}l\nonumber\\
    &= -\frac{\xi + 1}{l_\mathrm{min}}\int_{l_\mathrm{min}}^\infty \left(1 - \sqrt{1 - \left(\frac{l_\mathrm{p}}{l}\right)^2}\right)\left(\frac{l}{l_\mathrm{min}}\right)^\xi\ \mathrm{d}l\nonumber\\
    &= 1 + \left(\xi+1\right)\int_{l_\mathrm{min}}^\infty \sqrt{\left(\frac{l}{l_\mathrm{min}}\right)^2 - \left(\frac{l_\mathrm{p}}{l_\mathrm{min}}\right)^2}\left(\frac{l}{l_\mathrm{min}}\right)^{\xi - 1}\ \frac{\mathrm{d}l}{l_\mathrm{min}}\nonumber\\
    &= 1 + \left(\xi+1\right)\int_1^\infty \sqrt{\eta^2 - \left(\frac{l_\mathrm{p}}{l_\mathrm{min}}\right)^2}\eta^{\xi - 1}\ \mathrm{d}\eta.
\end{align}
In the second case, we split up the integral in two:
\begin{align}
    F_{L_\mathrm{p}}\left(l_\mathrm{p}\right) &= \int_{l_\mathrm{min}}^{l_\mathrm{p}} \mathbb{P}\left(\sin{\Theta} \leq \frac{l_\mathrm{p}}{l}\right) f_L\left(l\right)\ \mathrm{d}l\nonumber\\
    &+ \int_{l_\mathrm{p}}^\infty \mathbb{P}\left(\sin{\Theta} \leq \frac{l_\mathrm{p}}{l}\right) f_L\left(l\right)\ \mathrm{d}l\nonumber\\
    &= \int_{l_\mathrm{min}}^{l_\mathrm{p}} f_L\left(l\right)\ \mathrm{d}l + \int_{l_\mathrm{p}}^\infty \left(1-\sqrt{1-\left(\frac{l_\mathrm{p}}{l}\right)^2}\right) f_L\left(l\right)\ \mathrm{d}l\nonumber\\
    &= 1 - \left(\frac{l_\mathrm{p}}{l_\mathrm{min}}\right)^{\xi + 1} + \left(\frac{l_\mathrm{p}}{l_\mathrm{min}}\right)^{\xi + 1}\nonumber\\
    &+ \left(\xi+1\right)\int_{l_\mathrm{p}}^\infty \sqrt{\left(\frac{l}{l_\mathrm{min}}\right)^2 - \left(\frac{l_\mathrm{p}}{l_\mathrm{min}}\right)^2} \left(\frac{l}{l_\mathrm{min}}\right)^{\xi-1}\ \frac{\mathrm{d}l}{l_\mathrm{min}}\nonumber\\
    &= 1 + \left(\xi+1\right)\int_\frac{l_\mathrm{p}}{l}^\infty \sqrt{\eta^2 - \left(\frac{l_\mathrm{p}}{l}\right)^2} \eta^{\xi-1}\ \mathrm{d}\eta\nonumber\\
    &= 1 + \left(\xi+1\right)\frac{\sqrt{\pi}}{4}\left(\frac{l_\mathrm{p}}{l_\mathrm{min}}\right)^{\xi+1}\frac{\Gamma\left(-\frac{\xi}{2}-\frac{1}{2}\right)}{\Gamma\left(-\frac{\xi}{2}+1\right)}.
\end{align}
In summary,
\begin{align}
    &F_{L_\mathrm{p}}\left(l_\mathrm{p}\right) = \begin{cases}
    0 & \text{if } l_\mathrm{p} \leq 0\\
    1 + \left(\xi+1\right)\int_1^\infty \sqrt{\eta^2 - \left(\frac{l_\mathrm{p}}{l_\mathrm{min}}\right)^2}\eta^{\xi-1}\ \mathrm{d}\eta & \text{if } 0 < l_\mathrm{p} \leq l_\mathrm{min}\\
    1 + \left(\xi+1\right)\frac{\sqrt{\pi}}{4}\left(\frac{l_\mathrm{p}}{l_\mathrm{min}}\right)^{\xi+1}\frac{\Gamma\left(-\frac{\xi}{2}-\frac{1}{2}\right)}{\Gamma\left(-\frac{\xi}{2}+1\right)} & \text{if } l_\mathrm{p} > l_\mathrm{min}.
    \end{cases}
\label{eq:CDFLengthProperProjectedRG}
\end{align}
Through differentiation,
\begin{align}
    f_{L_\mathrm{p}}\left(l_\mathrm{p}\right) = \begin{cases}
    0 & \text{if } l_\mathrm{p} \leq 0\\
    -\frac{\xi+1}{l_\mathrm{min}}\frac{l_\mathrm{p}}{l_\mathrm{min}}I\left(\xi-1,\frac{l_\mathrm{p}}{l_\mathrm{min}}\right) & \text{if } 0 < l_\mathrm{p} \leq l_\mathrm{min}\\
    \frac{\left(\xi+1\right)^2}{l_\mathrm{min}}\frac{\sqrt{\pi}}{4}\left(\frac{l_\mathrm{p}}{l_\mathrm{min}}\right)^\xi \frac{\Gamma\left(-\frac{\xi}{2}-\frac{1}{2}\right)}{\Gamma\left(-\frac{\xi}{2}+1\right)} & \text{if } l_\mathrm{p} > l_\mathrm{min}.
    \end{cases}
\label{eq:PDFLengthProperProjectedRG}
\end{align}
where
\begin{align}
    I\left(a,b\right)\coloneqq \int_1^\infty \frac{\eta^a\ \mathrm{d}\eta}{\sqrt{\eta^2-b^2}} \text{ for } a < 0, \vert b \vert < 1.
\end{align}
Significantly, for $l_\mathrm{p} > l_\mathrm{min}$, the projected length distribution follows a power law in $l_\mathrm{p}$ with the same exponent as the power law for the intrinsic length distribution: $f_{L_\mathrm{p}} \propto l_\mathrm{p}^\xi$ (just as $f_L \propto l^\xi$).\footnote{For $l_\mathrm{max} < \infty$, this statement does not hold exactly.}
We compare $f_L$ and $f_{L_\mathrm{p}}$ in Fig.~\ref{fig:marginals}.

\subsubsection{Distribution for giants}
\label{ap:projectedLengthDistributionGRGs}
To derive the projected length distribution for giants, we consider the distribution of the conditioned RV $L_\mathrm{p}\ \vert\ L_\mathrm{p} > l_\mathrm{p, GRG}$:
\begin{align}
    F_{L_\mathrm{p}\ \vert\ L_\mathrm{p} > l_\mathrm{p, GRG}}\left(l_\mathrm{p}\right) \coloneqq& \mathbb{P}\left(L_\mathrm{p} \leq l_\mathrm{p}\ \vert\ L_\mathrm{p} > l_\mathrm{p, GRG}\right)\nonumber\\
    =& 1 - \mathbb{P}\left(L_\mathrm{p} > l_\mathrm{p}\ \vert\ L_\mathrm{p} > l_\mathrm{p, GRG}\right)\nonumber\\
    =& 1 - \frac{\mathbb{P}\left(L_\mathrm{p} > l_\mathrm{p}, L_\mathrm{p} > l_\mathrm{p, GRG}\right)}{\mathbb{P}\left(L_\mathrm{p} > l_\mathrm{p, GRG}\right)}.
\end{align}
For $l_\mathrm{p} > l_\mathrm{p,GRG}$, this reduces to
\begin{align}
    F_{L_\mathrm{p}\ \vert\ L_\mathrm{p} > l_\mathrm{p, GRG}}\left(l_\mathrm{p}\right) =& 1 - \frac{1 - F_{L_\mathrm{p}}\left(l_\mathrm{p}\right)}{1 - F_{L_\mathrm{p}}\left(l_\mathrm{p, GRG}\right)}.
    \label{eq:CDFLengthProperProjectedGRG}
\end{align}
Furthermore assuming $l_\mathrm{p, GRG} > l_\mathrm{min}$, we twice use the bottom expression of Eq.~\ref{eq:CDFLengthProperProjectedRG} to obtain the final CDF expression.
As before, the corresponding PDF follows through differentation.
We find
\begin{align}
    F_{L_\mathrm{p}\ \vert\ L_\mathrm{p} > l_\mathrm{p, GRG}}\left(l_\mathrm{p}\right) =& \begin{cases}
    0 & \text{if } l_\mathrm{p} \leq l_\mathrm{p,GRG}\\
    1 - \left(\frac{l_\mathrm{p}}{l_\mathrm{p, GRG}}\right)^{\xi+1} & \text{if } l_\mathrm{p} > l_\mathrm{p, GRG}.
    \end{cases}\\
    f_{L_\mathrm{p}\ \vert\ L_\mathrm{p} > l_\mathrm{p, GRG}}\left(l_\mathrm{p}\right) =& \begin{cases}
    0 & \text{if } l_\mathrm{p} \leq l_\mathrm{p, GRG}\\
    -\frac{\xi+1}{l_\mathrm{p, GRG}}\left(\frac{l_\mathrm{p}}{l_\mathrm{p, GRG}}\right)^\xi & \text{if } l_\mathrm{p} > l_\mathrm{p, GRG}.
    \end{cases}
\end{align}
Thus, the associated survival function is
\begin{align}
    \mathbb{P}\left(L_\mathrm{p} > l_\mathrm{p}\ \vert\ L_\mathrm{p} > l_\mathrm{p,GRG}\right) = \left(\frac{l_\mathrm{p}}{l_\mathrm{p,GRG}}\right)^{\xi+1}.
    \label{eq:SFLengthProperProjectedGRG}
\end{align}
The mean projected proper length of giants follows from the PDF by direct computation:
\begin{align}
    \mathbb{E}[L_\mathrm{p}\ \vert\ L_\mathrm{p} > l_\mathrm{p,GRG}] \coloneqq& \int_{-\infty}^\infty f_{L_\mathrm{p}\ \vert\ L_\mathrm{p}>l_\mathrm{p,GRG}}\left(l_\mathrm{p}\right) \cdot l_\mathrm{p}\ \mathrm{d}l_\mathrm{p}\nonumber\\
    =& l_\mathrm{p,GRG}\frac{\xi+1}{\xi+2}.
\end{align}
\subsection{Deprojection factor}
\label{ap:deprojectionFactor}
Consider a radio galaxy (RG) with a projected proper length $l_\mathrm{p}$.
Let the inclination angle $\theta$ denote the angle between the RG's central axis and the line of sight.
The RG's inclination angle, projected proper length, and intrinsic proper length $l$ relate via $l_\mathrm{p} = l \sin{\theta}$.
Switching to random variable (RV) notation by using capital letters, the intrinsic proper length (which is the most physically relevant quantity) follows from the projected proper length (which can be measured) and the inclination angle (which is typically unknown), according to
\begin{align}
    L = \frac{1}{\sin{\Theta}} L_\mathrm{p}.
\label{eq:projectionRVs}
\end{align}

\subsubsection{Without lobes}
Calling the deprojection factor $D \coloneqq \left(\sin{\Theta}\right)^{-1}$, we now calculate the distribution of $D$ for $f_\Theta\left(\theta\right) = \frac{1}{2}\sin{\theta}; \theta \in [0, \pi]$.
The result is a continuous univariate distribution without parameters supported on the semi-infinite interval $\left(1, \infty\right)$.
Let $F_D: \mathbb{R} \to [0, 1]$ be the cumulative density function (CDF) of $D$.
Then, for $d > 1$,
\begin{align}
    F_D\left(d\right) &\coloneqq \mathbb{P}\left(D \leq d\right)
    = \mathbb{P}\left(\sin{\Theta} \geq \frac{1}{d}\right)\nonumber\\
    &= \mathbb{P}\left(\arcsin{\frac{1}{d}} \leq \Theta \leq \pi - \arcsin{\frac{1}{d}}\right)\nonumber\\
    &= F_\Theta\left(\pi - \arcsin{\frac{1}{d}}\right) - F_\Theta\left(\arcsin{\frac{1}{d}}\right)\nonumber\\
    &= \cos{\arcsin{\frac{1}{d}}} = \sqrt{1 - \frac{1}{d^2}}.
\end{align}
Meanwhile, $F_D\left(d\right) = 0$ for $d \leq 1$.
The quantile function $F_D^{-1}: [0, 1] \to [1, \infty)$ follows from solving $F_D\left(d\right) = p$ for $d$:
\begin{align}
    F_D^{-1}\left(p\right) = \frac{1}{\sqrt{1-p^2}}.
\end{align}
Thus, the minimum factor is $F_D^{-1}\left(p=0\right) = 1$, the median $F_D^{-1}\left(p=\frac{1}{2}\right) = \frac{2}{\sqrt{3}}$, $F_D^{-1}\left(p=\frac{1}{2}\sqrt{2}\right) = \sqrt{2}$, $F_D^{-1}\left(p=\frac{1}{2}\sqrt{3}\right) = 2$, and factors can grow arbitrarily large: $F_D^{-1}\left(p\right) \to \infty$ as $p \to 1$.
We conclude that half of all RGs have an intrinsic proper length more than $\frac{2}{\sqrt{3}}$ their projected proper length.

\begin{figure*}
\centering
\begin{subfigure}{.49\textwidth}
    \centering
    \includegraphics[width=\textwidth]{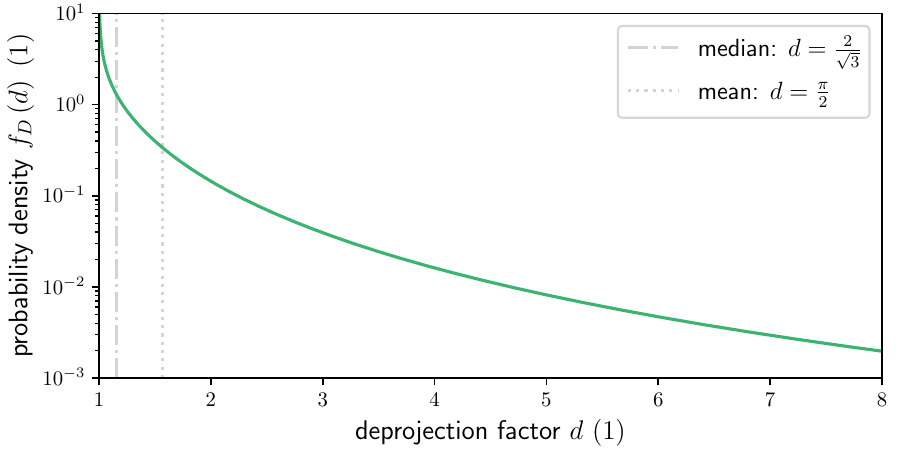}
\end{subfigure}
\begin{subfigure}{.49\textwidth}
    \centering
    \includegraphics[width=\textwidth]{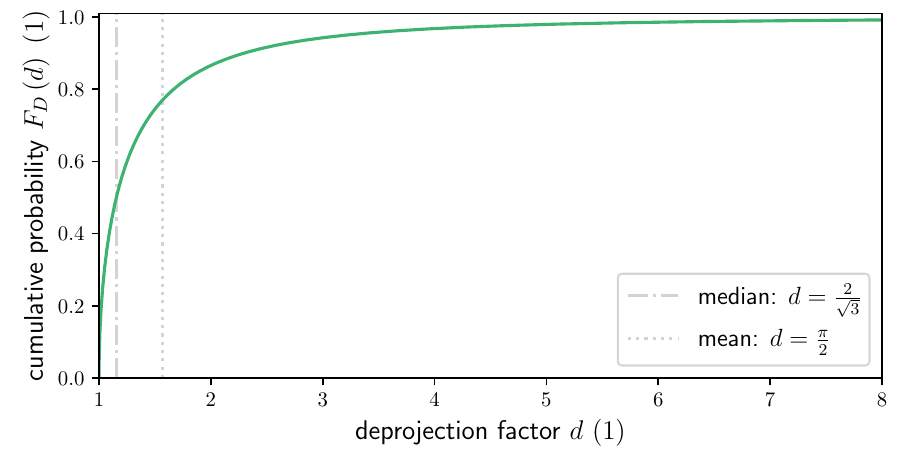}
\end{subfigure}
\begin{subfigure}{.49\textwidth}
    \centering
    \includegraphics[width=\textwidth]{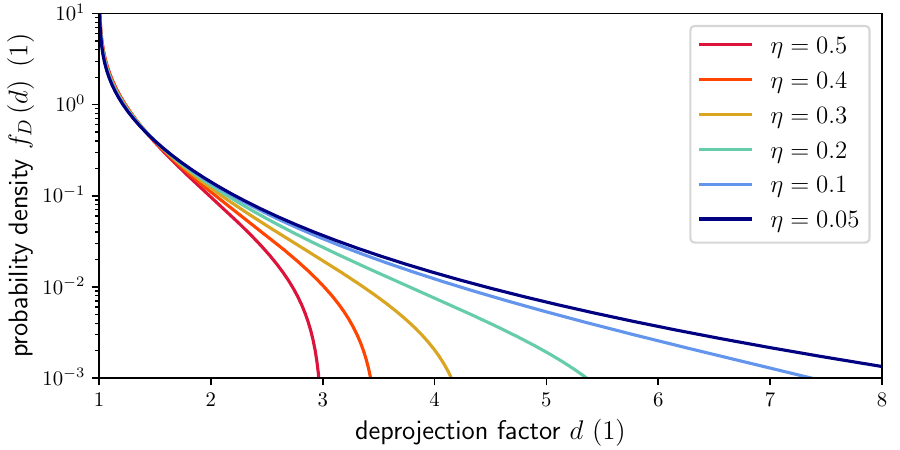}
\end{subfigure}
\begin{subfigure}{.49\textwidth}
    \centering
    \includegraphics[width=\textwidth]{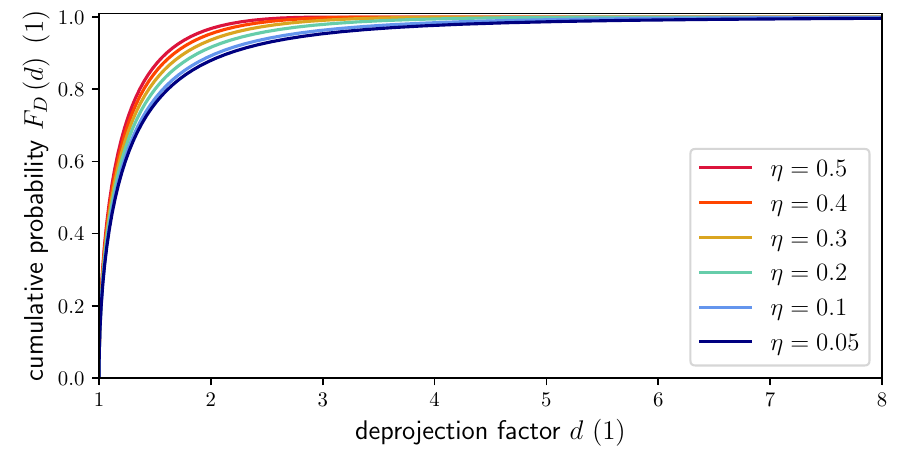}
\end{subfigure}
    \caption{
    PDFs \textit{(left column)} and CDFs \textit{(right column)} of the deprojection factor RV $D$.
    These functions quantify how much longer RGs are than their projected lengths suggest.
    \textit{Top row:} model without lobes.
    Of the three canonical measures of central tendency, only the median and mean exist (and equal $\frac{2}{\sqrt{3}}$ and $\frac{\pi}{2}$, respectively).
    \textit{Bottom row:} model with spherical lobes.
    The distribution of $D$ now depends on a parameter: the ratio $\eta$ between the lobe diameter $2R$ and the distance between the lobe centres $L$.
    The model without lobes is recovered in the limit $\ratio \to 0$.}
    \label{fig:deprojectionFactor}
\end{figure*}

By differentiating $F_D$ to $d$ we obtain $f_D: \mathbb{R} \to \mathbb{R}_{\geq 0}$, the probability density function (PDF) of $D$:
\begin{align}
f_D\left(d\right) = \begin{cases}
\frac{1}{d^2\sqrt{d^2-1}} &\text{if $d > 1$;}\\
0 &\text{if $d \leq 1$.}
\end{cases}
\end{align}
The mean of $D$ is $\mathbb{E}\left(D\right) = \frac{\pi}{2}$; the variance of $D$ is undefined, as the corresponding integral diverges.
Because $f_D\left(d\right)$ has no maximum, the mode is undefined too.
The PDF and CDF of $D$ are shown in the upper left and right panels of Fig.~\ref{fig:deprojectionFactor}, respectively.

\subsubsection{With lobes}
\begin{figure}
\begin{subfigure}{\columnwidth}
    \centering
    \includegraphics[width=\columnwidth]{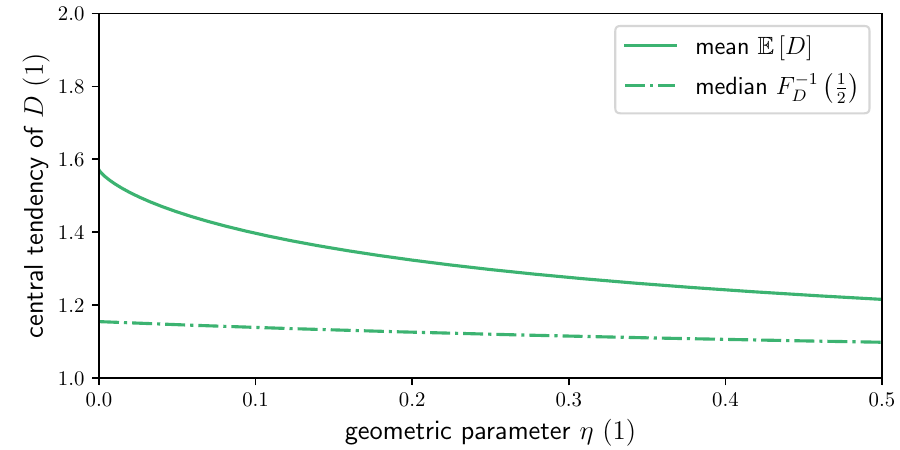}
\end{subfigure}
\begin{subfigure}{\columnwidth}
    \centering
    \includegraphics[width=\columnwidth]{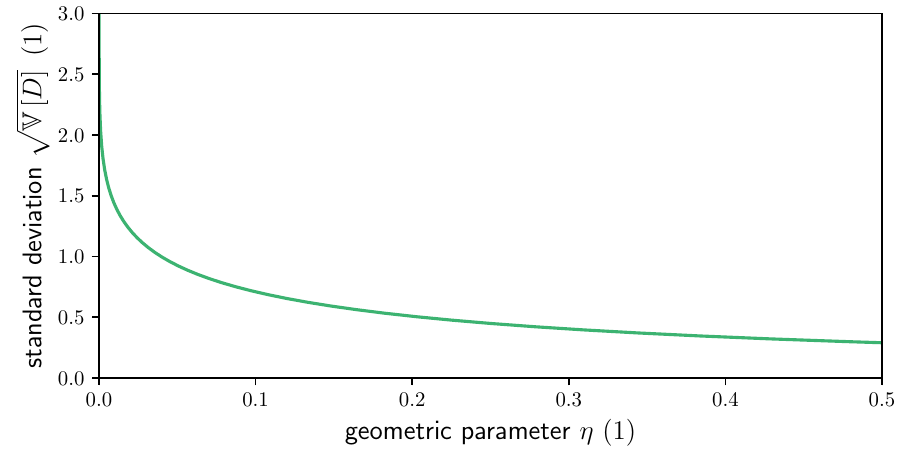}
\end{subfigure}
    \caption{
    Summary statistics for the deprojection factor RV $D$ under the spherical lobe model.
    \textit{Top:} the $\ratio$-dependency of two measures of central tendency of $D$.
    For $\ratio = 0$, the mean $\mathbb{E}\left[D\right] = \frac{\pi}{2}$ and the median $F_D^{-1}\left(\frac{1}{2}\right) = \frac{2}{\sqrt{3}}$.
    \textit{Bottom:} the $\ratio$-dependency of the standard deviation of $D$.
    As $\ratio \to 0+$, $\sqrt{\mathbb{V}\left[D\right]} \to \infty$.}
    \label{fig:deprojectionFactorMoments}
\end{figure}

The simplest model that includes a pair of lobes approximates them as spheres of radius $R$, whose centres are connected by a line segment of length $L$.
Regardless of the viewing angle, the spheres retain their size, in contrast to the line segment connecting them.
Calling $\ratio \coloneqq \frac{2R}{L} \in \mathbb{R}_{\geq 0}$, we have
\begin{align}
    D \coloneqq \frac{L + 2R}{L\sin{\Theta} + 2R} = \frac{1 + \ratio}{\sin{\Theta} + \ratio}.
\end{align}
Again assuming $f_\Theta\left(\theta\right) = \frac{1}{2}\sin{\theta}; \theta \in [0, \pi]$, we repeat the derivation and find
\begin{align}
F_D\left(d\right) &= \begin{cases}
0 & \text{if $d \leq 1$;}\\
\sqrt{1-\left(\frac{1 + \ratio}{d} - \ratio\right)^2} & \text{if $1 < d < \frac{1}{\ratio} + 1$;}\\
1 & \text{if $d \geq \frac{1}{\ratio} + 1$.}
\end{cases}\\
    f_D\left(d\right) &= \begin{cases}
    0 & \text{if $d \leq 1$;}\\
\frac{\left(1 + \ratio\right)\left(1+\ratio-\ratio d\right)}{d^2\sqrt{d^2-\left(1 + \ratio - \ratio d\right)^2}} &\text{if $1 < d < \frac{1}{\ratio} + 1$;}\\
0 &\text{if $d \geq \frac{1}{\ratio} + 1$.}
\end{cases}
\end{align}
The quantile function becomes
\begin{align}
F_D^{-1}\left(p\right) = \frac{1+\ratio}{\sqrt{1 - p^2}+\ratio}.
\end{align}
The minimum factor remains $F_D^{-1}\left(p = 0\right) = 1$, but a maximum factor now exists: $F_D^{-1}\left(p = 1\right) = \frac{1}{\ratio} + 1$.
The median of $D$ is
\begin{align}
    F_D^{-1}\left(p=\frac{1}{2}\right) = \frac{1 + \ratio}{\frac{1}{2}\sqrt{3} + \ratio},
\end{align}
which tends to $\frac{2}{\sqrt{3}}$ for $\ratio \to 0$ (as before), and to $1$ for $\ratio \to \infty$: projection ceases to be an appreciable effect when the lobes are much larger than the line segment connecting their centres.
If $\ratio \neq 0$, $D$ has finite support, and thus the mean, variance, and higher moments exist.
The $n$-th non-central moment is
\begin{align}
    \mathbb{E}\left[D^n\right] = \left(1+\ratio\right)^2 I_{n-2}\left(\ratio\right) - \ratio \left(1+\ratio\right) I_{n-1}\left(\ratio\right),
\end{align}
where
\begin{align}
    I_n\left(\ratio\right) \coloneqq \int_1^{\frac{1}{\ratio}+1} \frac{d^n\ \mathrm{d}d}{\sqrt{d^2 - \left(1+\ratio-\ratio d\right)^2}}.
\end{align}
In particular,
\begin{align}
    I_{-1}\left(\ratio\right) = \frac{\pi}{2\left(1 + \ratio\right)},\ \ \ I_0\left(\ratio\right) = \frac{\ln{\left(\sqrt{1-\ratio^2}+1\right)}-\ln{\ratio}}{\sqrt{1-\ratio^2}},\nonumber\\
    I_1\left(\ratio\right) = \frac{1}{\left(1 - \ratio\right) \ratio}\left(1 - \frac{2\ratio^2 \mathrm{arcsinh}\sqrt{\frac{1-\ratio}{2\ratio}}}{\sqrt{1-\ratio^2}}\right).
\end{align}
The expectation value of $D$ is
\begin{align}
\mathbb{E}\left[D\right] &= \left(1+\ratio\right)^2 I_{-1}\left(\ratio\right) - \ratio \left(1+\ratio\right) I_0\left(\ratio\right)\nonumber\\
&= \left(1+\ratio\right)\left(\frac{\pi}{2} - \frac{\ratio}{\sqrt{1-\ratio^2}}\ln{\left(\frac{\sqrt{1-\ratio^2}}{\ratio} + \frac{1}{\ratio}\right)}\right).
\label{eq:deprojectionFactorLobesExpectation}
\end{align}
The variance of $D$ follows from combining
\begin{align}
\mathbb{E}\left[D^2\right] = \left(1+\ratio\right)^2 I_0\left(\ratio\right) - \ratio \left(1+\ratio\right) I_1\left(\ratio\right)
\end{align}
and the identity $\mathbb{V}\left[D\right] = \mathbb{E}\left[D^2\right] - \mathbb{E}^2\left[D\right]$.
The mode remains undefined.
For typical values of $\ratio$, we show the PDF and CDF of $D$ in the bottom left and right panels of Fig.~\ref{fig:deprojectionFactor}, respectively.
Figure~\ref{fig:deprojectionFactorMoments} shows the median, mean, and standard deviation of $D$ as a function of $\ratio$.

\subsection{Intrinsic proper length posterior and its moments}
\label{ap:posterior}
Our next objective is to find the posterior PDF $f_{L\vert L_\mathrm{p} = l_\mathrm{p}}\left(l\right)$ through Bayes' theorem:
\begin{align}
    f_{L\vert L_\mathrm{p} = l_\mathrm{p}}\left(l\right) = \frac{f_{L_\mathrm{p}\vert L = l}\left(l_\mathrm{p}\right) f_L\left(l\right)}{f_{L_\mathrm{p}}\left(l_\mathrm{p}\right)}.
\end{align}
Our line of attack will be to first calculate the likelihood CDF $F_{L_\mathrm{p}\vert L = l}(l_\mathrm{p})$, and then through differentiation the likelihood PDF $f_{L_\mathrm{p}\vert L = l}(l_\mathrm{p})$.
Of course, we always consider $l > 0$, and
\begin{align}
    F_{L_\mathrm{p}\vert L = l}\left(l_\mathrm{p}\right) &\coloneqq \mathbb{P}\left(L_\mathrm{p} \leq l_\mathrm{p}\ \vert\ L = l\right)
    = \mathbb{P}\left(l\sin{\Theta} \leq l_\mathrm{p}\right)\nonumber\\
    &= \begin{cases}
    0 & \text{if } l_\mathrm{p} \leq 0\\
    2\ \mathbb{P}\left(\Theta \leq \arcsin{\frac{l_\mathrm{p}}{l}}\right) & \text{if } 0 < l_\mathrm{p} \leq l\\
    1 & \text{if}\ l_\mathrm{p} > l,\\
    \end{cases}
\end{align}
where we have invoked Eq.~\ref{eq:projectionRVs} and the fact that $f_\Theta\left(\theta\right)$ is symmetric in $\theta = \frac{\pi}{2}$ for the case $0 < l_\mathrm{p} < l$.
Concretising this case further yields
\begin{align}
    F_{L_\mathrm{p}\vert L = l}\left(l_\mathrm{p}\right) &= \ 2\ F_\Theta\left(\arcsin{\frac{l_\mathrm{p}}{l}}\right)
    = 1 - \cos{\arcsin{\frac{l_\mathrm{p}}{l}}}\nonumber\\
    &= 1 - \sqrt{1-\left(\frac{l_\mathrm{p}}{l}\right)^2}.
\end{align}
Through differentiation,
\begin{align}
    f_{L_\mathrm{p}\vert L = l}\left(l_\mathrm{p}\right) =
    \begin{cases}
    0 & \text{if } l_\mathrm{p} \leq 0\\
    \frac{l_\mathrm{p}}{l^2}\frac{1}{\sqrt{1-\left(\frac{l_\mathrm{p}}{l}\right)^2}} & \text{if } 0 < l_\mathrm{p} < l\\
    0 & \text{if } l_\mathrm{p} \geq l.
    \end{cases}
\end{align}
Having found the likelihood, the posterior PDF follows directly through Bayes' theorem and the PDFs computed hitherto.
In concreto, when $0 < l_\mathrm{p} \leq l_\mathrm{min}$,
\begin{align}
    f_{L\vert L_\mathrm{p} = l_\mathrm{p}}\left(l\right) =
    \begin{cases}
    0 & \text{if } l \leq l_\mathrm{min}\\
    \frac{1}{I\left(\xi-1,\frac{l_\mathrm{p}}{l_\mathrm{min}}\right)}\frac{1}{l_\mathrm{p}}\frac{1}{\sqrt{\left(\frac{l}{l_\mathrm{p}}\right)^2-1}}\left(\frac{l}{l_\mathrm{min}}\right)^{\xi-1} & \text{if } l > l_\mathrm{min}.
    \end{cases}
\end{align}
whereas for $l_\mathrm{p} > l_\mathrm{min}$,
\begin{align}
    f_{L\vert L_\mathrm{p} = l_\mathrm{p}}\left(l\right) =
    \begin{cases}
    0 & \text{if } l \leq l_\mathrm{p}\\
    -\frac{\xi}{2^{1+\xi}\pi}\frac{\Gamma^2\left(-\frac{\xi}{2}\right)}{\Gamma\left(-\xi\right)}\frac{1}{l_\mathrm{p}}\frac{1}{\sqrt{\left(\frac{l}{l_\mathrm{p}}\right)^2 - 1}}\left(\frac{l}{l_\mathrm{p}}\right)^{\xi-1} & \text{if } l > l_\mathrm{p}.
    \end{cases}
\end{align}
We note that in both cases, for $l \gg l_\mathrm{p}$, $f_{L\vert L_\mathrm{p} = l_\mathrm{p}}\left(l\right) \propto \left(\frac{l}{l_\mathrm{p}}\right)^{\xi-2}$: the posterior probability density follows a power law in $l$ with exponent $\xi - 2$.\\\\
For $l_\mathrm{p} \leq l_\mathrm{min}$, the posterior mean is
\begin{align}
    \mathbb{E}\left[L\ \vert\ L_\mathrm{p} = l_\mathrm{p}\right] = l_\mathrm{min} \frac{I\left(\xi, \frac{l_\mathrm{p}}{l_\mathrm{min}}\right)}{I\left(\xi - 1, \frac{l_\mathrm{p}}{l_\mathrm{min}}\right)}.
\end{align}
The posterior variance $\mathbb{V}\left[L\ \vert\ L_\mathrm{p} = l_\mathrm{p}\right]$ follows from considering the second non-central moment:
\begin{align}
    \mathbb{E}\left[L^2\ \vert\ L_\mathrm{p} = l_\mathrm{p}\right] = l_\mathrm{min}^2 \frac{I\left(\xi + 1, \frac{l_\mathrm{p}}{l_\mathrm{min}}\right)}{I\left(\xi - 1, \frac{l_\mathrm{p}}{l_\mathrm{min}}\right)},
\end{align}
so that
\begin{align}
\mathbb{V}\left[L\ \vert\ L_\mathrm{p} = l_\mathrm{p}\right] &= \mathbb{E}\left[L^2\ \vert\ L_\mathrm{p} = l_\mathrm{p}\right] - \mathbb{E}^2\left[L\ \vert\ L_\mathrm{p} = l_\mathrm{p}\right]\nonumber\\
&= l_\mathrm{min}^2\left(\frac{I\left(\xi + 1, \frac{l_\mathrm{p}}{l_\mathrm{min}}\right)}{I\left(\xi - 1, \frac{l_\mathrm{p}}{l_\mathrm{min}}\right)} - \frac{I^2\left(\xi, \frac{l_\mathrm{p}}{l_\mathrm{min}}\right)}{I^2\left(\xi - 1, \frac{l_\mathrm{p}}{l_\mathrm{min}}\right)}\right).
\end{align}
For $l_\mathrm{p} > l_\mathrm{min}$,
\begin{align}
    \mathbb{E}\left[L\ \vert\ L_\mathrm{p} = l_\mathrm{p}\right] &= l_\mathrm{p} \cdot \frac{-\xi}{2^{\xi+1}\pi}\frac{\Gamma^2\left(-\frac{\xi}{2}\right)}{\Gamma\left(-\xi\right)} \int_1^\infty \frac{\eta^\xi\ \mathrm{d}\eta}{\sqrt{\eta^2 - 1}}\nonumber\\
    &= l_\mathrm{p} \cdot \frac{-\xi}{2^{2\xi+3}\pi} \frac{\Gamma^4\left(-\frac{\xi}{2}\right)}{\Gamma^2\left(-\xi\right)}.
\end{align}
Proceeding analogously,
\begin{align}
    \mathbb{E}\left[L^2\ \vert\ L_\mathrm{p} = l_\mathrm{p}\right] = l_\mathrm{p}^2 \frac{\xi}{\xi+1},
\end{align}
so that
\begin{align}
    \mathbb{V}\left[L\ \vert\ L_\mathrm{p} = l_\mathrm{p}\right] = l_\mathrm{p}^2\left(\frac{\xi}{\xi + 1} - \frac{\xi^2}{2^{4\xi + 6}\pi^2}\frac{\Gamma^8\left(-\frac{\xi}{2}\right)}{\Gamma^4\left(-\xi\right)}\right).
\end{align}
In this case, both the mean and standard deviation of $L\ \vert\ L_\mathrm{p} = l_\mathrm{p}$ are proportional to $l_\mathrm{p}$.
In the table below, we list the mean and standard deviation in multiples of $l_\mathrm{p}$ for several values of $\xi$.
Since we assume $\xi < -1$, the mean and variance are guaranteed to exist.
The existence of higher-order moments is $\xi$-dependent; in concreto, the highest defined order is $\lceil -\xi \rceil$.

We prove that $L_\mathrm{p}$ and $D$ are not independent by contradiction.
If we assume that $L_\mathrm{p}$ and $D$ are independent, then $\mathbb{E}[L_\mathrm{p} D] = \mathbb{E}[L_\mathrm{p}] \mathbb{E}[D]$, or $\mathbb{E}[L] = \mathbb{E}[L]\mathbb{E}[\sin{\Theta}]\mathbb{E}[D]$ by the independence of $L$ and $\sin{\Theta}$.
In other words, if $L_\mathrm{p}$ and $D$ are independent, then $1 = \mathbb{E}[\sin{\Theta}]\mathbb{E}[D]$.
However, $\mathbb{E}[\sin{\Theta}] = \frac{\pi}{4}$ and $\mathbb{E}[D] = \frac{\pi}{2}$; because $1 \neq \frac{\pi^2}{8}$, the initial assertion must be wrong.

\subsection{GRG inclination angle}
\label{ap:GRGInclinationAngle}
We derive the inclination angle distribution for giants.
The probability that a GRG has inclination angle $\theta$ is
\begin{align}
    f_{\Theta\ \vert\ L_\mathrm{p} > l_\mathrm{p,GRG}}(\theta)\ \mathrm{d}\theta &= \frac{\mathbb{P}\left(\Theta = \theta, L \sin{\Theta} > l_\mathrm{p,GRG}\right)}{\mathbb{P}\left(L_\mathrm{p} > l_\mathrm{p,GRG}\right)}\nonumber\\
    &= \frac{f_\Theta(\theta)\ \mathrm{d}\theta \cdot \mathbb{P}\left(L > \frac{l_\mathrm{p,GRG}}{\sin{\theta}}\right)}{\mathbb{P}\left(L_\mathrm{p} > l_\mathrm{p,GRG}\right)},
\end{align}
where we make use of the fact that the numerator's joint probability factorises because $L$ and $\Theta$ are independent RVs.
We thus find
\begin{align}
    f_{\Theta\ \vert\ L_\mathrm{p} > l_\mathrm{p,GRG}}(\theta) = \frac{\left(1 - F_L\left(\frac{l_\mathrm{p,GRG}}{\sin{\theta}}\right)\right) f_\Theta\left(\theta\right)}{1 - F_{L_\mathrm{p}}(l_\mathrm{p,GRG})}.
\end{align}
Under the Paretian assumption for $L$, we have
\begin{align}
    f_{\Theta\ \vert\ L_\mathrm{p} > l_\mathrm{p,GRG}}(\theta) = \frac{2}{-(\xi+1)\sqrt{\pi}}\frac{\Gamma\left(-\frac{\xi}{2}+1\right)}{\Gamma\left(-\frac{\xi}{2}-\frac{1}{2}\right)}\sin^{-\xi}{\theta}.
\end{align}
\subsection{GRG angular length}
\label{ap:angularLength}
The GRG angular length RV $\Phi\ \vert\ L_\mathrm{p} > l_\mathrm{p,GRG}$ relates to the GRG projected proper length RV $L_\mathrm{p}\ \vert\ L_\mathrm{p} > l_\mathrm{p,GRG}$ and the comoving distance RV $R$ as
\begin{align}
    \Phi\ \vert\ L_\mathrm{p} > l_\mathrm{p,GRG} = L_\mathrm{p}\ \vert\ L_\mathrm{p} > l_\mathrm{p,GRG} \cdot  \frac{1+z\left(R\right)}{R}.
    \label{eq:GRGAngularLength}
\end{align}
The model predicts the distribution of GRG angular lengths in the Local Universe up to comoving distance $r_\mathrm{max}$.
If the GRG number density is constant in the Local Universe,
\begin{align}
    F_R\left(r\right) &= \begin{cases}
    0 & \text{if } r \leq 0\\
    \left(\frac{r}{r_\mathrm{max}}\right)^3 & \text{if } 0 < r < r_\mathrm{max}\\
    1 & \text{if } r \geq r_\mathrm{max}
    \end{cases}\\
    f_R\left(r\right) &= \begin{cases}
    0 & \text{if } r \leq 0\\
    \frac{3}{r_\mathrm{max}}\left(\frac{r}{r_\mathrm{max}}\right)^2 & \text{if } 0 < r < r_\mathrm{max}\\
    0 & \text{if } r \geq r_\mathrm{max}.
    \end{cases}
\end{align}
Because GRG life cycles are shorter than the age of the Universe, $L_\mathrm{p}\ \vert\ L_\mathrm{p} > l_\mathrm{p,GRG}$ and $R$ are independent RVs.
In a non-expanding universe, $z(R) = 0$, and the distribution of $\Phi\ \vert\ L_\mathrm{p} > l_\mathrm{p,GRG}$ can be calculated analytically.
From a well-known ratio distribution identity,
\begin{align}
    f_{\Phi\ \vert\ L_\mathrm{p} > l_\mathrm{p,GRG}}(\phi) &= \int_{-\infty}^\infty |r| \cdot f_{L_\mathrm{p}\ \vert\ L_\mathrm{p}>l_\mathrm{p,GRG}}(\phi r) \cdot f_R(r)\ \mathrm{d}r\nonumber\\
    &= \int_0^{r_\mathrm{max}} f_{L_\mathrm{p}\ \vert\ L_\mathrm{p}>l_\mathrm{p,GRG}}(\phi r) \cdot 3 \left(\frac{r}{r_\mathrm{max}}\right)^3\ \mathrm{d}r.
\end{align}
The integrand is non-zero only when $\phi r \geq l_\mathrm{p,GRG}$, suggesting a lower integration limit of $\frac{l_\mathrm{p,GRG}}{\phi}$.
The integral vanishes altogether when $\frac{l_\mathrm{p,GRG}}{\phi} \geq r_\mathrm{max}$.
Calling $\phi_\mathrm{GRG} \coloneqq \frac{l_\mathrm{p,GRG}}{r_\mathrm{max}}$, we have $f_{\Phi\ \vert\ L_\mathrm{p} > l_\mathrm{p,GRG}}(\phi) = 0$ for $\phi \leq \phi_\mathrm{GRG}$.
For $\phi > \phi_\mathrm{GRG}$,
\begin{align}
    f_{\Phi\ \vert\ L_\mathrm{p} > l_\mathrm{p,GRG}}(\phi) &= \int_{\frac{l_\mathrm{p,GRG}}{\phi}}^{r_\mathrm{max}} -\frac{\xi+1}{l_\mathrm{p,GRG}}\left(\frac{\phi r}{l_\mathrm{p,GRG}}\right)^\xi \cdot 3 \left(\frac{r}{r_\mathrm{max}}\right)^3\ \mathrm{d}r\nonumber\\
    &= -3 \frac{\xi + 1}{\xi + 4}\cdot \frac{1}{\phi_\mathrm{GRG}}\left(\left(\frac{\phi}{\phi_\mathrm{GRG}}\right)^\xi - \left(\frac{\phi}{\phi_\mathrm{GRG}}\right)^{-4}\right),
    \label{eq:GRGAngularLengthPDF}
\end{align}
where we assume $\xi \neq -4$.
This PDF depends on $\xi$, $l_\mathrm{p,GRG}$, and $r_\mathrm{max}$ only.
The CDF follows from direct integration:
\begin{align}
    &F_{\Phi\ \vert\ L_\mathrm{p} > l_\mathrm{p,GRG}}(\phi)= \nonumber\\
    &\begin{cases}
    0 & \text{if } \phi \leq \phi_\mathrm{GRG}\\
    1 + \frac{1}{\xi + 4}\left(-3 \left(\frac{\phi}{\phi_\mathrm{GRG}}\right)^{\xi + 1} - (\xi + 1)\left(\frac{\phi}{\phi_\mathrm{GRG}}\right)^{-3}\right) & \text{if } \phi > \phi_\mathrm{GRG},
    \end{cases}
\end{align}
just as the mean:
\begin{align}
    \mathbb{E}\left[\Phi\ \vert\ L_\mathrm{p} > l_\mathrm{p,GRG}\right] = 3 \frac{\xi + 1}{\xi + 4}\left(\frac{1}{\xi + 2} + \frac{1}{2}\right) \cdot \phi_\mathrm{GRG},
\end{align}
which exists only for $\xi < -2$.
By solving
\begin{align}
    \frac{\mathrm{d} f_\Phi\ \vert\ L_\mathrm{p} > l_\mathrm{p,GRG}}{\mathrm{d} \phi}\left(\phi_\mathrm{mode}\right) = 0
\end{align}
for the GRG angular length mode $\phi_\mathrm{mode}$, we find
\begin{align}
    \phi_\mathrm{mode} = \left(\frac{-4}{\xi}\right)^{\frac{1}{\xi + 4}} \cdot \phi_\mathrm{GRG}.
\end{align}
There is no explicit expression for the associated median.

For a general non-Euclidean universe, no simple analytic form for the GRG angular length PDF appears to exist.
We find an approximation valid at low redshifts by considering the Maclaurin polynomial of degree $1$ for $z(r)$, the relation between comoving distance and cosmological redshift.
One finds $z(r) \approx \frac{H_0}{c}r$.
As a result, for the Local Universe, Eq.~\ref{eq:GRGAngularLength} becomes
\begin{align}
    \Phi\ \vert\ L_\mathrm{p} > l_\mathrm{p,GRG} &\approx L_\mathrm{p}\ \vert\ L_\mathrm{p} > l_\mathrm{p,GRG} \cdot \frac{1+\frac{H_0}{c} R}{R}\nonumber\\
    &= L_\mathrm{p}\ \vert\ L_\mathrm{p} > l_\mathrm{p,GRG} \cdot \left(\frac{1}{R} + \frac{1}{d_H}\right),
\end{align}
where we use that the Hubble distance $d_H \coloneqq \frac{c}{H_0}$.
From a well-known inverse distribution identity, 
\begin{align}
    f_{\frac{1}{R}}(k) = \begin{cases}
    0 & \text{if } k \leq \frac{1}{r_\mathrm{max}}\\
    \frac{3}{r_\mathrm{max}^3 k^4} & \text{if } k > \frac{1}{r_\mathrm{max}},
    \end{cases}
\end{align}
so that
\begin{align}
    f_{\frac{1}{R} + \frac{1}{d_H}}(k) = \begin{cases}
    0 & \text{if } k \leq \frac{1}{r_\mathrm{max}} + \frac{1}{d_H}\\
    \frac{3}{r_\mathrm{max}^3 \left(k - \frac{1}{d_H}\right)^4} & \text{if } k > \frac{1}{r_\mathrm{max}} + \frac{1}{d_H}.
    \end{cases}
\end{align}
Combining a well-known product distribution identity with the fact that $L_\mathrm{p}\ \vert\ L_\mathrm{p} > l_\mathrm{p,GRG}$ and $\frac{1}{R} + \frac{1}{d_H}$ are independent RVs, we find
\begin{align}
    f_{\Phi\ \vert\ L_\mathrm{p} > l_\mathrm{p,GRG}}(\phi) = \frac{-3\left(\xi+1\right)\phi^\xi}{r_\mathrm{max}^3 l_\mathrm{p,GRG}^{\xi+1}} \int_{\frac{1}{r_\mathrm{max}} + \frac{1}{d_H}}^{\frac{\phi}{l_\mathrm{p,GRG}}} \frac{\mathrm{d}k}{k^{\xi+1}\left(k - \frac{1}{d_H}\right)^4}
\end{align}
when $\phi > \phi_\mathrm{GRG} + \frac{l_\mathrm{p,GRG}}{d_H}$.
When $\phi \leq \phi_\mathrm{GRG} + \frac{l_\mathrm{p,GRG}}{d_H}$, $f_{\Phi\ \vert\ L_\mathrm{p} > l_\mathrm{p,GRG}}(\phi) = 0$.
It is easy to verify that in the Euclidean limit, $\frac{1}{d_H} \to 0$, this expression reduces to that of Eq.~\ref{eq:GRGAngularLengthPDF}.
\subsection{Maximum likelihood estimation of the tail index}
\label{ap:MLE}
How can we estimate $\xi$ from a sample of $N$ giants?
Let $\xi_\mathrm{MLE}$ be the maximum likelihood estimate (MLE) of $\xi$.
This RV is a function of $N$ IID RVs $\{L_{\mathrm{p},1}, ..., L_{\mathrm{p},N}\} \sim L_\mathrm{p}\ \vert\ L_\mathrm{p} > l_\mathrm{p, GRG}$.
Define the following likelihood and log-likelihood functions:
\begin{align}
    \mathcal{L}\left(\xi\right) &\coloneqq \prod_{i=1}^N f_{L_\mathrm{p}\ \vert\ L_\mathrm{p} > l_\mathrm{p, GRG}}\left(L_\mathrm{p, i}\right)\\
    \tilde{\mathcal{L}}\left(\xi\right) &\coloneqq \ln{\left(\mathcal{L}\left(\xi\right) l_\mathrm{p, GRG}^N\right)}\nonumber\\
    &= N\ln{\left(-\left(\xi+1\right)\right)} + \xi \sum_{i=1}^N \ln{\frac{L_\mathrm{p,i}}{l_\mathrm{p,GRG}}}.
\end{align}
We note the necessity to include a factor $l_\mathrm{p,GRG}^N$ in the definition of the log-likelihood to avoid a dimensionality error.
The second derivative of $\tilde{\mathcal{L}}$ to $\xi$ is
\begin{align}
    \frac{\mathrm{d}^2 \tilde{\mathcal{L}}}{\mathrm{d}\xi^2} = -\frac{N}{\left(\xi+1\right)^2} < 0.
\end{align}
Thus, if there exists a solution to the equation $\frac{\mathrm{d}\tilde{\mathcal{L}}}{\mathrm{d}\xi} = 0$, it must correspond to a global \emph{maximum} of the likelihood and log-likelihood functions:
\begin{align}
    \frac{\mathrm{d}\tilde{\mathcal{L}}}{\mathrm{d}\xi}\left(\xi = \xi_\mathrm{MLE}\right) = 0, \text{or }
    \xi_\mathrm{MLE} &= -\frac{N}{\sum_{i=1}^N \ln{\frac{L_\mathrm{p,i}}{l_\mathrm{p,GRG}}}} - 1.
\end{align}
\subsection{Observed projected proper length}
\subsubsection{General considerations}
\label{ap:completenessGeneral}
The model can be extended to incorporate observational selection effects.
The relevant effects to consider vary from RG search campaign to RG search campaign, although some formulae apply in all cases.
We derive these here.

To keep our extensions simple, we assume that the survey sensitivity is sufficient to detect RGs up to some redshift $z_\mathrm{max}$ only.
In addition we assume that the projected proper length distribution does not evolve between $z = z_\mathrm{max}$ and $z = 0$.
We let $p_\mathrm{obs}(l_\mathrm{p},z)$ denote the probability that an RG of projected proper length $l_\mathrm{p}$ at cosmological redshift $z$ is detected during the campaign.
Also, $r$ is the radial comoving distance and $n$ is the total RG number density, counting the intrinsic number of RGs (irrespective of length) per unit of comoving volume.
The observed number of RGs with projected proper length between $l_\mathrm{p}$ and $l_\mathrm{p} + \mathrm{d}l_\mathrm{p}$ throughout a survey covering a solid angle $\Omega$ is $\mathrm{d}N_{L_\mathrm{p},\mathrm{obs}}\left(l_\mathrm{p}, \Omega\right)$, where
\begin{align}
    &\mathrm{d}N_{L_\mathrm{p},\mathrm{obs}}\left(l_\mathrm{p},\Omega\right)\nonumber\\
    &= \frac{\Omega}{4\pi}\int_0^{z_\mathrm{max}} n \cdot f_{L_\mathrm{p}}\left(l_\mathrm{p}\right) \mathrm{d}l_\mathrm{p} \cdot p_\mathrm{obs}\left(l_\mathrm{p},z\right) 4\pi r^2\left(z\right) \frac{\mathrm{d}r}{\mathrm{d}z}\ \mathrm{d}z.
\label{eq:observedProjectedProperLengthNumberDensity}
\end{align}
The total number of RGs with projected proper length between $l_\mathrm{p}$ and $l_\mathrm{p} + \mathrm{d}l_\mathrm{p}$ throughout a survey with solid angle $\Omega$ is $\mathrm{d}N_{L_\mathrm{p}}\left(l_\mathrm{p},\Omega\right)$, where
\begin{align}
    \mathrm{d}N_{L_\mathrm{p}}\left(l_\mathrm{p},\Omega\right) = \frac{\Omega}{4\pi}\int_0^{z_\mathrm{max}} n \cdot f_{L_\mathrm{p}}\left(l_\mathrm{p}\right) \mathrm{d}l_\mathrm{p} \cdot 4\pi r^2\left(z\right) \frac{\mathrm{d}r}{\mathrm{d}z}\ \mathrm{d}z.
\end{align}
We define the completeness $C(l_\mathrm{p}, z_\mathrm{max})$ to be
\begin{align}
    C(l_\mathrm{p}, z_\mathrm{max}) &\coloneqq \frac{\mathrm{d}N_{L_\mathrm{p},\mathrm{obs}}\left(l_\mathrm{p}, \Omega\right)}{\mathrm{d}N_{L_\mathrm{p}}\left(l_\mathrm{p},\Omega\right)}\nonumber\\
    &= \frac{\int_0^{z_\mathrm{max}}p_\mathrm{obs}\left(l_\mathrm{p},z\right) r^2\left(z\right) E^{-1}\left(z\right)\ \mathrm{d}z}{\int_0^{z_\mathrm{max}}r^2\left(z\right)E^{-1}\left(z\right)\ \mathrm{d}z}.
\end{align}
The completeness only depends on the function $p_\mathrm{obs}(l_\mathrm{p},z)$ and our choice of $z_\mathrm{max}$.

Let the RV $\mathcal{O}$ denote whether an RG picked at random within $z < z_\mathrm{max}$ is detected during the search campaign.
We have $\mathcal{O}\sim\mathrm{Bernoulli}\left(C(L_\mathrm{p}, z_\mathrm{max})\right)$; the parameter that determines the distribution of $\mathcal{O}$ is itself an RV.
It immediately follows that $\mathbb{P}\left(\mathcal{O} = 1\ \vert\ L_\mathrm{p} = l_\mathrm{p}\right) = C(l_\mathrm{p}, z_\mathrm{max})$.

If $L_\mathrm{p}$ would be discrete,
\begin{align}
    \mathbb{P}\left(L_\mathrm{p} = l_\mathrm{p}\ \vert\ \mathcal{O} = 1\right) &= \frac{\mathbb{P}\left(\mathcal{O} = 1\ , L_\mathrm{p} = l_\mathrm{p}\right)}{\mathbb{P}\left(\mathcal{O}=1\right)}\nonumber\\
    &= \frac{\mathbb{P}\left(\mathcal{O} = 1\ \vert\ L_\mathrm{p} = l_\mathrm{p}\right)\mathbb{P}\left(L_\mathrm{p} = l_\mathrm{p}\right)}{\sum_{l_\mathrm{p}} \mathbb{P}\left(\mathcal{O} = 1\ \vert\ L_\mathrm{p} = l_\mathrm{p}\right)\mathbb{P}\left(L_\mathrm{p} = l_\mathrm{p}\right)}.
\end{align}
Let $L_\mathrm{p,obs}$ be the observed projected proper length RV; that is $L_\mathrm{p,obs} \coloneqq L_\mathrm{p}\ \vert\ \mathcal{O} = 1$.
Its PDF is given by the continuous analogon of the preceding equation:
\begin{align}
    f_{L_\mathrm{p,obs}}\left(l_\mathrm{p}\right) = \frac{C\left(l_\mathrm{p}, z_\mathrm{max}\right)f_{L_\mathrm{p}}\left(l_\mathrm{p}\right)}{\int_0^\infty C\left(l_\mathrm{p}, z_\mathrm{max}\right) f_{L_\mathrm{p}}\left(l_\mathrm{p}\right)\ \mathrm{d}l_\mathrm{p}}.
\end{align}
Obviously, $F_{L_\mathrm{p,obs}}\left(l_\mathrm{p}\right) = \int_0^{l_\mathrm{p}}f_{L_\mathrm{p,obs}}\left(l_\mathrm{p}'\right)\ \mathrm{d}l_\mathrm{p}'$.
The CDF $F_{L_\mathrm{p,obs}\vert L_\mathrm{p,obs} > l_\mathrm{p,GRG}}\left(l_\mathrm{p}\right)$ follows from an analogon of Eq.~\ref{eq:CDFLengthProperProjectedGRG}.

We note that multiplying $p_\mathrm{obs}(l_\mathrm{p},z)$ with an $l_\mathrm{p}$- and $z$-independent factor changes $C(l_\mathrm{p},z_\mathrm{max})$ by the same factor, but leaves $f_{L_\mathrm{p,obs}}$ unaltered.

\subsubsection{Fuzzy angular length threshold}
\label{ap:completenessAngularLength}
Here we illustrate a simple extension.
When performing visual searches for GRG candidates, a natural criterion is to only inspect sources with an angular length that is larger than some threshold.
Researchers determine the threshold based on the amount of time available to them: lower thresholds will lead to more complete samples, but will take more time to collect.
For humans, it is hard to estimate a source's angular length precisely by eye; as a result, some of the GRG candidates included in the project's GRG candidate catalogue will be sources with an angular length below the threshold, whilst others will be sources with an angular length above the threshold.
We idealise this situation by asserting that sources with angular length $\phi_\mathrm{min}$ or below are included in the catalogue with probability 0 (i.e. never), and that sources with an angular length $\phi_\mathrm{max}$ or above are included in the catalogue with probability 1 (i.e. always).
We assume a linear increase in probability as a function of $\phi$ for intermediate angular lengths: a source with angular length $\phi_\mathrm{min} < \phi < \phi_\mathrm{max}$ is included in the catalogue with probability
\begin{align}
    p_\mathrm{obs}\left(l_\mathrm{p},z\right) = \min{\left\{\max{\left\{\frac{\phi\left(l_\mathrm{p},z\right) - \phi_\mathrm{min}}{\phi_\mathrm{max}-\phi_\mathrm{min}}, 0\right\}}, 1\right\}}.
\label{eq:probabilityObservingAngularThresholdAppendix}
\end{align}
In a flat Friedmann--Lema\^itre--Robertson--Walker (FLRW) universe, the angular length of an RG with projected proper length $l_\mathrm{p}$ at cosmological redshift $z$ is
\begin{align}
    \phi\left(l_\mathrm{p},z\right) = \frac{l_\mathrm{p}\left(1+z\right)}{r\left(z\right)}.
\end{align}

\subsubsection{Surface brightness limitations}
\label{ap:completenessSurfaceBrightness}
\paragraph{Fanaroff--Riley class II}
If the surface brightness $B_\nu$ of the lobes is proportional to some negative power $\zeta$ of the GRG's proper length $L$, that is
\begin{align}
    B_\nu = b_{\nu,\mathrm{ref}} \left(\frac{L}{l_\mathrm{ref}}\right)^\zeta,
\end{align}
then $B_\nu$ is Pareto distributed --- just like $L$:
\begin{align}
    F_{B_\nu}\left(b\right) &= \begin{cases}
    0 & \text{if } b \leq 0\\
    \left(\frac{b}{b_{\nu,\mathrm{ref}}}\right)^\frac{\xi+1}{\zeta} & \text{if } 0 < b < b_{\nu,\mathrm{ref}}\\
    1 & \text{if } b \geq b_{\nu,\mathrm{ref}},
    \end{cases}\\
    f_{B_\nu}\left(b\right) &= \begin{cases}
    0 & \text{if } b \leq 0\\
    \frac{\xi + 1}{\zeta}\frac{1}{b_{\nu,\mathrm{ref}}}\left(\frac{b}{b_{\nu,\mathrm{ref}}}\right)^{\frac{\xi+1}{\zeta}-1} & \text{if } 0 < b <  b_{\nu,\mathrm{ref}}\\
    0 & \text{if } b \geq b_{\nu,\mathrm{ref}}.
    \end{cases}
\end{align}
If RGs are self-similar in shape, then their lobe volumes are proportional to $L^3$.
RGs appear to retain constant lobe luminosity density over most of their lifetime, so that their lobe monochromatic emission coefficients \citep{Rybicki11986} are proportional to $L^{-3}$.
As line-of-sight lengths through the lobes are proportional to $L$, surface brightness is proportional to $L^{-2}$.
These arguments thus suggest $\zeta = -2$.

It is a poor approximation to assume that all giants of proper length $l_\mathrm{ref}$ have the same surface brightness $b_{\nu,\mathrm{ref}}$: observations suggest a variability of several orders of magnitude.
A better description is
\begin{align}
    B_\nu = b_{\nu,\mathrm{ref}} \left(\frac{L}{l_\mathrm{ref}}\right)^\zeta S,
\end{align}
where $S$ is a lognormal RV whose median is 1.
The PDF of $S$ is then determined by parameter $\sigma_\mathrm{ref}$:
\begin{align}
    f_S\left(s\right) = \frac{1}{\sqrt{2\pi}\sigma_\mathrm{ref} s}\exp{\left(-\frac{\ln^2s}{2\sigma_\mathrm{ref}^2}\right)}.
\end{align}
The surface brightness CDF and PDF now become
\begin{align}
    F_{B_\nu}\left(b\right) &= \begin{cases}
    0 & \text{if } b \leq 0\\
    F_S\left(\frac{b}{b_{\nu,\mathrm{ref}}}\right) + \left(\frac{b}{b_{\nu,\mathrm{ref}}}\right)^\frac{\xi+1}{\zeta}\int_\frac{b}{b_{\nu,\mathrm{ref}}}^\infty s^{-\frac{\xi+1}{\zeta}}f_S\left(s\right)\mathrm{d}s & \text{if } b > 0,
    \end{cases}\nonumber\\
    \\
    f_{B_\nu}\left(b\right) &= \begin{cases}
    0 & \text{if } b \leq 0\\
    \frac{\xi + 1}{\zeta}\frac{1}{b_{\nu,\mathrm{ref}}}\left(\frac{b}{b_{\nu,\mathrm{ref}}}\right)^{\frac{\xi+1}{\zeta}-1} \int_\frac{b}{b_{\nu,\mathrm{ref}}}^\infty s^{-\frac{\xi+1}{\zeta}} f_S\left(s\right) \mathrm{d}s & \text{if } b > 0.
    \end{cases}\nonumber
\end{align}
We note that $B_\nu$ is not exactly Pareto distributed anymore.

In a relativistic rather than Euclidean universe, surface brightness is not constant with distance.
To describe RGs beyond $z = 0$, we introduce a final model refinement:
\begin{align}
B_\nu = \frac{b_{\nu,\mathrm{ref}} \cdot S}{\left(1+Z\right)^{3-\alpha}}\left(\frac{L}{l_\mathrm{ref}}\right)^\zeta,
\label{eq:surfaceBrightnessRV}
\end{align}
where the RV $Z$ denotes cosmological redshift and $\alpha$ is the spectral index of the lobes.
We interpret $b_{\nu,\mathrm{ref}} \cdot S$ as the (lognormally distributed) lobe surface brightness for RGs of intrinsic proper length $l_\mathrm{ref}$ at $z = 0$.
In this case, the CDF and PDF of $B_\nu$ are most easily determined through sampling.
To sample $Z$, we can first sample the comoving distance RV $R$ instead, and subsequently use $Z = z_\mathfrak{M}\left(R\right)$.
We stress that this conversion depends on cosmological parameters $\mathfrak{M}$.

To forward model a survey's surface brightness selection effect, we must compute 
\begin{align}
    p_\mathrm{obs}\left(l_\mathrm{p},z\right) &= \mathbb{P}\left(B_\nu > b_{\nu,\mathrm{th}}\ \vert\ L_\mathrm{p} = l_\mathrm{p},\ Z = z\right)\nonumber\\
    &= 1 - F_{B_\nu\ \vert\ L_\mathrm{p} = l_\mathrm{p},\ Z = z}\left(b_{\nu,\mathrm{th}}\right),
\end{align}
where $b_{\nu,\mathrm{th}} > 0$ is the surface brightness threshold.
Typically, $b_{\nu,\mathrm{th}}$ is comparable to the survey's RMS noise.
What is $F_{B_\nu\ \vert\ L_\mathrm{p} = l_\mathrm{p},\ Z = z}\left(b\right)$?
In the simplest case, devoid of $S$- and $Z$-dependence,
\begin{align}
    F_{B_\nu\ \vert\ L_\mathrm{p} = l_\mathrm{p},\ Z = z}\left(b\right) &= \mathbb{P}\left(b_{\nu,\mathrm{ref}}\left(\frac{L_\mathrm{p}}{\sin{\Theta} \cdot l_\mathrm{ref}}\right)^\zeta \leq b\ \vert\ L_\mathrm{p} = l_\mathrm{p}\right)\nonumber\\
    &= \mathbb{P}\left(\sin^{-\zeta}\Theta \leq \frac{b}{b_{\nu,\mathrm{ref}}}\left(\frac{l_\mathrm{p}}{l_\mathrm{ref}}\right)^{-\zeta}\right)\nonumber\\
    &= \mathbb{P}\left(\sin\Theta \leq \tilde{s}^{-\frac{1}{\zeta}}\right)\nonumber\\
    &= \begin{cases}
    0 & \text{if } b \leq 0\\
    2F_\Theta\left(\arcsin{\left(\tilde{s}^{-\frac{1}{\zeta}}\right)}\right) & \text{if } 0 < b < b_{\nu,\mathrm{ref}}\left(\frac{l_\mathrm{p}}{l_\mathrm{ref}}\right)^\zeta\\
    1 & \text{if } b \geq b_{\nu,\mathrm{ref}}\left(\frac{l_\mathrm{p}}{l_\mathrm{ref}}\right)^\zeta
    \end{cases}\nonumber\\
    &= 1 - \sqrt{1 - \min{\left\{\tilde{s}^{-\frac{2}{\zeta}}\left(\max{\left\{b, 0\right\}}\right), 1\right\}}},
\end{align}
where
\begin{align}
    \tilde{s} = \tilde{s}\left(b\right) \coloneqq \frac{b}{b_{\nu,\mathrm{ref}}}\left(\frac{l_\mathrm{p}}{l_\mathrm{ref}}\right)^{-\zeta}.
\end{align}
Therefore,
\begin{align}
    p_\mathrm{obs}\left(l_\mathrm{p},z\right) = \sqrt{1 - \min{\left\{\tilde{s}^{-\frac{2}{\zeta}}\left(b_{\nu,\mathrm{th}}\right), 1\right\}}}.
\end{align}
In the most refined case, for $b > 0$,
\begin{align}
    &F_{B_\nu\ \vert\ L_\mathrm{p} = l_\mathrm{p},\ Z = z}\left(b\right)\nonumber\\
    &= \mathbb{P}\left(\frac{b_{\nu,\mathrm{ref}} \cdot S}{\left(1+Z\right)^{3-\alpha}}\left(\frac{L_\mathrm{p}}{\sin{\Theta} \cdot l_\mathrm{ref}}\right)^\zeta \leq b\ \vert\ L_\mathrm{p} = l_\mathrm{p},\ Z = z\right)\nonumber\\
    &= \mathbb{P}\left(\sin^{-\zeta}\Theta \cdot S \leq \frac{b}{b_{\nu,\mathrm{ref}}}\left(\frac{l_\mathrm{p}}{l_\mathrm{ref}}\right)^{-\zeta}\left(1+z\right)^{3 - \alpha}\right)\nonumber\\
    &= \int_0^\infty \mathbb{P}\left(\sin^{-\zeta}\Theta \leq \frac{b}{b_{\nu,\mathrm{ref}}}\frac{1}{s}\left(\frac{l_\mathrm{p}}{l_\mathrm{ref}}\right)^{-\zeta}\left(1+z\right)^{3-\alpha}\right) f_S\left(s\right) \mathrm{d}s\nonumber\\
    &= \int_0^\infty \mathbb{P}\left(\sin\Theta \leq \left(\frac{\tilde{s}}{s}\right)^{-\frac{1}{\zeta}}\right) f_S\left(s\right) \mathrm{d}s\nonumber\\
    &= \int_0^{\tilde{s}}f_S\left(s\right)\mathrm{d}s + \int_{\tilde{s}}^\infty 2F_\Theta\left(\arcsin{\left(\left(\frac{\tilde{s}}{s}\right)^{-\frac{1}{\zeta}}\right)}\right)f_S\left(s\right)\mathrm{d}s\nonumber\\
    &= \int_0^{\tilde{s}}f_S\left(s\right)\mathrm{d}s + \int_{\tilde{s}}^\infty f_S\left(s\right)\mathrm{d}s\nonumber\\
    &- \int_{\tilde{s}}^\infty\cos{\arcsin{\left(\left(\frac{\tilde{s}}{s}\right)^{-\frac{1}{\zeta}}\right)}} f_S\left(s\right)\mathrm{d}s\nonumber\\
    &= 1 - \int_{\tilde{s}}^\infty \sqrt{1 - \left(\frac{\tilde{s}}{s}\right)^{-\frac{2}{\zeta}}}f_S\left(s\right)\mathrm{d}s,
\end{align}
where
\begin{align}
    \tilde{s} = \tilde{s}\left(b\right) \coloneqq \frac{b}{b_{\nu,\mathrm{ref}}}\left(\frac{l_\mathrm{p}}{l_\mathrm{ref}}\right)^{-\zeta}\left(1+z\right)^{3-\alpha}.
\end{align}
The minimum value of $S$ for which an RG of projected length $l_\mathrm{p}$ at redshift $z$ is detectable, is $s_\mathrm{min}(l_\mathrm{p},z) \coloneqq \tilde{s}(b_{\nu,\mathrm{th}})$.
For brevity, we simply write $s_\mathrm{min}$ instead.
We find
\begin{align}
    p_\mathrm{obs}\left(l_\mathrm{p},z\right) =& \int_{s_\mathrm{min}}^\infty\sqrt{1 - \left(\frac{s_\mathrm{min}}{s}\right)^{-\frac{2}{\zeta}}}f_S\left(s\right) \mathrm{d}s.
\end{align}
The following approximation might facilitate the numerical evaluation of this integral.
We note that for $s \gg s_\mathrm{min}$, the square root factor in the integral approaches $1$.
Now split up the original integral in two parts, where $\eta$ governs the approximation's accuracy:
\begin{align}
    p_\mathrm{obs} &\approx \hat{p}_\mathrm{obs}\\
    &= \int_{s_\mathrm{min}}^{\eta s_\mathrm{min}}\sqrt{1 - \left(\frac{s_\mathrm{min}}{s}\right)^{-\frac{2}{\zeta}}}f_S\left(s\right) \mathrm{d}s + \int_{\eta s_\mathrm{min}}^\infty f_S\left(s\right)\mathrm{d}s\nonumber\\
    &= \int_{s_\mathrm{min}}^{\eta s_\mathrm{min}}\sqrt{1 - \left(\frac{s_\mathrm{min}}{s}\right)^{-\frac{2}{\zeta}}}f_S\left(s\right) \mathrm{d}s + 1 - F_S\left(\eta s_\mathrm{min}\right)\nonumber,
\end{align}
where we substitute numerically integrating to infinity for an evaluation of the CDF of the lognormally distributed RV $S$.
The approximation error is bounded from above:
\begin{align}
    \hat{p}_\mathrm{obs} - p_\mathrm{obs} &= \int_{\eta s_\mathrm{min}}^\infty \left(1 - \sqrt{1 - \left(\frac{s_\mathrm{min}}{s}\right)^{-\frac{2}{\zeta}}}\right) f_S\left(s\right)\mathrm{d}s\nonumber\\
    &< \left(1 - \sqrt{1-\left(\frac{s_\mathrm{min}}{\eta s_\mathrm{min}}\right)^{-\frac{2}{\zeta}}}\right)\int_{\eta s_\mathrm{min}}^\infty f_S\left(s\right)\mathrm{d}s\nonumber\\
    &= \left(1 - \sqrt{1 - \eta^\frac{2}{\zeta}}\right)\left(1-F_S\left(\eta s_\mathrm{min}\right)\right)\nonumber\\
    &< 1 - \sqrt{1-\eta^\frac{2}{\zeta}}.
\end{align}
Let us assume $\zeta = -2$.
For $\eta = 100$, the approximation error is at most $0.005$, and for $\eta = 1000$, the approximation error is at most $0.0005$.

\paragraph{Fanaroff--Riley class I}
The simplest correction in which FRI RGs retain a well-defined notion of length assumes a linearly decreasing surface brightness, from some value $b_\nu(0)$ at the core to zero at the RG's two endpoints.
For a symmetric FRI RG, the surface brightness profile along one of the jets is $b_\nu: \mathbb{R}_{\geq 0} \to \mathbb{R}_{\geq 0}$, and depends on the projected proper distance from the core $r_\mathrm{p}$ as
\begin{align}
    b_\nu(r_\mathrm{p}) = \begin{cases}
    b_\nu(0)\left(1 - \frac{2 r_\mathrm{p}}{l_\mathrm{p}}\right) & \text{if } 0 \leq r_\mathrm{p} < \frac{l_\mathrm{p}}{2}\\
    0 & \text{if } r_\mathrm{p} \geq \frac{l_\mathrm{p}}{2}.
    \end{cases}
\end{align}
Now we consider an FRI GRG, whose projected proper length $l_\mathrm{p} > l_\mathrm{p,GRG}$ would only be observed in full in the absence of noise.
In actual observations, this GRG is detected as a GRG if and only if
\begin{align}
    b_\nu\left(\frac{l_\mathrm{p,GRG}}{2}\right) > b_{\nu,\mathrm{th}},\ \ \ \text{or } b_\nu(0)\left(1 - \frac{l_\mathrm{p,GRG}}{l_\mathrm{p}}\right) > b_{\nu,\mathrm{th}}.
    \label{eq:surfaceBrightnessFRIGRGObservationCondition}
\end{align}
Under our assumption of a linear surface brightness profile, the mean surface brightness along the jet axis $\langle b_\nu \rangle = b_\nu\left(\frac{1}{2}\cdot\frac{l_\mathrm{p}}{2}\right)$.
As this must be half of the surface brightness at the core, 
\begin{align}
    \langle b_\nu \rangle = b_\nu\left(\frac{l_\mathrm{p}}{4}\right) = \frac{b_\nu(0)}{2}.
    \label{eq:surfaceBrightnessFRIGRGMean}
\end{align}
Combining Eqs.~\ref{eq:surfaceBrightnessFRIGRGObservationCondition} and \ref{eq:surfaceBrightnessFRIGRGMean}, we find that the GRG is detected as such if
\begin{align}
    \langle b_\nu \rangle > \frac{b_{\nu,\mathrm{th}}}{2\left(1 - \frac{l_\mathrm{p,GRG}}{l_\mathrm{p}}\right)}.
\end{align}
Now regarding $\langle b_\nu \rangle$ as an RV and recognising that it might behave exactly as in Eq.~\ref{eq:surfaceBrightnessRV}, we find that the surface brightness selection effect for FRI giants may be modelled as
\begin{align}
    p_\mathrm{obs}(l_\mathrm{p},z) = \mathbb{P}\left(B_\nu > \frac{b_{\nu,\mathrm{th}}}{2\left(1 - \frac{l_\mathrm{p,GRG}}{l_\mathrm{p}}\right)}\ \vert\ L_\mathrm{p} = l_\mathrm{p}, Z = z\right).
\end{align}
We see that the full formulaic structure of $p_\mathrm{obs}(l_\mathrm{p},z)$ is the same for FRI and FRII giants, except that for FRI giants a change
\begin{align}
    b_{\nu,\mathrm{th}} \to \frac{b_{\nu,\mathrm{th}}}{2\left(1 - \frac{l_\mathrm{p,GRG}}{l_\mathrm{p}}\right)}
\end{align}
is necessary.
There is no change for $l_\mathrm{p} = 2 l_\mathrm{p,GRG}$.

\subsection{GRG number density}
\label{ap:GRGComovingNumberDensity}
A statistic of major interest is the number density of giants in the contemporary Universe.
Let $n_\mathrm{GRG}$ be the comoving GRG number density, so that
\begin{align}
    n_\mathrm{GRG} = n\ \mathbb{P}\left(L_\mathrm{p} > l_\mathrm{p,GRG}\right) = n \left(1 - F_{L_\mathrm{p}}\left(l_\mathrm{p,GRG}\right)\right).
\label{eq:GRGComovingNumberDensityDefinition}
\end{align}
\begin{align}
    N_\mathrm{GRG,obs}\left(\Omega,z_\mathrm{max}\right) \coloneqq \int_{l_\mathrm{p,GRG}}^\infty \mathrm{d}N_{L_\mathrm{p,obs}}\left(l_\mathrm{p}, \Omega\right).
\end{align}
After invoking Eq.~\ref{eq:observedProjectedProperLengthNumberDensity} and isolating $n$, we obtain
\begin{align}
    n = \frac{\frac{4\pi}{\Omega} N_\mathrm{GRG,obs}\left(\Omega, z_\mathrm{max}\right)}{\int_{l_\mathrm{p,GRG}}^\infty\int_0^{z_\mathrm{max}} f_{L_\mathrm{p}}\left(l_\mathrm{p}\right)p_\mathrm{obs}\left(l_\mathrm{p},z\right) 4\pi r^2\left(z\right)\frac{\mathrm{d}r}{\mathrm{d}z}\ \mathrm{d}z\ \mathrm{d}l_\mathrm{p}}.
\label{eq:RGComovingNumberDensity}
\end{align}
Combining Eqs.~\ref{eq:CDFLengthProperProjectedRG}, \ref{eq:PDFLengthProperProjectedRG}, \ref{eq:GRGComovingNumberDensityDefinition}, and \ref{eq:RGComovingNumberDensity} for $l_\mathrm{p,GRG} > l_\mathrm{min}$, we arrive at
\begin{align}
    &n_\mathrm{GRG}\nonumber\\
    &= -\frac{l_\mathrm{p,GRG}^{\xi+1}}{\xi+1} \cdot \frac{\frac{4\pi}{\Omega}N_\mathrm{GRG,obs}\left(\Omega,z_\mathrm{max}\right)}{\int_{l_\mathrm{p,GRG}}^\infty l_\mathrm{p}^\xi \int_0^{z_\mathrm{max}} p_\mathrm{obs}\left(l_\mathrm{p},z\right) 4\pi r^2\left(z\right)\frac{\mathrm{d}r}{\mathrm{d}z}\ \mathrm{d}z\ \mathrm{d}l_\mathrm{p}}.
\end{align}
From observations, we know $\Omega$ and can --- for a given $z_\mathrm{max}$ --- simply count $N_\mathrm{GRG,obs}\left(\Omega,z_\mathrm{max}\right)$.
Moreover, we can fit $\xi$ and the parameters that occur in $p_\mathrm{obs}(l_\mathrm{p},z)$ (e.g. $\frac{1}{2}(\phi_\mathrm{max} - \phi_\mathrm{min})$, $b_{\nu,\mathrm{ref}}$, and $\sigma_\mathrm{ref}$) to the ECDF of $L_\mathrm{p,obs}\ \vert\ L_\mathrm{p,obs} > l_\mathrm{p,GRG}$.
We note that $n_\mathrm{GRG}$ does not depend on $l_\mathrm{min}$, which drops out through the division.
However, $n_\mathrm{GRG}$ does depend on cosmological parameters through the relation between cosmological redshift $z$ and radial comoving distance $r$.

\subsection{GRG lobe volume-filling fraction}
\label{ap:VFF}
Assuming self-similar growth, the combined proper volume $V$ of an RG's lobes and its intrinsic proper length $l$ obey $V \propto l^3$.
The constant of proportionality varies per RG and depends on the shape of the lobes; we treat it as an RV $\Upsilon \coloneqq \frac{V}{L^3}$.
Then the proper VFF of GRG lobes $\mathrm{VFF}_\mathrm{GRG}(z) = \mathrm{VFF}_\mathrm{GRG}(z = 0) \cdot (1 + z)^3$, where
\begin{align}
\mathrm{VFF}_\mathrm{GRG}(z=0) \coloneqq&\ n_\mathrm{GRG} \cdot \mathbb{E}[V\ \vert\ L_\mathrm{p} > l_\mathrm{p,GRG}]\nonumber\\
=&\ n_\mathrm{GRG} \cdot \mathbb{E}[\frac{V}{L^3} \cdot L^3\ \vert\ L_\mathrm{p} > l_\mathrm{p,GRG}]\nonumber\\
=&\ n_\mathrm{GRG} \cdot \mathbb{E}[\Upsilon \cdot L^3\ \vert\ L_\mathrm{p} > l_\mathrm{p,GRG}].
\end{align}
Assuming that RGs grow self-similarly, so that shape does not reveal length, $\Upsilon$ and $L^3$ are conditionally independent given $L_\mathrm{p} > l_\mathrm{p,GRG}$: $\Upsilon \Perp L^3\ \vert\ L_\mathrm{p} > l_\mathrm{p,GRG}$.
As a result,
\begin{align}
\mathrm{VFF}_\mathrm{GRG}(z=0) &= n_\mathrm{GRG} \cdot \mathbb{E}[\Upsilon\ \vert\ L_\mathrm{p} > l_\mathrm{p,GRG}] \cdot \mathbb{E}[L^3\ \vert\ L_\mathrm{p} > l_\mathrm{p,GRG}]\nonumber\\
&= n_\mathrm{GRG} \cdot \mathbb{E}[\Upsilon] \cdot \mathbb{E}[L^3\ \vert\ L_\mathrm{p} > l_\mathrm{p,GRG}].
\end{align}
To obtain this last line, we once more exploit self-similarity: $\Upsilon\ \vert\ (L_\mathrm{p} > l_\mathrm{p,GRG}) = \Upsilon$.
We can approximate $\mathbb{E}[\Upsilon]$ by taking the mean of some $\frac{V}{l^3}$ deduced from observations.
A technical complication arises from the fact that $\mathbb{E}[L^3\ \vert\ L_\mathrm{p} > l_\mathrm{p,GRG}]$ does not exist for $\xi \geq -4$ under our model.
This is an artefact of the Pareto distribution assumption for $L$, which unrealistically features support over an infinitely long part of the real line: $\{l \in \mathbb{R}\ \vert\ l > l_\mathrm{min}\}$.
This causes the expectation value integral to diverge for $\xi \geq -4$.
An approximation to $\mathrm{VFF}_\mathrm{GRG}(z=0)$ that works for all $\xi$ is the lower bound
\begin{align}
    \mathrm{VFF}_\mathrm{GRG}(z=0) > n_\mathrm{GRG} \cdot \mathbb{E}[\Upsilon] \cdot \mathbb{E}^3[L\ \vert\ L_\mathrm{p} > l_\mathrm{p,GRG}],
\end{align}
which follows from Jensen's inequality.
Here
\begin{align}
    \mathbb{E}[L\ \vert\ L_\mathrm{p}>l_\mathrm{p,GRG}] = l_\mathrm{p,GRG} \frac{\Gamma\left(-\frac{\xi}{2}-1\right)\Gamma\left(-\frac{\xi}{2}+1\right)}{\Gamma\left(-\frac{\xi}{2}-\frac{1}{2}\right)\Gamma\left(-\frac{\xi}{2}+\frac{1}{2}\right)}.
\end{align}
Alternative approximation formulae, which use $\Upsilon_\mathrm{p} \coloneqq \frac{V}{L_\mathrm{p}^3}$, are
\begin{align}
    \mathrm{VFF}_\mathrm{GRG}(z=0) > n_\mathrm{GRG} \cdot \mathbb{E}[\Upsilon_\mathrm{p}] \cdot \mathbb{E}^3[L_\mathrm{p}\ \vert\ L_\mathrm{p} > l_\mathrm{p,GRG}]
\label{eq:VFF1}
\end{align}
and
\begin{align}
\mathrm{VFF}_\mathrm{GRG}(z=0) > n_\mathrm{GRG} \cdot \mathbb{E}[\Upsilon_\mathrm{p}] \cdot m_{L_\mathrm{p}^3\ \vert\ L_\mathrm{p} > l_\mathrm{p,GRG}},
\label{eq:VFF2}
\end{align}
where
\begin{align}
m_{L_\mathrm{p}^3\ \vert\ L_\mathrm{p} > l_\mathrm{p,GRG}} = l_\mathrm{p,GRG}^3 \cdot 2^{\frac{3}{-(\xi+1)}}
\end{align}
is the median of the cubed projected proper length for giants.
An advantage of these latter expressions is that there are more data available to estimate $\mathbb{E}[\Upsilon_\mathrm{p}]$ than there are to estimate $\mathbb{E}[\Upsilon]$.

\subsection{Unification model constraints from quasar and non-quasar giants}
\label{ap:unificationModel}
The quasar GRG probability $p_\mathrm{Q}$ is
\begin{align}
    p_\mathrm{Q} \coloneqq& \frac{\mathbb{P}\left(L_\mathrm{p,obs} \geq l_\mathrm{p,GRG},\ \sin{\Theta} \leq \sin{\theta_\mathrm{max}}\right)}{\mathbb{P}\left(L_\mathrm{p,obs} \geq l_\mathrm{p,GRG}\right)}\nonumber\\
    =& \mathbb{P}\left(\sin{\Theta} \leq \sin{\theta_\mathrm{max}}\ \vert\ L_\mathrm{p,obs} \geq l_\mathrm{p,GRG}\right)\nonumber\\
    =& \int_0^{\sin{\theta_\mathrm{max}}}\frac{\mathbb{P}\left(L_\mathrm{p,obs} \geq l_\mathrm{p,GRG}\ \vert\ \sin{\Theta} = x\right) f_{\sin{\Theta}}(x)\ \mathrm{d}x}{\mathbb{P}\left(L_\mathrm{p,obs} \geq l_\mathrm{p,GRG}\right)}\nonumber\\
    =& \int_0^{\sin{\theta_\mathrm{max}}}\frac{\mathbb{P}\left(L_\mathrm{p} \geq l_\mathrm{p,GRG}\ \vert\ \sin{\Theta} = x\right) f_{\sin{\Theta}}(x)\ \mathrm{d}x}{\mathbb{P}\left(L_\mathrm{p} \geq l_\mathrm{p,GRG}\right)}\nonumber\\
    =& \int_0^{\sin{\theta_\mathrm{max}}}\frac{\left(1 - F_L\left(\frac{l_\mathrm{p,GRG}}{x}\right)\right) f_{\sin{\Theta}}(x)\ \mathrm{d}x}{1 - F_{L_\mathrm{p}}(l_\mathrm{p,GRG})}.
\end{align}
Now using Eqs.~\ref{eq:CDFLengthProperIntrinsicRG} and \ref{eq:CDFLengthProperProjectedRG} and $f_{\sin{\Theta}}(x) = \frac{x}{\sqrt{1-x^2}}$ over the domain of integration,
\begin{align}
    p_\mathrm{Q} = \frac{4 \Gamma\left(-\frac{\xi}{2}+1\right)}{-\left(\xi+1\right)\sqrt{\pi}\ \Gamma\left(-\frac{\xi}{2}-\frac{1}{2}\right)}\int_0^{\sin{\theta_\mathrm{max}}}\frac{x^{-\xi}\ \mathrm{d}x}{\sqrt{1-x^2}}.
\end{align}

\subsection{Extreme giants in a sample}
\label{ap:extremeGRGs}
An interesting feature of the model is its ability to predict the occurrence of giants with extreme projected proper lengths in a GRG sample of, say, size $N$.
Now consider some $l_\mathrm{p} > l_\mathrm{p, GRG}$ --- what is the probability $p_{>l_\mathrm{p}}$ that an observed GRG will have a projected proper length exceeding $l_\mathrm{p}$?
Proceeding as in the derivation of Eq.~\ref{eq:CDFLengthProperProjectedGRG}, we find
\begin{align}
    p_{>l_\mathrm{p}}\left(l_\mathrm{p}\right) \coloneqq& \mathbb{P}\left(L_\mathrm{p,obs} > l_\mathrm{p}\ \vert\ L_\mathrm{p,obs} > l_\mathrm{p, GRG}\right)\nonumber\\
    =& \frac{1 - F_{L_\mathrm{p,obs}}\left(l_\mathrm{p}\right)}{1 - F_{L_\mathrm{p,obs}}\left(l_\mathrm{p, GRG}\right)}.
\end{align}
In the absence of selection effects, $p_{>l_\mathrm{p}}\left(l_\mathrm{p}\right)$ is given by Eq.~\ref{eq:SFLengthProperProjectedGRG}.
The number of giants with extreme projected proper lengths $N_{>l_\mathrm{p}} \sim \mathrm{Binom}(N, p_{>l_\mathrm{p}}(l_\mathrm{p}))$.

Interesting questions can be answered readily.
For example, the probability that the sample contains at least one GRG with projected proper length $l_\mathrm{p}$ or larger, is
\begin{align}
    \mathbb{P}\left(N_{>l_\mathrm{p}} \geq 1\right) = 1 - \mathbb{P}\left(N_{>l_\mathrm{p}} = 0\right) = 1 - \left(1 - p_{>l_\mathrm{p}}\left(l_\mathrm{p}\right)\right)^N.
\end{align}


\section{Additional images}
In this appendix, as in Fig.~\ref{fig:LoTSSDR2GRGs1}, we show newly discovered giants.
These cover the projected length range $l_\mathrm{p} \in [0.7\ \mathrm{Mpc}, 4\ \mathrm{Mpc})$.
\begin{figure*}[p]
    \centering
    \begin{subfigure}{\columnwidth}
    \includegraphics[width=\columnwidth]{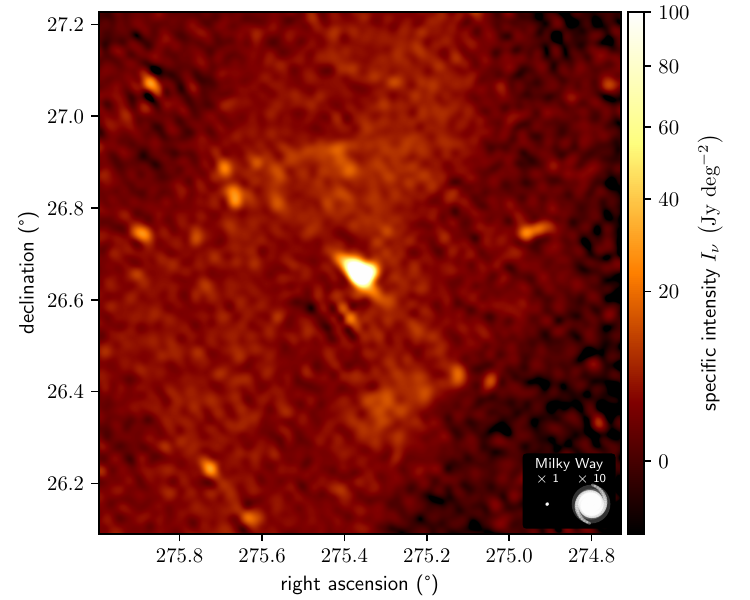}
    \end{subfigure}
    \begin{subfigure}{\columnwidth}
    \includegraphics[width=\columnwidth]{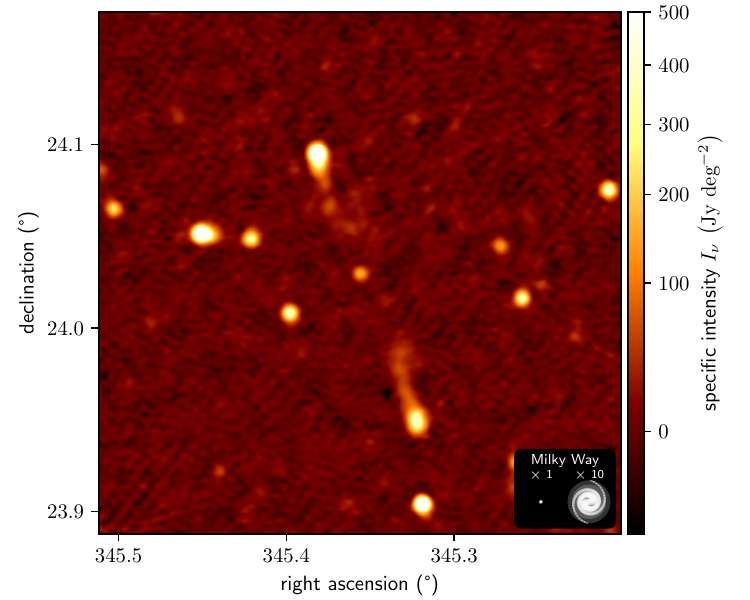}
    \end{subfigure}
    \begin{subfigure}{\columnwidth}
    \includegraphics[width=\columnwidth]{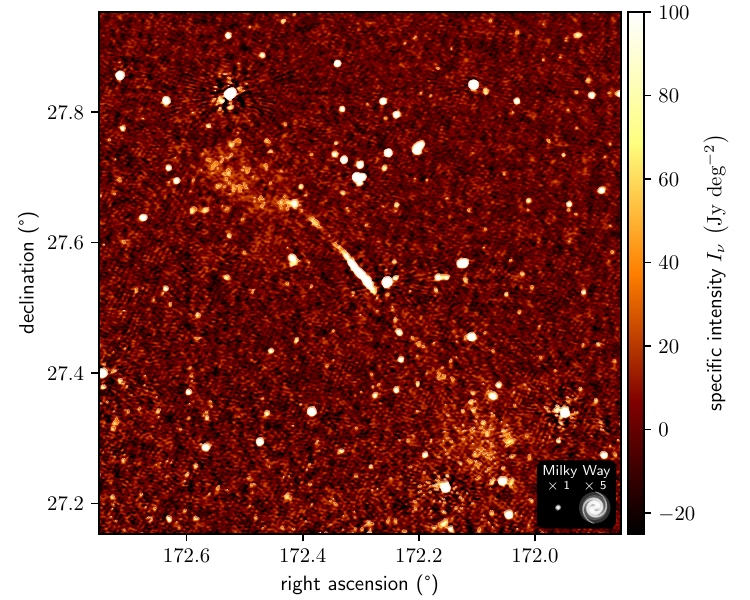}
    \end{subfigure}
    \begin{subfigure}{\columnwidth}
    \includegraphics[width=\columnwidth]{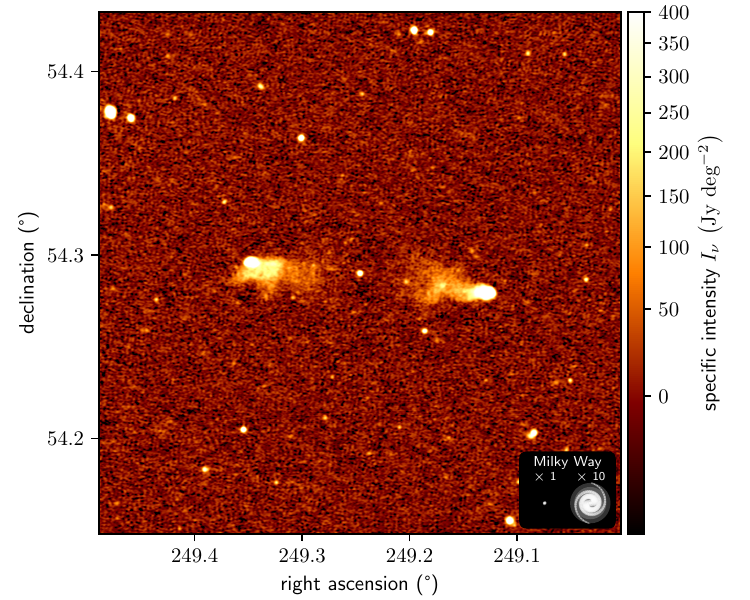}
    \end{subfigure}
    \begin{subfigure}{\columnwidth}
    \includegraphics[width=\columnwidth]{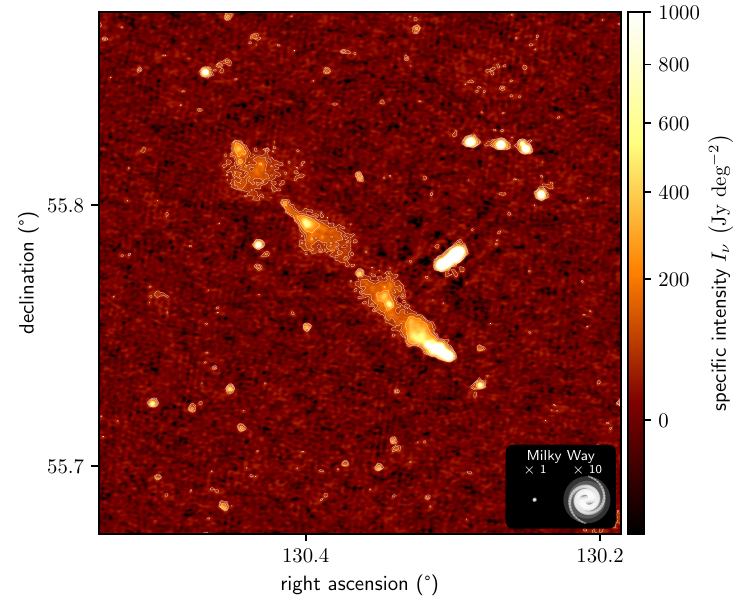}
    \end{subfigure}
    \begin{subfigure}{\columnwidth}
    \includegraphics[width=\columnwidth]{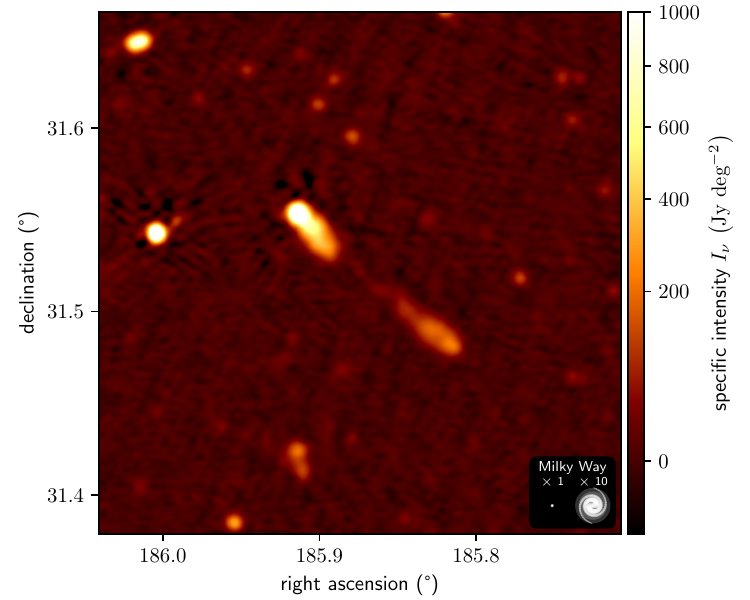}
    \end{subfigure}
    \caption{
    Details of the LoTSS DR2--estimated specific intensity function $I_\nu\left(\hat{r}\right)$ at central observing frequency $\nu_\mathrm{obs} = 144\ \mathrm{MHz}$ and resolutions $\theta_\mathrm{FWHM} \in \{6'',\ 20''\,\ 90''\}$, centred around the hosts of newly discovered giants.
    Row-wise from left to right, from top to bottom, the projected proper length $l_\mathrm{p}$ is $3.9\ \mathrm{Mpc}$, $3.5\ \mathrm{Mpc}$, $3.3\ \mathrm{Mpc}$, $3.3\ \mathrm{Mpc}$, $3.3\ \mathrm{Mpc}$, and $3.2\ \mathrm{Mpc}$; in the same order, $\theta_\mathrm{FWHM}$ is $90''$, $20''$, $20''$, $6''$, $6''$, and $20''$.
    The giants in the top-left and middle-left panels appear larger in the sky than the Moon.
    Contours signify 3, 5, and 10 sigma-clipped standard deviations above the sigma-clipped median.
    For scale, we show the stellar Milky Way disk (with a diameter of 50 kpc) generated using the \citet{Ringermacher12009} formula, alongside a 5 or 10 times inflated version.
    }
    \label{fig:LoTSSDR2GRGs2}
\end{figure*}\noindent
\begin{figure*}[p]
    \centering
    \begin{subfigure}{\columnwidth}
    \includegraphics[width=\columnwidth]{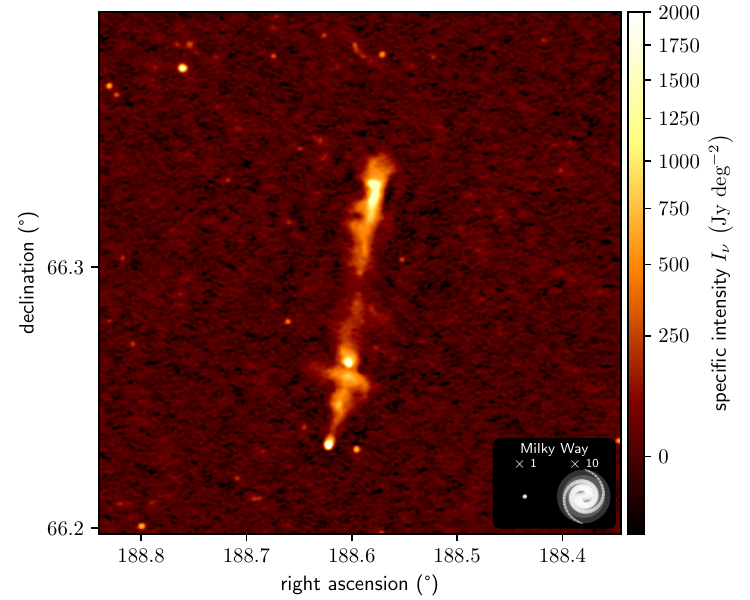}
    \end{subfigure}
    \begin{subfigure}{\columnwidth}
    \includegraphics[width=\columnwidth]{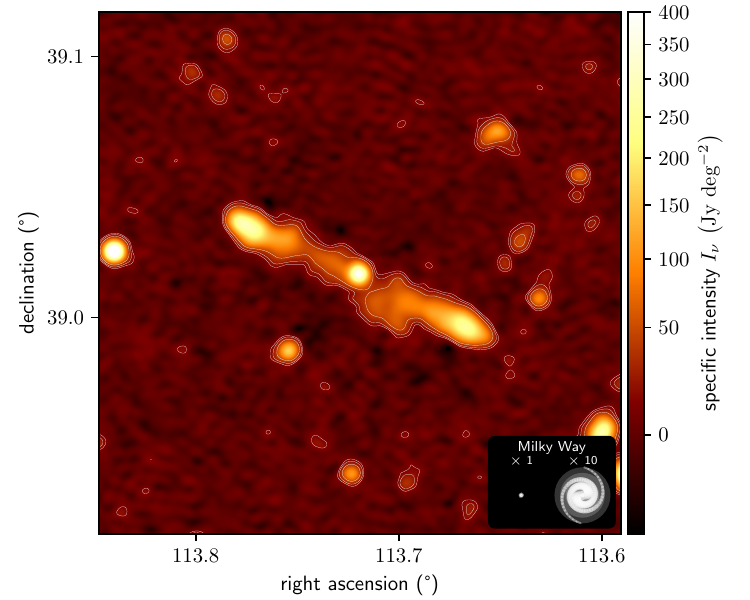}
    \end{subfigure}
    \begin{subfigure}{\columnwidth}
    \includegraphics[width=\columnwidth]{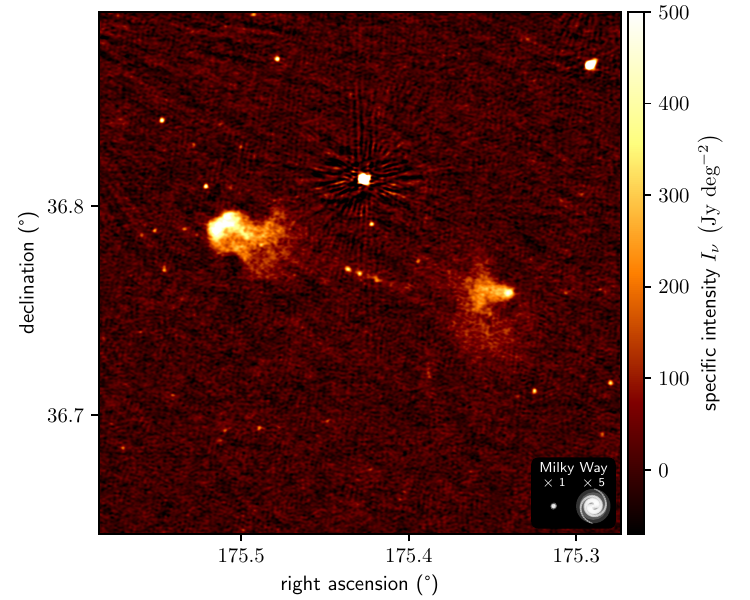}
    \end{subfigure}
    \begin{subfigure}{\columnwidth}
    \includegraphics[width=\columnwidth]{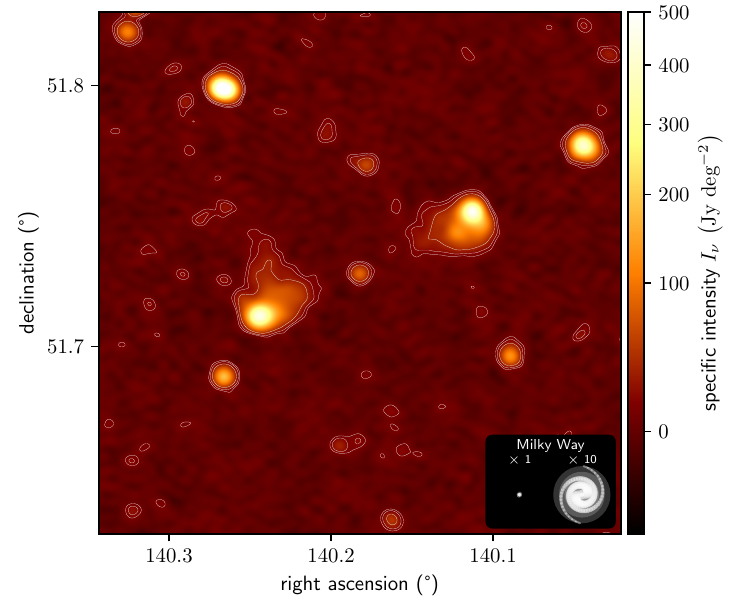}
    \end{subfigure}
    \begin{subfigure}{\columnwidth}
    \includegraphics[width=\columnwidth]{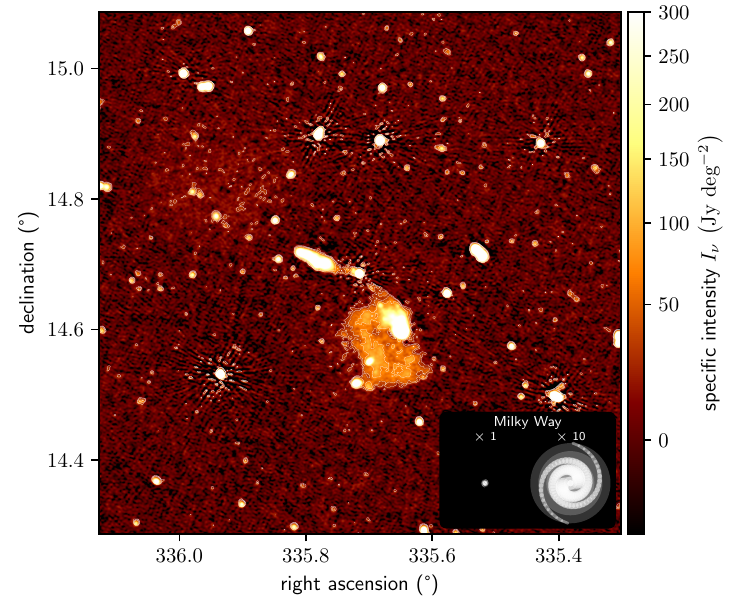}
    \end{subfigure}
    \begin{subfigure}{\columnwidth}
    \includegraphics[width=\columnwidth]{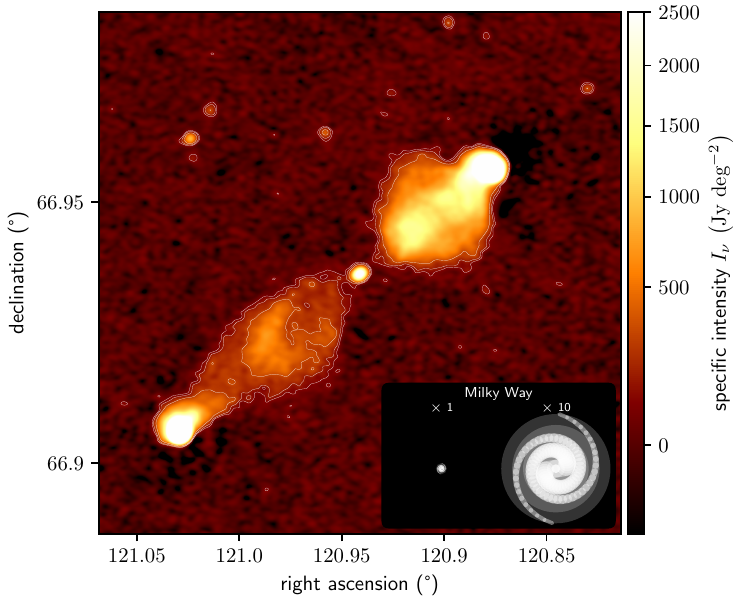}
    \end{subfigure}
    \caption{
    Details of the LoTSS DR2--estimated specific intensity function $I_\nu\left(\hat{r}\right)$ at central observing frequency $\nu_\mathrm{obs} = 144\ \mathrm{MHz}$ and resolutions $\theta_\mathrm{FWHM} \in \{6'',\ 20''\}$, centred around the hosts of newly discovered giants.
    Row-wise from left to right, from top to bottom, the projected proper length $l_\mathrm{p}$ is $2.8\ \mathrm{Mpc}$, $2.6\ \mathrm{Mpc}$, $2.6\ \mathrm{Mpc}$, $2.2\ \mathrm{Mpc}$, $2.1\ \mathrm{Mpc}$, and $2.1\ \mathrm{Mpc}$; in the same order, $\theta_\mathrm{FWHM}$ is $6''$, $20''$, $6''$, $20''$, $20''$, and $6''$.
    The GRG in the bottom-left panel appears larger in the sky than the Moon.
    Contours signify 3, 5, and 10 sigma-clipped standard deviations above the sigma-clipped median.
    For scale, we show the stellar Milky Way disk (with a diameter of 50 kpc) generated using the \citet{Ringermacher12009} formula, alongside a 5 or 10 times inflated version.
    }
    \label{fig:LoTSSDR2GRGsMix1}
\end{figure*}\noindent
\begin{figure*}[p]
    \centering
    \begin{subfigure}{\columnwidth}
    \includegraphics[width=\columnwidth]{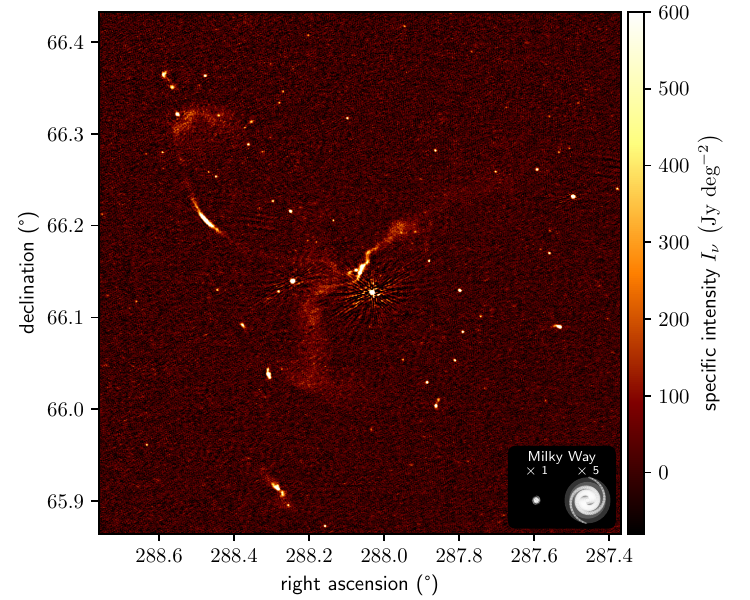}
    \end{subfigure}
    \begin{subfigure}{\columnwidth}
    \includegraphics[width=\columnwidth]{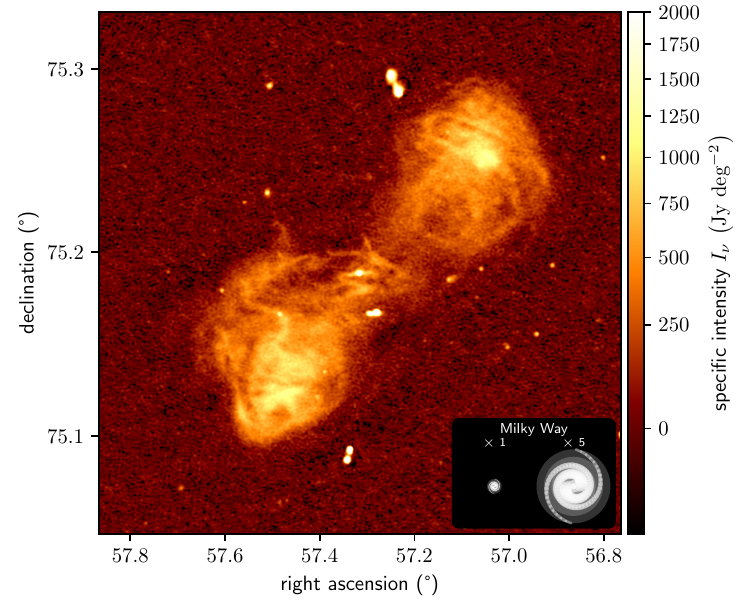}
    \end{subfigure}
    \begin{subfigure}{\columnwidth}
    \includegraphics[width=\columnwidth]{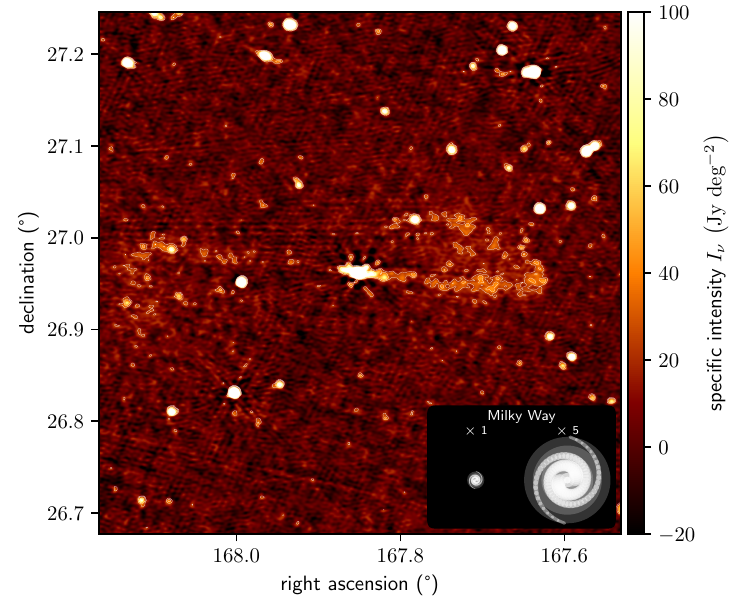}
    \end{subfigure}
    \begin{subfigure}{\columnwidth}
    \includegraphics[width=\columnwidth]{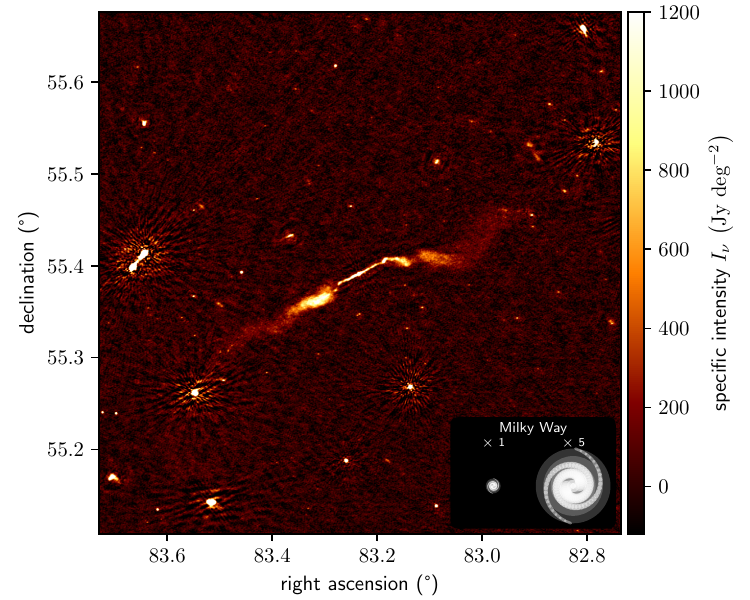}
    \end{subfigure}
    \begin{subfigure}{\columnwidth}
    \includegraphics[width=\columnwidth]{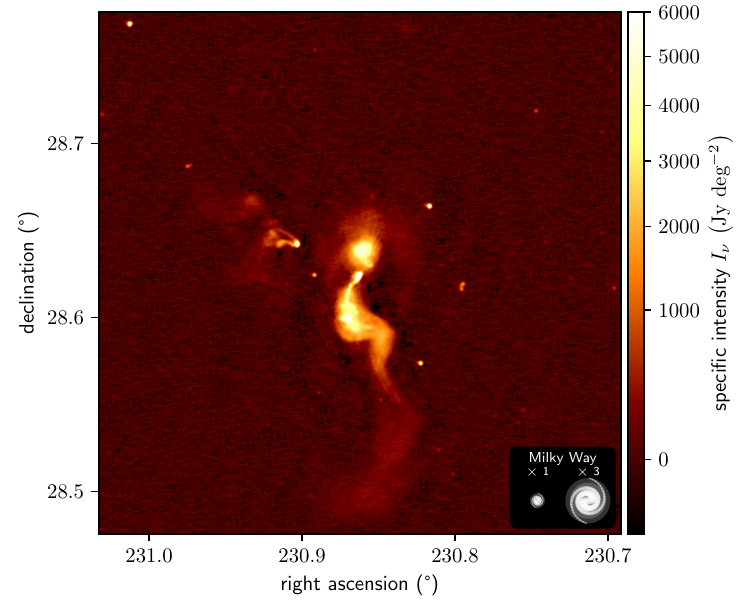}
    \end{subfigure}
    \begin{subfigure}{\columnwidth}
    \includegraphics[width=\columnwidth]{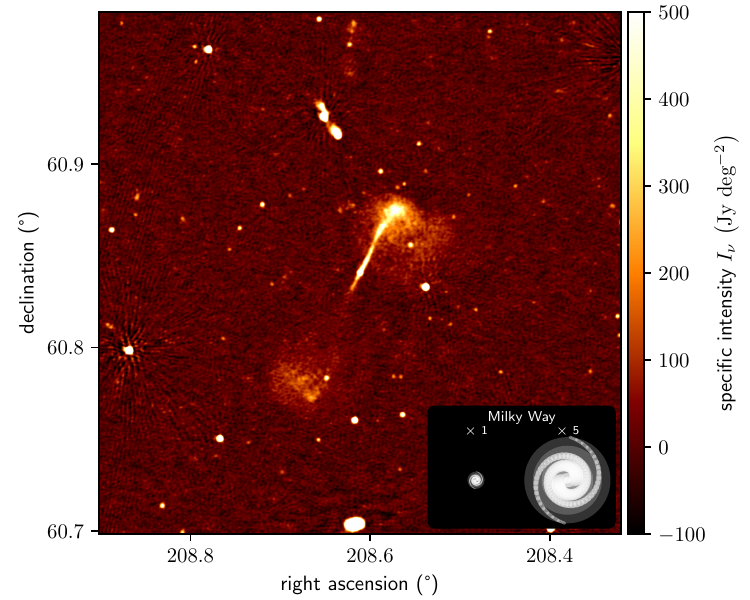}
    \end{subfigure}
    \caption{
    Details of the LoTSS DR2--estimated specific intensity function $I_\nu\left(\hat{r}\right)$ at central observing frequency $\nu_\mathrm{obs} = 144\ \mathrm{MHz}$ and resolutions $\theta_\mathrm{FWHM} \in \{6'',\ 20''\}$, centred around the hosts of newly discovered giants.
    Row-wise from left to right, from top to bottom, the projected proper length $l_\mathrm{p}$ is $1.6\ \mathrm{Mpc}$, $1.5\ \mathrm{Mpc}$, $1.3\ \mathrm{Mpc}$, $1.2\ \mathrm{Mpc}$, $1.1\ \mathrm{Mpc}$, and $0.7\ \mathrm{Mpc}$; in the same order, $\theta_\mathrm{FWHM}$ is $6''$, $6''$, $20''$, $6''$, $6''$, and $6''$.
    Contours signify 3, 5, and 10 sigma-clipped standard deviations above the sigma-clipped median.
    For scale, we show the stellar Milky Way disk (with a diameter of 50 kpc) generated using the \citet{Ringermacher12009} formula, alongside a 3 or 5 times inflated version.
    }
    \label{fig:LoTSSDR2GRGsMix2}
\end{figure*}\noindent

\section{Stellar and supermassive black hole masses}
\label{ap:masses}
\begin{figure}
    \centering
    \begin{subfigure}{\columnwidth}
    \includegraphics[width=\columnwidth]{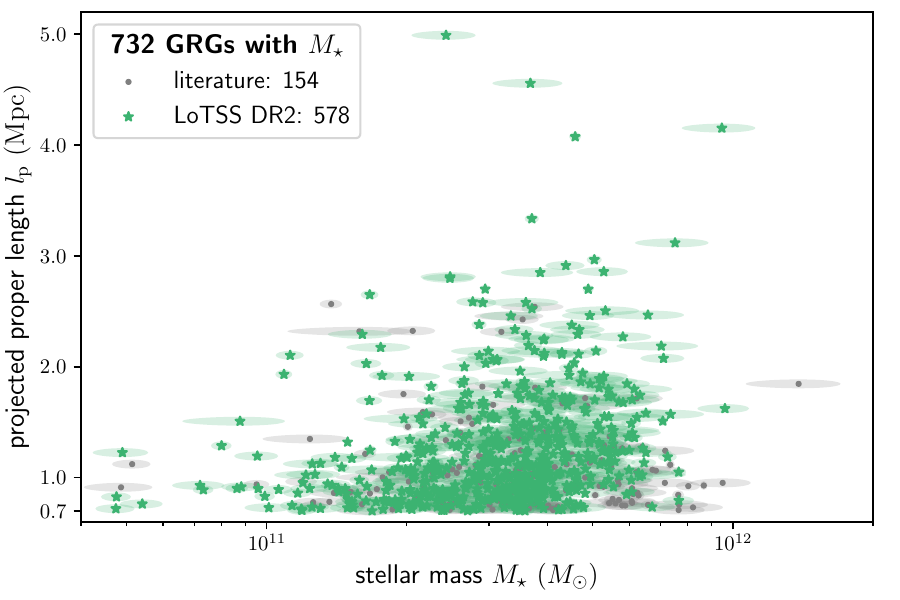}
    \end{subfigure}
    \begin{subfigure}{\columnwidth}
    \includegraphics[width=\columnwidth]{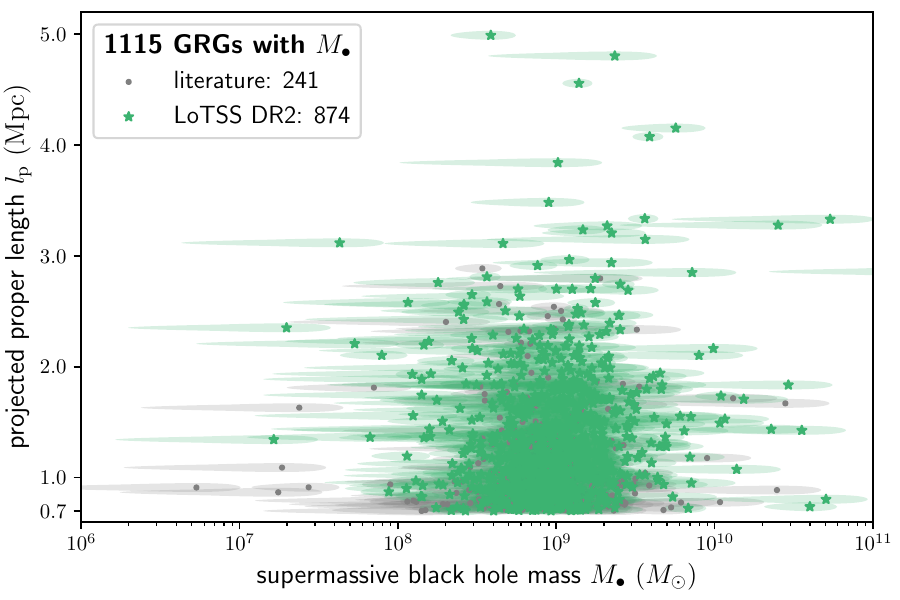}
    \end{subfigure}
    \caption{
    Observed relations between host stellar mass $M_\star$ and GRG projected length $l_\mathrm{p}$ \textit{(top)} and between host SMBH mass $M_\bullet$ and GRG projected length $l_\mathrm{p}$ \textit{(bottom)}.
    Our LoTSS DR2 sample confirms that luminous giants typically have $M_\star \in 10^{11}$--$10^{12}\ M_\odot$ and $M_\bullet \in 10^8$--$10^{10}\ M_\odot$.
    The sample effects an almost fivefold increase in the number of giants with SDSS-derived host stellar masses and SMBH masses.
    We do not show or count giants for which only a nearest host candidate could be determined.
    }
    \label{fig:stellarMass}
\end{figure}

In Fig.~\ref{fig:stellarMass}, we present the SDSS-derived relations between host stellar mass and projected proper length, and between host supermassive black hole mass and projected proper length, for all giants in our final catalogue.
We obtained the data as in Section~3.7 of \citet{Oei12022Alcyoneus}.

\section{Surface brightness prior}
\label{ap:surfaceBrightnessPrior}
Consider a radio galaxy at cosmological redshift $z$ of intrinsic proper length $l$ bounded by spherical lobes of radius $R$ and spectral index $\alpha$.
If a fraction $f_l$ of the radio galaxy's central axis lies within the lobes, then
\begin{align}
R = \frac{1}{4} f_l \cdot l.
\end{align}
If a fraction $f_{L_\nu}$ of the radio galaxy's total luminosity density $L_\nu$ comes from the lobes, then the monochromatic emission coefficient
\begin{align}
    j_\nu = \frac{L_\nu}{4\pi}\frac{f_{L_\nu}}{2}\frac{1}{V},
\end{align}
where the lobe volume $V = \frac{4}{3}\pi R^3$.
A formula of practical value features projected proper length $l_\mathrm{p}$ instead of $l$.
Given the approximate nature of our approach, we therefore simply assume $l \approx \mathbb{E}[D](\eta(f_l)) \cdot l_\mathrm{p}$, with the deprojection factor expectation $\mathbb{E}[D]$ given in Eq.~\ref{eq:deprojectionFactorLobesExpectation}; $\eta(f_l) = \frac{f_l}{2 - f_l}$.
The maximum surface brightness of the lobes as seen by an observer is
\begin{align}
    b_{\nu,\mathrm{max}} = \frac{3 f_{L_\nu} \cdot L_\nu}{\pi^2 \cdot \mathbb{E}[D](\eta(f_l)) \cdot f_l^2 \cdot l_\mathrm{p}^2 \cdot (1 + z)^{3-\alpha}},
\end{align}
valid for the line of sight that pierces through a lobe along a diameter.
The average line of sight length $d$ within a lobe is smaller than $d_\mathrm{max} = 2R$, though, and given by
\begin{align}
    \langle d \rangle \coloneqq \frac{\int_0^R 2\sqrt{R^2 - x^2} 2\pi x\ \mathrm{d}x}{\pi R^2} = \frac{4}{3}R.
\end{align}
Therefore, the mean surface brightness of the lobes as seen by an observer is
\begin{align}
    \langle b_\nu \rangle = \frac{\langle d \rangle}{d_\mathrm{max}} b_{\nu,\mathrm{max}} = \frac{2}{3} b_{\nu,\mathrm{max}}.
\end{align}

\section{Likelihood function}
\begin{figure*}[h!]
\centering
\includegraphics[width=\textwidth]{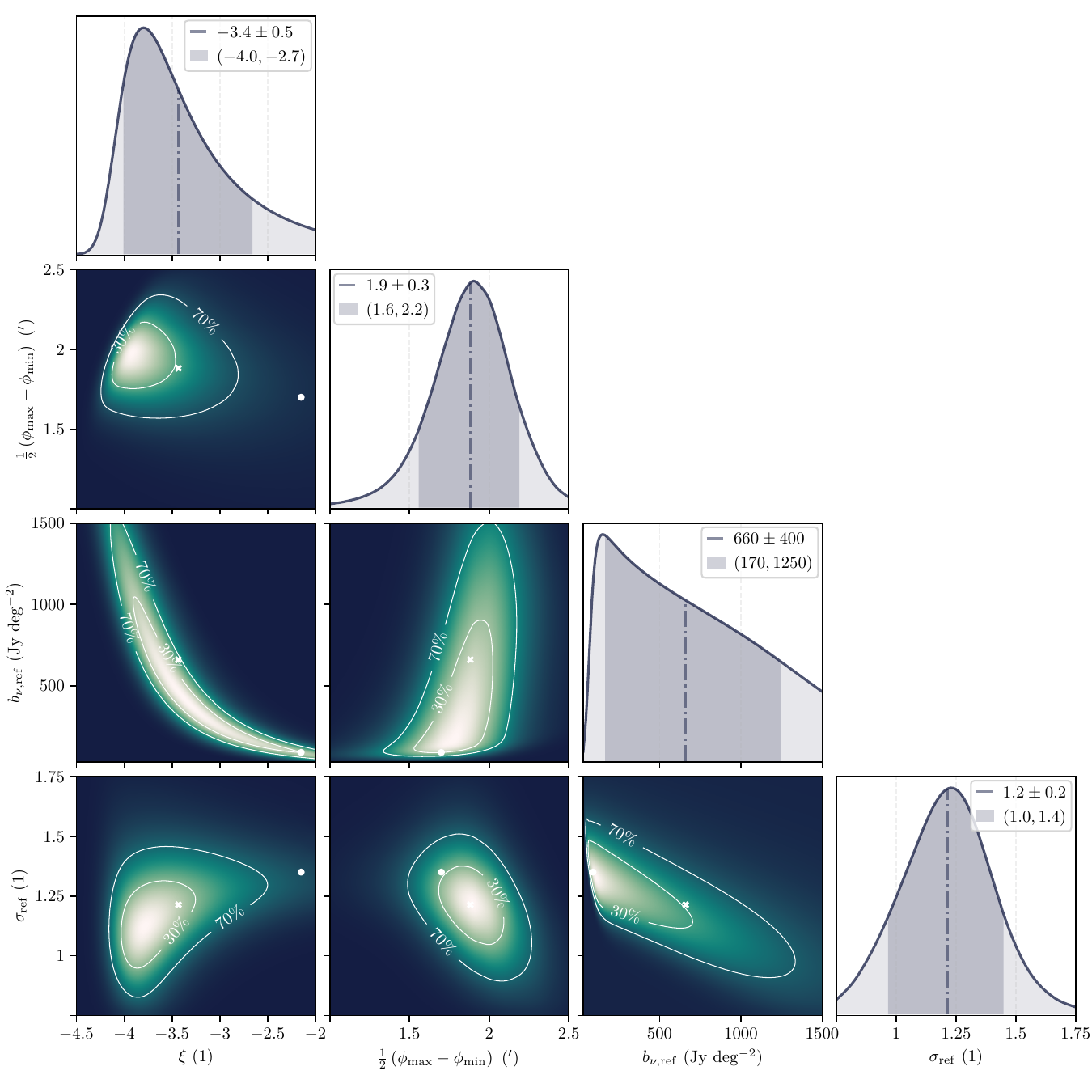}
\caption{
Joint likelihood function over $\xi$ --- the parameter of interest --- and $\frac{1}{2}(\phi_\mathrm{max} - \phi_\mathrm{min})$, $b_{\nu,\mathrm{ref}}$ and $\sigma_\mathrm{ref}$ --- the selection effect parameters, based on 1473 projected lengths of LoTSS DR2 giants up to $z_\mathrm{max} = 0.5$.
    We show all two-parameter marginals of the likelihood function, with contours enclosing $30\%$ and $70\%$ of total probability.
    We mark the maximum likelihood parameters (white circle) and the likelihood mean parameters (white cross).
    The single-parameter marginals again show the estimated mean, now marked by a vertical line, alongside shaded median-centred $80\%$ credible intervals.
    The likelihood function is the posterior for a uniform prior.
    To compare the likelihood function to the posterior actually chosen, see Fig.~\ref{fig:posterior}.
    }
\label{fig:likelihood}
\end{figure*}

In Table~\ref{tab:likelihood}, we present maximum likelihood and likelihood mean and standard deviation estimates for the inference described in Section~\ref{sec:inference}; one may compare the results to those in Table~\ref{tab:posterior}.
In Fig.~\ref{fig:likelihood}, we visually summarise the likelihood function; one may compare to Fig.~\ref{fig:posterior}.

The strong degeneracy between $\xi$ and $b_{\nu,\mathrm{ref}}$, directly apparent from the central two-parameter marginal in the leftmost column of Fig.~\ref{fig:likelihood}, translates to a ridge of essentially constant likelihood that extends from $\xi \approx -4$ to $\xi \approx -2$.
Compared to a non-degenerate case, this makes the maximum likelihood parameters both intrinsically less meaningful and more prone to numerical approximation errors.
\begin{center}
\captionof{table}{
Maximum likelihood estimate (MLE) and likelihood mean and standard deviation (SD) estimates of the free parameters in intrinsic GRG length distribution inference.\protect\footnotemark}
$z_\mathrm{max} = 0.5$:
\vspace{1mm}\\
\begin{tabular}{c c c}
\hline
parameter & MLE & likelihood mean and SD\\
 [3pt] \hline\arrayrulecolor{lightgray}
$\xi$ & $-2.15$ & $-3.4 \pm 0.5$\\
\hline
$\frac{1}{2}(\phi_\mathrm{max} - \phi_\mathrm{min})$ & $1.7'$ & $1.9\pm 0.3'$\\
\hline
$b_{\nu,\mathrm{ref}}$ & $90\ \mathrm{Jy\ deg^{-2}}$ & $660 \pm 400\ \mathrm{Jy\ deg^{-2}}$\\
\hline
$\sigma_\mathrm{ref}$ & $1.35$ & $1.2 \pm 0.2$
\end{tabular}
\vspace{2mm}\\
$z_\mathrm{max} = 0.25$:
\vspace{1mm}\\
\begin{tabular}{c c c}
\hline
parameter & MLE & likelihood mean and SD\\
 [3pt] \hline\arrayrulecolor{lightgray}
$\xi$ & $-2.2$ & $-3.5 \pm 0.4$\\
\hline
$\frac{1}{2}(\phi_\mathrm{max} - \phi_\mathrm{min})$ & $2.25'$ & $1.7\pm 0.3'$\\
\hline
$b_{\nu,\mathrm{ref}}$ & $150\ \mathrm{Jy\ deg^{-2}}$ & $1050 \pm 560\ \mathrm{Jy\ deg^{-2}}$\\
\hline
$\sigma_\mathrm{ref}$ & $1.25$ & $1.4 \pm 0.4$
\end{tabular}
\label{tab:likelihood}
\end{center}
\footnotetext{
The model assumes $\xi$ is constant for $z \in [0, z_\mathrm{max}]$.
We determined the likelihood function twice: for $z_\mathrm{max} = 0.5$, using 1473 giants, and for $z_\mathrm{max} = 0.25$, using 811 giants.
We caution that, in this case, the MLE parameters are a poor measure of central tendency.
}

\section{Properties of newly discovered giants}
\label{sec:supplementaryMaterialData}
Table~\ref{tab:GRGProperties} provides properties of the 50 projectively longest giants discovered during this work's LoTSS DR2 search cam\-paign.
We share these data, alongside those for the other 2010 (98\%) giants in our sample, in Flexible Image Transport System (FITS) format through the Centre de Donn\'ees astronomiques de Strasbourg (CDS).
One can either use anonymous File Transfer Protocol (FTP) to \texttt{\href{ftp://cdsarc.cds.unistra.fr}{ftp://cdsarc.cds.unistra.fr}} (130.79.128.5) or visit \texttt{\href{https://cdsarc.cds.unistra.fr/cgi-bin/qcat?J/A+A/}{https://cdsarc.cds.unistra.fr/cgi-bin/qcat?J/A+A/}}.

For our final catalogue, which also includes literature giants, please contact the authors.

\end{appendix}
\end{document}